\documentclass[aps, prr, twocolumn,superscriptaddress,floatfix]{revtex4-2}
\usepackage{graphicx,amsfonts,amssymb,amsmath,mathrsfs,hyperref,bm,dsfont,mathtools}
\usepackage{stackengine}
\usepackage{relsize}
\usepackage{comment}
\usepackage{physics}
\usepackage{ulem}

\usepackage[utf8]{inputenc}
\usepackage[T1]{fontenc}
\usepackage[english]{babel}
\usepackage{todonotes}
\usepackage{braket}
\usepackage[export]{adjustbox}

\newif\ifhyper
\hypertrue
\ifhyper
\hypersetup{
	citecolor = {red},
	colorlinks = {true}, % false
	linkcolor = {blue},
	urlcolor = {blue} % magenta
}
\fi

\def\be{\begin{equation}}
	\def\ee{\end{equation}}
\def\bea{\begin{eqnarray}}
	\def\eea{\end{eqnarray}}

\newcommand{\jfull}{$J_1$-$J_2$-$J_3$ }
\newcommand{\jtwo}{$J_1$-$J_2$ }
\newcommand{\jthree}{$J_1$-$J_3$ }

\newcommand{\resonatingplaquette}[1][20pt]{\hspace{0.05cm}\includegraphics[height=#1,valign=c]{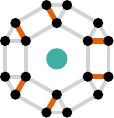}\hspace{0.05cm}}

\newcommand{\resonatingplaquettetwo}[1][20pt]{\hspace{0.05cm}\includegraphics[height=#1,valign=c]{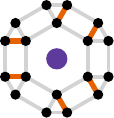}\hspace{0.05cm}}

\newcommand{\plaquettetwo}[1][20pt]{\hspace{0.05cm}\includegraphics[height=#1,valign=c]{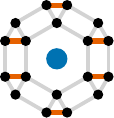}\hspace{0.05cm}}

\newcommand{\plaquettethree}[1][20pt]{\hspace{0.05cm}\includegraphics[height=#1,valign=c]{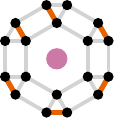}\hspace{0.05cm}}

\newcommand{\plaquettefour}[1][20pt]{\hspace{0.05cm}\includegraphics[height=#1,valign=c]{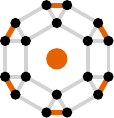}\hspace{0.05cm}}

\newcommand{\plaquettefive}[1][20pt]{\hspace{0.05cm}\includegraphics[height=#1,valign=c]{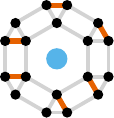}\hspace{0.05cm}}

\newcommand{\plaquette}[1][20pt]{\hspace{0.05cm}\includegraphics[height=#1,valign=c]{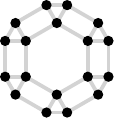}\hspace{0.05cm}}

\newcommand{\plaquetteplain}[1][20pt]{\hspace{0.05cm}\includegraphics[height=#1,valign=c]{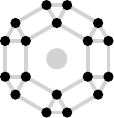}\hspace{0.05cm}}

\newcommand{\schiffchena}[1][8pt]{\hspace{0.05cm}\includegraphics[angle=90,height=#1,valign=c]{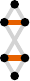}\hspace{0.05cm}}

\newcommand{\schiffchenb}[1][8pt]{\hspace{0.05cm}\includegraphics[angle=90,height=#1,valign=c]{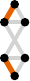}\hspace{0.05cm}}

\newcommand{\schiffchenc}[1][8pt]{\hspace{0.05cm}\includegraphics[angle=90,height=#1,valign=c]{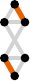}\hspace{0.05cm}}

\newcommand{\schiffchen}[1][8pt]{\hspace{0.05cm}\includegraphics[angle=90,height=#1,valign=c]{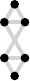}\hspace{0.05cm}}

\definecolor{black}{HTML}{000000}
\definecolor{light gray}{HTML}{D3D3D3}
\definecolor{lightgray}{HTML}{D3D3D3}
\definecolor{gray}{HTML}{808080}
\definecolor{gray}{HTML}{909090}
\definecolor{gray}{HTML}{B8B8B8}
\definecolor{red}{HTML}{E66100}
\definecolor{sky blue}{HTML}{56B4E9}
\definecolor{bluegreen}{HTML}{40B0A6} %
\definecolor{yellow}{HTML}{F0E442}
\definecolor{blue}{HTML}{0072B2}
\definecolor{vermillion}{HTML}{D55E00}
\definecolor{reddishpurple}{HTML}{CC79A7}
\definecolor{dark green}{HTML}{117733}
\definecolor{beige}{HTML}{DDCC77}
\definecolor{orange}{HTML}{E66100}
\definecolor{purple}{HTML}{5D3A9B} %purple
\definecolor{lila}{HTML}{BF40BF} %purple

\newcommand{\circlebluegreen}{\begin{tikzpicture}[baseline=-.45ex]\fill[bluegreen] circle(0.1);\end{tikzpicture}}

\newcommand{\circleorange}{\begin{tikzpicture}[baseline=-.45ex]\fill[orange] circle(0.1);\end{tikzpicture}}
\newcommand{\circlepink}{\begin{tikzpicture}[baseline=-.45ex]\fill[reddishpurple] circle(0.1);\end{tikzpicture}}
\newcommand{\circleblue}{\begin{tikzpicture}[baseline=-.45ex]\fill[blue] circle(0.1);\end{tikzpicture}}

\newcommand{\circlelightgray}{\begin{tikzpicture}[baseline=-.45ex]\fill[light gray] circle(0.1);\end{tikzpicture}}

\newcommand{\circleskyblue}{\begin{tikzpicture}[baseline=-.45ex]\fill[sky blue] circle(0.1);\end{tikzpicture}}

\newcommand{\scirclebluegreen}{\adjustbox{valign=c}{\begin{tikzpicture}[baseline=-.6ex]\fill[bluegreen] circle(0.1);\end{tikzpicture}}}
\newcommand{\scirclepurple}{\begin{tikzpicture}[baseline=-.6ex]\fill[purple] circle(0.1);\end{tikzpicture}}
\newcommand{\scircleorange}{\begin{tikzpicture}[baseline=-.6ex]\fill[orange] circle(0.1);\end{tikzpicture}}
\newcommand{\scirclepink}{\begin{tikzpicture}[baseline=-.6ex]\fill[reddishpurple] circle(0.1);\end{tikzpicture}}
\newcommand{\scircleskyblue}{\begin{tikzpicture}[baseline=-.6ex]\fill[sky blue] circle(0.1);\end{tikzpicture}}

\newcommand{\scirclepurpleborder}{\adjustbox{valign=c}{\begin{tikzpicture}[baseline=-.6ex]\draw[purple,line width=1] circle(0.1);\end{tikzpicture}}}
\newcommand{\sdashedpink}{\adjustbox{valign=c}{\begin{tikzpicture}[baseline=-.6ex]\draw[dashed,line width=2.5pt, reddishpurple] (0,0) to (0.56,0);\end{tikzpicture}}}
\newcommand{\sdashedbluegreen}{\adjustbox{valign=c}{\begin{tikzpicture}[baseline=-.6ex]\draw[dashed,line width=2.5pt, bluegreen] (0,0) to (0.56,0);\end{tikzpicture}}}
\newcommand{\scirclelightgray}{\begin{tikzpicture}[baseline=-.45ex]\fill[light gray] circle(0.1);\end{tikzpicture}}

\newcommand{\scirclelilaopa}{\begin{tikzpicture}[baseline=-.6ex]\fill[lila, fill opacity = 0.5] circle(0.1);\end{tikzpicture}}
\newcommand{\scirclegreenborder}{\adjustbox{valign=c}{\begin{tikzpicture}[baseline=-.6ex]\draw[dark green,line width=1] circle(0.1);\end{tikzpicture}}}

\begin{document}
	
	\title{Order-by-disorder in the antiferromagnetic $J_1$-$J_2$-$J_3$ transverse-field Ising model on the ruby lattice}
	%Order-by-disorder in the antiferromagnetic long-range transverse-field Ising model on the ruby lattice \patrickcomment{Back to the roots? long-range -> $J_1$-$J_2$-$J_3$}}
\author{Antonia Duft}
\affiliation{Friedrich-Alexander-Universit\"at Erlangen-N\"urnberg (FAU), Department Physik, Staudtstra{\ss}e 7, D-91058 Erlangen, Germany}
\author{Jan A. Koziol}
\affiliation{Friedrich-Alexander-Universit\"at Erlangen-N\"urnberg (FAU), Department Physik, Staudtstra{\ss}e 7, D-91058 Erlangen, Germany}
\author{Patrick Adelhardt}
\affiliation{Friedrich-Alexander-Universit\"at Erlangen-N\"urnberg (FAU), Department Physik, Staudtstra{\ss}e 7, D-91058 Erlangen, Germany}
\author{Matthias Mühlhauser}
\affiliation{Friedrich-Alexander-Universit\"at Erlangen-N\"urnberg (FAU), Department Physik, Staudtstra{\ss}e 7, D-91058 Erlangen, Germany}
\author{Kai P. Schmidt}
\affiliation{Friedrich-Alexander-Universit\"at Erlangen-N\"urnberg (FAU), Department Physik, Staudtstra{\ss}e 7, D-91058 Erlangen, Germany}

\begin{abstract}
	We investigate the quantum phase diagram of the $J_1$-$J_2$-$J_3$ antiferromagnetic transverse-field Ising model on the ruby lattice. In the low-field limit we derive an effective quantum dimer model, analyzing how the extensive ground-state degeneracy at zero field is lifted by an order-by-disorder scenario. We support our analysis by studying the gap-closing of the high-field phase using series expansions. For $J_2>J_3$, we find a columnar phase at low fields, followed by a clock-ordered phase stabilized by resonating plaquettes at intermediate field values, and an emergent 3d-XY quantum phase transition to the polarized high-field phase. For $J_3>J_2$, an order-by-disorder mechanism stabilizes a distinct $\bm{k}=(0,0)$ order and a quantum phase transition in the 3d-Ising universality class is observed. Further, we discuss the possible implementation of the columnar- and clock-ordered phase in existing Rydberg atom quantum simulators. When taking into account the full algebraically decaying long-range interactions on the ruby lattice, we find that long-range interactions favor the same ground state as the quantum fluctuations induced by a transverse field, which could make the ruby lattice a promising candidate for the realization of a clock-ordered phase.
	\end{abstract}

\maketitle

%Introduction
%%%%%%%%%%%%%%%%%%%%%%%%%%%%%%%%%%%%%%%%%%%%%%%%%%%%%%%%%%%%%%%%%%%%%%%%%%%%%%%%%%%%%%%%%%%%
\section{Introduction}
Extensively degenerate ground-state spaces due to frustration pose a
formidable resource for emergent exotic quantum phenomena. Frustration
arises either from conflicting spin interactions like, most prominently,
in Kitaev’s honeycomb model \cite{Kitaev2006} or due to the lattice
geometry like in antiferromagnetic quantum magnets on the triangular
\cite{Wannier1950,Wannier1973,Moessner2000,Moessner2001,Isakov2003,Powalski2013},
Kagome
\cite{Kano1953,Moessner2000,Moessner2001,Nikolic2005,Powalski2013}, or
pyrochlore \cite{Balents2010,Roechner2016} lattice containing loops of
odd length. Perturbing extensively degenerate ground-state spaces may
result in several distinct scenarios. First, a symmetry-broken order can emerge for infinitesimal perturbations
(order-by-disorder) \cite{Villain1980,Moessner2000,Moessner2001,Isakov2003,Powalski2013}.
Second, a direct realization of a symmetry unbroken phase may occur (disorder-by-disorder). This
phase can either be trivial \cite{Kano1953,Priour2001,Powalski2013,Malitz2013} or
an exotic quantum spin liquid \cite{Castelnovo2008,Jaubert2009,Fennell2009,Balents2010,Roechner2016,Savary2017}.

A distinct strand of recent research focuses on quantum phenomena in many-body systems with long-range interactions \cite{Defenu2023}. Such systems are relevant for a wide range of quantum-optical platforms including cold atoms \cite{Gross2017,Browaeys2020,Chomaz2022} and ions \cite{Friedenauer2008,Kim2009,Kim2010,Islam2011,Schneider2012,Britton2012,Islam2013,Jurcevic2014,Bohnet2016}. In particular, Rydberg atom quantum simulators are a promising platform to study frustrated Ising quantum spin systems with long-range interactions \cite{Lukin2001,Jaksch2000,Sachdev2002,Fendley2004,Labuhn2016,Schauss2018,Browaeys2020,Scholl2021,Semeghini2021}. Recent theoretical \cite{Verresen2021,Samajdar2023} and experimental \cite{Semeghini2021} studies of Rydberg atoms on the ruby lattice demonstrate a rich quantum phase diagram including a $\mathbb{Z}_2$ quantum spin liquid \cite{Verresen2021,Semeghini2021}. Further, unfrustrated quantum systems with long-range interactions are known to display exotic quantum-critical properties like continuously varying critical exponents \cite{Dutta2001,Fey2016,Defenu2017,Fey2019,Zhu2019,Adelhardt2020,Koziol2021,Langheld2022,Adelhardt2023,Song2023,Adelhardt2024}. Geometric frustration and long-range interactions typically compete with each other, i.\,e., quantum fluctuations favor different ground states compared to the ones that benefit from the long-range interactions \cite{Koziol2023}. An important example is the clock-ordered phase in the nearest-neighbor transverse-field Ising model on the triangular lattice resulting from an order-by-disorder scenario at low fields, which is destroyed by long-range Ising interactions giving rise to a stripe phase \cite{Smerald2016,Smerald2018,Saadatmand2018,Fey2019,Koziol2019,Koziol2023,Adelhardt2024}.

In this article, we investigate the antiferromagnetic $J_1$-$J_2$-$J_3$ transverse-field Ising model (TFIM) on the ruby lattice. In the absence of a magnetic field the model shows an extensively degenerate ground-state space due to geometric frustration. We study the breakdown of this degeneracy in the presence of a small field by deriving the leading order effective  Hamiltonians in two limiting cases. First, we consider the $J_1$-$J_2$ case, where we derive an effective low-field description analogous to the paradigmatic Rokhsar Kivelson (RK) quantum dimer model (QDM) on the honeycomb lattice \cite{Rokhsar1988, Read1990, Moessner2001new, Fradkin2004}. Using the RK QDM picture, we discuss the existence of a columnar phase at small fields and argue for the stability of a clock-ordered phase at intermediate transverse fields which melts with the corresponding 3d-XY phase transition to the high-field phase \cite{Moessner2000,Moessner2001,Isakov2003}. We confirm the 3d-XY phase transition to the high-field phase from high-field series expansions \cite{Knetter2000,Knetter2003,Powalski2013,Muehlhauser2022}.
Second, we repeat our investigations for the $J_1$-$J_3$ case and find a diagonal effective low-field Hamiltonian resulting in an order-by-disorder scenario to a gapped $\bm{k}=(0,0)$ order with a 3d-Ising quantum phase transition to the high-field phase. Further, we infer the phase diagram for the general $J_1$-$J_2$-$J_3$ model from the two previously considered limiting cases. In the end, we discuss the possible implementation of the columnar and clock-ordered phase in Rydberg atom quantum simulators. On the ruby lattice, the algebraically decaying long-range interactions present in Rydberg atom platforms favor the same order as the quantum fluctuations induced by the transverse field which sets it apart from realizations on other lattice geometries like the triangular lattice where long-range interactions and quantum fluctuations compete and favor mutually exclusive states of matter \cite{Smerald2016,Smerald2018,Saadatmand2018,Fey2019,Koziol2019,Koziol2023,Adelhardt2024,Koziol2024}. \\

%$J_1$-$J_2-J_3$ model
%%%%%%%%%%%%%%%%%%%%%%%%%%%%%%%%%%%%%%%%%%%%%%%%%%%%%%%%%%%%%%%%%%%%%%%%%%%%%%%%%%%%%%%%%%%%
%\noindent{\bf{Model}}\\ 
\section{Model}
\begin{figure}[t]
\centering
\includegraphics[width=\columnwidth]{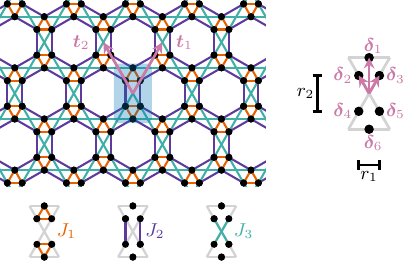}
\caption{Illustration of the ruby lattice with $\rho=r_2/r_1=\sqrt{3}$, showing the geometry of the three nearest-neighbor couplings $J_1$, $J_2$ and $J_3$, the elementary lattice vectors $ \bm{t}_1$ and $ \bm{t}_2$ and the positions $ \bm{\delta}_1, \ldots , \bm{\delta}_6$ of the six sites in the unit cell. For this aspect ratio, the geometry of the ruby lattice is equivalent to placing atoms on the links of a Kagome lattice. }
\label{fig:link-kagome}
\end{figure}

We study the antiferromagnetic $J_1$-$J_2$-$J_3$ TFIM on the ruby lattice. The Hamiltonian is given by
\begin{align}
\mathcal{H} &= \sum_{\langle i,j \rangle_{k=1,2,3}}\hspace{-0.1cm} J_k\, \sigma_{i}^z\sigma_{j}^z + h \sum_{i} \sigma_{i}^x 
\label{eq:j1j2j3_ham}
\end{align}
with Pauli matrices $\sigma_{i}^{x/z}$ describing spins-$1/2$ located on lattice sites $i$. 
The three couplings $J_1, J_2, J_3\geq0$ define the interaction strengths between sites as depicted in Fig.~\ref{fig:link-kagome}, such that the coupling $J_1\geq0$ denotes interactions between nearest-neighbor sites, while $J_2\geq0$ and \mbox{$J_3\geq0$} denote interactions between the second- and third-nearest-neighbor sites for all aspect ratios \mbox{$\rho=r_2/r_1>1$} (see Fig.~\ref{fig:link-kagome}). The transverse-field amplitude is denoted by $h$. For $h\gg J$ the ground state of the model is a trivial $x$-polarized phase. 

The ruby lattice can be understood as a triangular lattice with elementary lattice vectors $\bm{t}_1$ and $\bm{t}_2$ and six sites per unit cell at positions $\bm{\delta}_1, \ldots ,  \bm{\delta}_6$ (compare Fig.~\ref{fig:link-kagome}). We set the distance between sites interacting via $J_1$ to $r_1=1$.
Note that the aspect ratio $\rho=\sqrt{3}$ chosen for all illustrations corresponds to placing the atoms on the links of a Kagome lattice. 

%%%%%%%%%%%%%%%%%%%%%%%%%%%%%%%%%%%%%%%%%%%%%%%%%%%%%%%%%%%%%%%%%%%%%%%%%%%%%%%%%%%%%%%%%%%%
%%%%%%%%%%%%%%%%%%%%%%%%%%%%%%%%%%%%%%%%%%%%%%%%%%%%%%%%%%%%%%%%%%%%%%%%%%%%%%%%%%%%%%%%%%%%
In the limiting cases of two vanishing couplings $J_i$ the ruby lattice decomposes into isolated structures. For $J_2=J_3=0$ the lattice reduces to isolated triangles, for $J_1=J_3=0$ to isolated hexagons and for $J_1=J_2=0$ to decoupled chains. As long as $J_1 > 0$ the system is geometrically frustrated and there is an extensive number of ground states for $h=0$. \\

\section{Discussion of the quantum phase diagram}
We start by discussing the quantum phase diagram of two limiting cases of the model, specifically $J_1$-$J_2$ and $J_1$-$J_3$. From this we then analyze the full $J_1$-$J_2$-$J_3$ TFIM. 
%%%%%%%%%%%%%%%%%%%%%%%%%%%%%%%%%%%%%%%%%%%%%%%%%%%%%%%%%%%%%%%%%%%%%%%%%%%%%%%%%%%%%%%%%%%%
%\noindent{\bf{$J_1$-$J_2$ case}}\\ 
\subsection{$J_1$-$J_2$ case}
The subsequent low-field analysis of the case $J_1,J_2>0$ and $J_3=0$ is independent of the ratio $J_2/J_1$.
%%%%%%%%%%%%%%%%%%%%%%%%%%%%%%%%%%%%%%%%%%%%%%%%%%%%%%%%%%%%%%%%%%%%%%%%%%%%%%%%%%%%%%%%%%%%
\begin{figure}[t]
\centering
\includegraphics[width=1.\columnwidth]{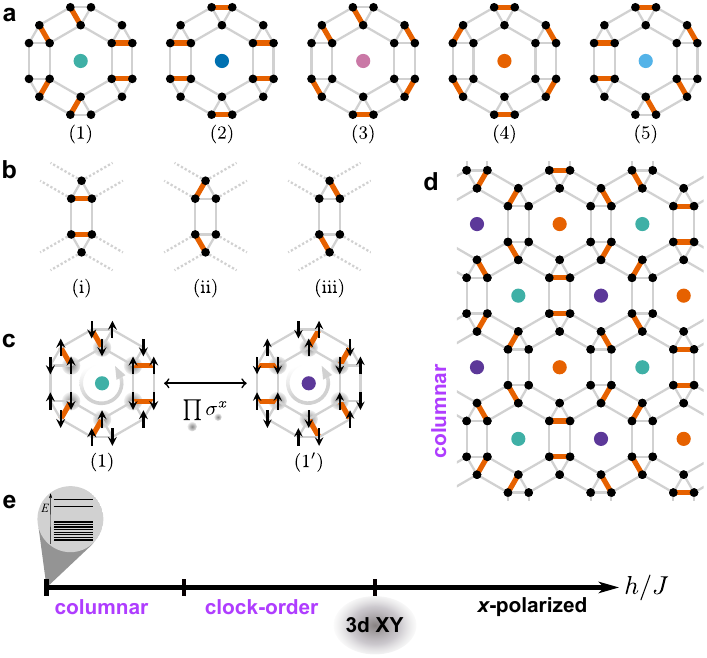}
\caption{$J_1$-$J_2$ model. (\textbf{\textsf{a}}) Illustration of hexagonal plaquettes minimizing the number of frustrated ferromagnetic $J_1$-bonds (orange; antiferromagnetic bonds in gray). (\textbf{\textsf{b}}) Unit-cell configurations appearing within the zero-field ground states. (\textbf{\textsf{c}}) Leading sixth-order off-diagonal process inducing resonances between type-(1) and -($1^\prime$) plaquettes. (\textbf{\textsf{d}}) Columnar ground state with $\bm{k}=(2\pi/3,-2\pi/3)$ in the low-field limit. (\textbf{\textsf{e}}) Sketch of the quantum phase diagram: The extensive degeneracy of the zero-field ground state breaks down to a columnar phase in an order-by-disorder scenario for infinitesimal transverse fields, followed by an intermediate clock-ordered phase and a 3d-XY phase transition into the trivial $x$-polarized phase.}
\label{fig:j1j2-case}
\end{figure}
%%%%%%%%%%%%%%%%%%%%%%%%%%%%%%%%%%%%%%%%%%%%%%%%%%%%%%%%%%%%%%%%%%%%%%%%%%%%%%%%%%%%%%%%%%%%
% --- zero field case ---
Ground states at $h=0$ minimize the number of frustrated $J_1$-bonds. All these states contain only hexagonal plaquettes of the five types represented (up to rotations) in Fig.~\ref{fig:j1j2-case}a. The resulting zero-field ground-state manifold is extensively degenerate. Note that in all those states, no $J_2$ bond is ferromagnetic and thus the ground-state manifold is independent from the ratio $J_1/J_2$. 

% --- low field ---
We analyze the leading-order low-field contributions by performing degenerate perturbation theory around $h=0$, including both diagonal and off-diagonal processes \cite{Takahashi1977} (see supplemental material in Ref.~\cite{supplementary}). 
% diagonal 
We find that the leading-order contribution lifting the ground-state degeneracy is a fourth-order diagonal energy correction selecting type-(1) and type-(4) plaquettes. For this we evaluate the diagonal energy corrections on the three unit cell configurations of the zero-field ground-state plaquettes, depicted in Fig.~\ref{fig:j1j2-case}b. The energy corrections in fourth order yield the decreasing energetic hierarchy of the unit cell configurations as $(\mathrm{i}) \to (\mathrm{ii}) \to (\mathrm{iii})$ for $J_1,J_2>0$. As each plaquette implies the same density of 1/3 of configurations (i) on the lattice, plaquettes (1) and (4) are beneficial as they contain zero of the energetically least favorable configuration (iii). 

To understand the low-field phase we construct the zero-field states containing exclusively type-(1) and (4) plaquettes, resulting in a $\bm{k}=(2\pi/3,-2\pi/3)$ placement of the type-(4) plaquettes as depicted in Fig.~\ref{fig:j1j2-case}d. Respective states are selected from the extensively degenerate ground-state manifold in an order-by-disorder scenario for $h>0$. Note, it is not possible to have a ground state with only type-(1) or type-(4) plaquettes with $\bm{k}=(0,0)$.

We further find that the leading-order off-diagonal contribution is a resonant sixth-order process acting on the inner sites of the hexagon of plaquettes of type-(1) mapping them to type-(1$^\prime$) plaquettes (see Fig.~\ref{fig:j1j2-case}c). The density of type-(1) and type-(1$^\prime$) plaquettes is maximized by the states already selected by the diagonal fourth-order corrections.  
The action of the transverse field on respective states in sixth order is analogous to the famous resonating first order process within the effective low-field description of the antiferromagnetic TFIM on the triangular lattice \cite{Moessner2000,Moessner2001,Isakov2003}. 
The effective low-field description of the ruby lattice in terms of plaquettes, as well as the effective dual low-field description on the triangular lattice, can be understood in terms of the Rokhsar Kivelson (RK) quantum dimer model (QDM) on the honeycomb lattice \cite{Rokhsar1988,Read1990,Moessner2000,Moessner2001,Moessner2001new,Isakov2003,Fradkin2004}.

In analogy to the RK QDM, we describe the low-field limit on the ruby lattice with the leading-order diagonal and off-diagonal contribution as
\begin{align}\label{eq:j1j2}
\mathcal{H}_\mathrm{eff} &= \bar{E}_0 +\frac{ h^4}{2} \sum_{\plaquette[8pt]} \quad \sum_{\mathclap{\substack{\protect\circlelightgray=\protect\circlebluegreen,\protect\circleblue,\protect\circlepink, \\ \quad \protect\circleorange,\protect\circleskyblue}}} \quad
\left( E_{J_2}^{(4)}({\adjustbox{valign=c}{\protect\circlelightgray}})\ket{\plaquetteplain}\bra{\plaquetteplain} \right)	+ \nonumber \\ 
&+  h^6 \sum_{\plaquette[8pt]}  E_\mathrm{res}^{(6)} \left(\ket{\resonatingplaquette}\bra{\resonatingplaquettetwo} +\mathrm{h.c.} \right) \,,
\end{align}
where the sum runs over all plaquettes within the lattice and $\bar{E}_0$ is a constant containing perturbative contributions which are equal for all ground states. For the sake of a compact notation, the first sum containing the diagonal terms runs over all five plaquette types defined in Fig.~\ref{fig:j1j2-case}a (up to rotations). 
The diagonal corrections $E_{J_2}^{(4)}({\adjustbox{valign=c}{\protect\circlelightgray}})$ of the five plaquette types and the amplitude of the resonating process $ E_\mathrm{res}^{(6)}$ are given in Ref.~\cite{supplementary}. 

Note, to associate Eq.~\eqref{eq:j1j2} exactly with the well-studied RK QDM on the honeycomb lattice \cite{Moessner2001new}, we have to regard $E_{J_2}^{(4)}({\adjustbox{valign=c}{\circleblue}})=E_{J_2}^{(4)}({\adjustbox{valign=c}{\circlepink}})=E_{J_2}^{(4)}({\adjustbox{valign=c}{\circleorange}})=E_{J_2}^{(4)}({\adjustbox{valign=c}{\circleskyblue}})$ \cite{supplementary}. 
This condition is needed to make estimates about the low-field quantum phase diagram based on the RK QDM. We note that our analysis of the high-field limit is consistent with the conclusions drawn from this mapping.
In the RK QDM the effective fourth-order contribution corresponds to the diagonal term $v=h^4(E_{J_2}^{(4)}({\adjustbox{valign=c}{\circlebluegreen}})-E_{J_2}^{(4)}({\adjustbox{valign=c}{\circlelightgray}}))/2$, which we evaluate for ${\adjustbox{valign=c}{\circlelightgray}}={\adjustbox{valign=c}{\circleblue}},{\adjustbox{valign=c}{\circlepink}},{\adjustbox{valign=c}{\circleorange}},{\adjustbox{valign=c}{\circleskyblue}}$, and the sixth-order off-diagonal contribution corresponds to the kinetic term $-t=h^6E_\mathrm{res}^{(6)} $ \cite{Rokhsar1988,Moessner2001new}. For $v/t<-0.2\pm0.05$ the RK QDM has a columnar ground state with a static order and a first-order phase transition to a resonating plaquette phase at this boundary \cite{Moessner2001new}.

Therefore, the ground state we find at small $h>0$ is associated with the columnar phase of the RK QDM \cite{Moessner2001new}. With increasing $h$ the ratio $v/t$ is tuned from $-\infty$ to $0$. Using our effective model and the aforementioned mapping we can estimate the transition to the resonating plaquette phase as a function of $h$, $J_1$ and $J_2$. We call this plaquette phase clock-ordered phase in analogy to the low-field clock-ordered phase of the antiferromagnetic TFIM on the triangular lattice which is described by the RK QDM with $v=0$ and therefore also a resonating plaquette phase \cite{Moessner2000,Moessner2001,Moessner2001new,Isakov2003}. Motivated by this analogy, one expects a 3d-XY quantum phase transition from the clock-ordered phase towards the $x$-polarized high-field phase \cite{Moessner2000,Moessner2001,Isakov2003}. Note, the $x$-polarized high-field phase is not part of the effective description in Eq.~\eqref{eq:j1j2} and there is the possibility that the clock-ordered phase occurs at $h$-values larger than the transition to the $x$-polarized phase.
However, we do not find this scenario by analyzing the transition from the columnar phase to the clock-ordered phase of the effective low-field model and the phase transition to the polarized phase, since the latter always occurs at larger field strengths (see Ref.~\cite{supplementary}). Thus, we expect an intermediate clock-ordered phase to be present.  

% --- high field --- 
To obtain values for the phase transition to the $x$-polarized phase we employ %We confirm this scenario by 
high-field series expansions using the method of perturbative continuous unitary transformations (pCUT) \cite{Knetter2000} with the help of linked-cluster expansions set up as a full graph decomposition \cite{Coester2015,Muehlhauser2022,Gelfand2000,Oitmaa2006} and DlogPadé extrapolations \cite{Baker1996,Guttmann1989} (for details see Ref.~\cite{supplementary}) to investigate the closing of the elementary excitation gap. We determine the series of the gap up to order 10 for the general $J_1$-$J_2$ case and order 11 for $J_1=J_2$.
With this analysis we determine the critical momentum $\bm{k}$, critical point $\lambda_c$, and gap exponent $z\nu$ of a potential quantum phase transition. We find the critical gap momentum at ${\bm{k} = (2\pi/3,-2\pi/3)}$ reflecting the periodicity of the low-field order and a closing of the excitation gap at some critical point $\lambda_c$ for all ratios of $J_1/J_2$ except for $J_1/J_2\rightarrow\infty\,(J_1/J_2\to 0)$ when the limit of decoupled triangles (hexagons) is approached and $\lambda_c$ shifts towards large coupling strengths. 
The associated critical exponents $z\nu$ ($z\nu=0.806\pm0.056$ for $J_1=J_2$) are in line with the conjectured 3d-XY universality class with $z=1$ and $\nu_\mathrm{3d-XY}=0.67169(7)$ \cite{Hasenbusch2019,Chester2020} within the limitations of the series expansion. This approach is known to slightly overestimate critical exponents. Especially the determination of 3d-XY exponents is challenging as observed for the antiferromagnetic TFIM on the triangular lattice \cite{Powalski2013, Fey2019}. In comparison to the triangular lattice the considered perturbative processes span even fewer unit cells on the ruby lattice so that an even larger deviation to the known critical exponents is reasonable. The critical point as well as the gap exponent as a function of $J_1/J_2$ are depicted in Ref.~\cite{supplementary}. \\

%$J_1$-$J_3$
%%%%%%%%%%%%%%%%%%%%%%%%%%%%%%%%%%%%%%%%%%%%%%%%%%%%%%%%%%%%%%%%%%%%%%%%%%%%%%%%%%%%%%%%%%%%
\subsection{$J_1$-$J_3$ case}
Next we consider the case $J_1,J_3>0$ and $J_2=0$. Again, the low-field scenario is independent of the ratio $J_1/J_3$.
%%%%%%%%%%%%%%%%%%%%%%%%%%%%%%%%%%%%%%%%%%%%%%%%%%%%%%%%%%%%%%%%%%%%%%%%%%%%%%%%%%%%%%%%%%%%
\begin{figure}[t]
\centering
\includegraphics[width=1.\columnwidth]{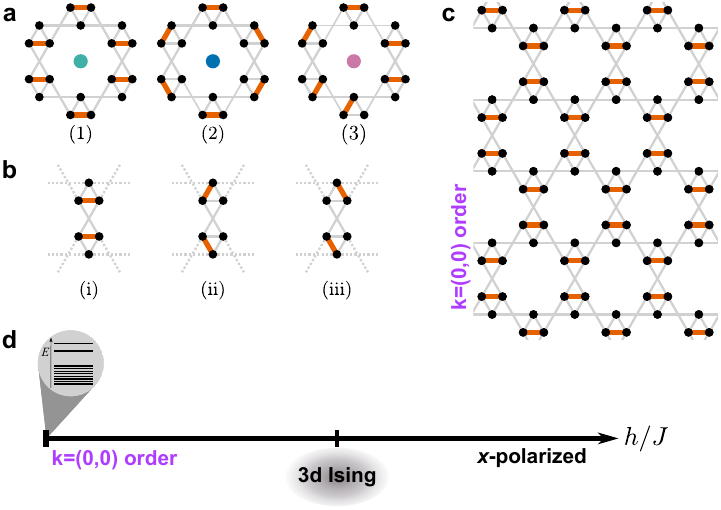}
\caption{$J_1$-$J_3$ model. (\textbf{\textsf{a}}) Illustration of hexagonal plaquettes minimizing the number of frustrated ferromagnetic $J_1$-bonds (orange; antiferromagnetic bonds in gray). (\textbf{\textsf{b}}) Unit-cell configurations appearing within the zero-field ground states. (\textbf{\textsf{c}}) Emerging $\bm{k}=(0,0)$ ordered ground-state in the low-field limit. (\textbf{\textsf{d}}) Sketch of the quantum phase diagram: The extensive degeneracy of the zero-field ground state breaks down to a $\bm{k}=(0,0)$ order in an order-by-disorder scenario for infinitesimal transverse fields, followed by a 3d-Ising phase transition into the trivial $x$-polarized phase.}
\label{fig:j1j3-case}
\end{figure}
%%%%%%%%%%%%%%%%%%%%%%%%%%%%%%%%%%%%%%%%%%%%%%%%%%%%%%%%%%%%%%%%%%%%%%%%%%%%%%%%%%%%%%%%%%%%
% --- zero field ---
At $h=0$, ground states contain only hexagonal plaquettes of the three types represented in Fig.~\ref{fig:j1j3-case}a (up to rotations). The manifold of ground states built from these three plaquette types is extensively degenerate. Since there is no ferromagnetic $J_3$ bond in a zero-field ground state, the zero-field results are independent of $J_3$. 
% --- low field ---
We investigate the low-field limit by performing degenerate perturbation theory around $h=0$. 
% diagonal
We demonstrate that the leading-order contribution lifting the ground-state degeneracy is a fourth-order process by evaluating the diagonal energy corrections for all zero-field ground states analogous to the $J_1$-$J_2$ case. We find the decreasing energetic hierarchy of the configurations depicted in Fig.~\ref{fig:j1j3-case}b as $(\mathrm{i})\to (\mathrm{iii})\to (\mathrm{ii})$.
% off-diagonal
In fact, there can be no finite-order off-diagonal process in the $J_1$-$J_3$ case: When performing a spin flip within a zero-field ground state, one introduces two domain walls in the $J_3$ chain the affected site is part of. Therefore, to map one zero-field ground state to another, one needs to flip at least all spins along the $J_3$ chain.
The effective Hamiltonian in the low-field limit reads
\begin{align}\label{eq:j1j3}
\mathcal{H}_\mathrm{eff} &= \bar{E}_0 +  h^4 \sum_{\schiffchen[5pt]} \left( E_{J_3}^{(4)}(\mathrm{i})\ket{\schiffchena}\bra{\schiffchena} + \right. \\ + & \left.E_{J_3}^{(4)}(\mathrm{ii}) \ket{\schiffchenb}\bra{\schiffchenb} + E_{J_3}^{(4)}(\mathrm{iii}) \ket{\schiffchenc}\bra{\schiffchenc} \right) \nonumber\,, 
\end{align}
where $\bar{E}_0$ is a constant containing perturbative contributions which are equal for all ground states. The amplitudes $E_{J_3}^{(4)}(\mathrm{i,ii,iii})$ are listed in Ref.~\cite{supplementary}. As each zero-field plaquette implies the same density of 1/3 configurations (i) on the lattice (see Fig.~\ref{fig:j1j3-case}a), we can derive the ground-state configuration for $h>0$ by maximizing the density of the energetically second-most beneficial configuration (iii). Thus, the plaquette of type-(1) is energetically most favorable, resulting in the ground-state configuration depicted in Fig.~\ref{fig:j1j3-case}c. We find a diagonal order-by-disorder scenario where the quantum fluctuations stabilize a $\bm{k} =(0,0)$ state adiabatically connected to this ground state for all $J_1,J_3>0$. From this low-field scenario a 3d-Ising quantum phase transition to the $x$-polarized high-field phase is expected and we confirm this with the subsequent high-field analysis analogous to the $J_1$-$J_2$ case. We determine the series of the gap up to order 10 for the general $J_1$-$J_3$ case and order 11 for $J_1=J_3$.

The critical gap momentum $\bm{k} =(0,0)$ found in series expansions around the high-field limit for all ratios $J_1/J_3$ verifies the low-field order. We find a quantum phase transition for all $J_1/J_3>0$. Similar to the $J_1$-$J_2$ case, $\lambda_c$ shifts to large couplings for $J_1/J_3\rightarrow\infty$, while $\lambda_c\to0.5$ as the limit of isolated chains is approached for $J_1/J_3\rightarrow 0$. By investigating the critical exponent $z\nu$ we find $z\nu=0.632\pm0.101$ for $J_1=J_3$ which coincides with the expected 3d-Ising universality (\mbox{$z\nu_\text{3d-Ising}=0.629971(4)$} \cite{Kos2016}) within error bars. The critical point and exponent as a function of $J_1/J_3$ are depicted in Ref.~\cite{supplementary}. Note that the extracted gap exponent is more accurate compared to the 3d-XY exponent in the $J_1$-$J_2$ model as expected. \\% 

%$J_1$-$J_2$-$J_3$
%%%%%%%%%%%%%%%%%%%%%%%%%%%%%%%%%%%%%%%%%%%%%%%%%%%%%%%%%%%%%%%%%%%%%%%%%%%%%%%%%%%%%%%%%%%%
%\noindent{\bf{$J_1$-$J_2$-$J_3$ case}}\\  
\subsection{$J_1$-$J_2$-$J_3$ case}
We continue by deriving the $J_1$-$J_2$-$J_3$ phase diagram from the two limiting cases considered so far. We define $J_2 = \cos\theta\, J_1$ and $J_3 = \sin\theta\,J_1$.
% --- zero field --- 
In the full $J_1$-$J_2$-$J_3$ model the zero-field ground states of the $J_1$-$J_2$ case and the $J_1$-$J_3$ case have the two distinct energies $-\frac{1}{3}J_1 -J_2 + \frac{1}{3}J_3$ and $-\frac{1}{3}J_1 + \frac{1}{3}J_2-J_3$ per site for $h=0$. 
Thus, for $J_2>J_3$ ($J_3>J_2$) the $J_1$-$J_2$ ($J_1$-$J_3$) ground states have a lower energy than the $J_1$-$J_3$ ($J_1$-$J_2$) ground states and realize ground states of the $J_1$-$J_2$-$J_3$ model. 
From the preceding low-field analysis we expect that for all $J_2>J_3$ there is a qualitatively similar effective order-by-disorder low-field description with a transition from a columnar phase to a clock-ordered one as the $J_1$-$J_2$ ground states also realize the zero-field ground states here.
Following the same argument, an order-by-disorder to the $\bm{k} =(0,0)$ ordered state takes place for all $J_3>J_2$. 

% --- high field --- 
The orders found by the low-field analysis are stable until the point $\lambda_c$ where a quantum phase transition towards the high-field phase takes place. Using series expansions as before, we find that the critical gap momenta correspond indeed to the expected momenta ${\bm{k} =(2\pi/3,-2\pi/3)}$ for $J_2>J_3$ and $\bm{k} =(0,0)$ for $J_3>J_2$. 
%%%%%%%%%%%%%%%%%%%%%%%%%%%%%%%%%%%%%%%%%%%%%%%%%%%%%%%%%%%%%%%%%%%%%%%%%%%%%%%%%%%%%%%%%%%%
\begin{figure}[t]
\centering
\includegraphics[width=1.\columnwidth]{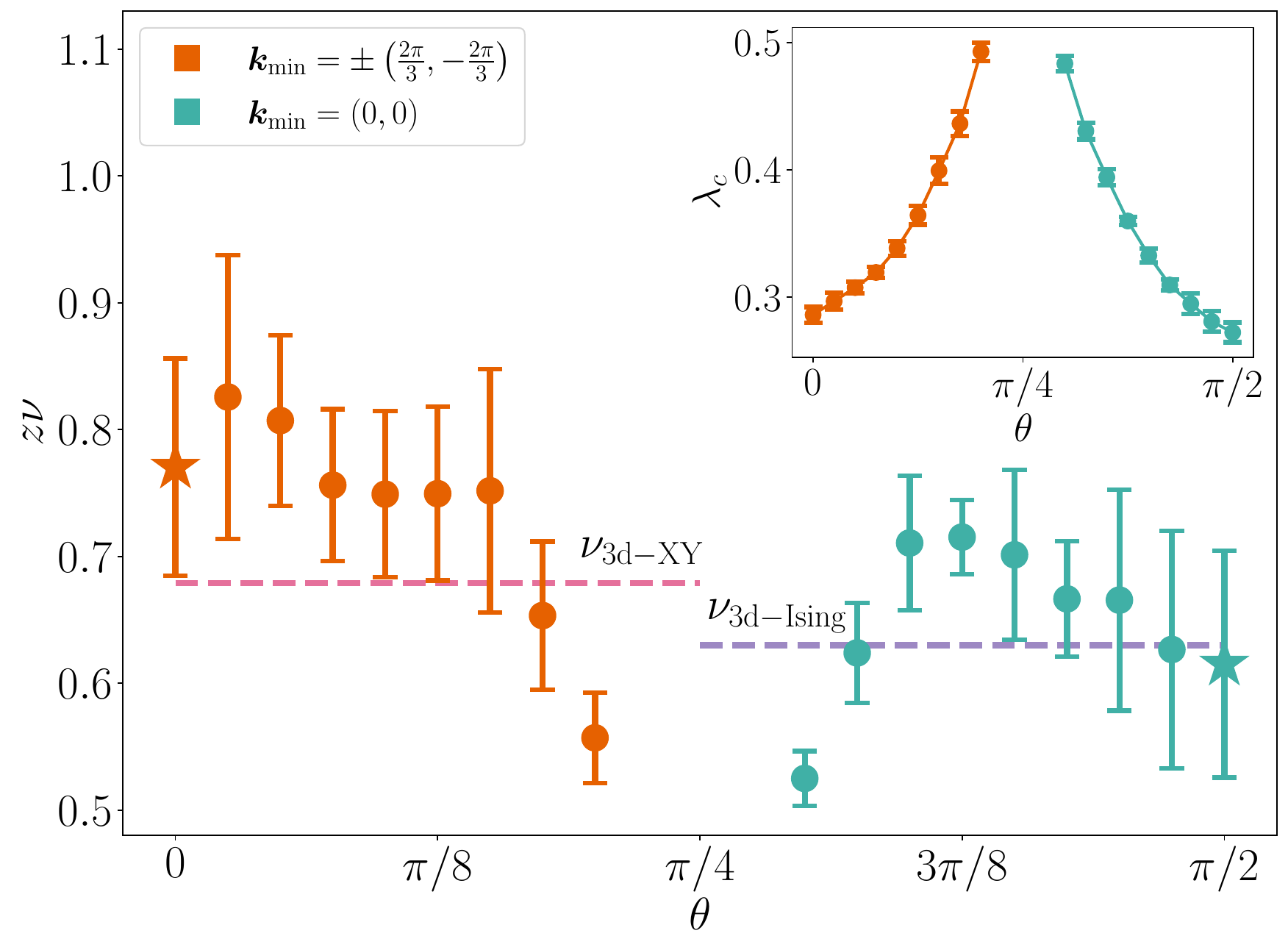}
\caption{Critical point $\lambda_c$ (inset) and gap exponent $z\nu$ in the $J_1$-$J_2$-$J_3$ TFIM from high-field series expansions tuning with $\theta$ from the $J_1$-$J_2$ to the $J_1$-$J_3$ case. The critical exponents match the two expected exhibit two universality classes 3d-XY for $J_2>J_3$ and 3d-Ising for $J_3>J_2$ with an accuracy similar to the results in the limiting cases $J_1$-$J_2$ to the $J_1$-$J_3$ which are marked by the respective stars. The depicted results resemble averages over several high-order extrapolants with the respective standard deviation.} 
\label{fig:criticality}
\end{figure}
%%%%%%%%%%%%%%%%%%%%%%%%%%%%%%%%%%%%%%%%%%%%%%%%%%%%%%%%%%%%%%%%%%%%%%%%%%%%%%%%%%%%%%%%%%%%
We extract the critical exponent $z\nu$ from the gap-closing at $\lambda_c$ as shown in Fig.~\ref{fig:criticality} as a function of $\theta$, i.\,e., sweeping from the $J_1$-$J_2$ to the $J_1$-$J_3$ case. 
We calculate the series expansion of the gap up to order 9 for the general $J_1$-$J_2$-$J_3$ case. We find a divergence of $\lambda_c$ approaching the symmetric $J_2=J_3$ case at $\theta=\pi/4$ from both sides. 

Regarding the gap exponent $z\nu$ we find a qualitative match with the expected quantum criticalities (3d-XY criticality for $J_2>J_3$ and 3d-Ising criticality for $J_3>J_2$). The agreement with the respective literature value is comparable to the one in the two previously analyzed limiting $J_1$-$J_2$ and $J_1$-$J_3$ cases. In particular, as discussed for the $J_1$-$J_2$ case, the DlogPadé approximants systematically overestimate the critical exponent for $J_2>J_3$ in a sizable manner. As $\lambda_c$ shifts to infinity for $J_2\approx J_3$, the estimation of the critical exponent becomes unreliable. 

% --- J2=J3 --- 
Indeed, the point $J_2=J_3$ is special in the $J_1$-$J_2$-$J_3$ TFIM. At ${h=0}$, the degeneracy of $J_1$-$J_2$ and $J_1$-$J_3$ ground states is further enhanced by multiple additional ground states (see Ref.~\cite{supplementary}). Further, the extrapolation of the high-field gap located at \mbox{$\bm{k}=(2\pi/3,-2\pi/3)$} does not indicate any gap-closing. 
One therefore may speculate about a disorder-by-disorder scenario so that the ground state at infinitesimally small $h$ is adiabatically connected to the high-field limit \cite{Priour2001, Powalski2013, Malitz2013} or that a quantum spin liquid is present which breaks no symmetry \cite{Roechner2016, Tarabunga2022}.
Note that proving a disorder-by-disorder scenario using solely series expansions about the high-field limit is not possible and further investigations using other methods are necessary \cite{Powalski2013, Roechner2016,Patil2023}.

Altogether, we have shown that the general $J_1$-$J_2$-$J_3$ TFIM features a plethora of intriguing ground-state phenomena including two order-by-disorder mechanisms and the realization of the RK QDM as an effective low-field model. \\

% Experimental realization 
%%%%%%%%%%%%%%%%%%%%%%%%%%%%%%%%%%%%%%%%%%%%%%%%%%%%%%%%%%%%%%%%%%%%%%%%%%%%%%%%%%%%%%%%%%%% 
\section{Experimental Realization}
The so far considered model contains a truncated version of the TFIM with algebraically decaying long-range interactions. Here, we show how  Rydberg atom quantum simulators \cite{Schauss2012,Zeiher2015,Schauss2015,Labuhn2016,Schauss2018,Lienhardt2018,Browaeys2020,Semeghini2021} can implement the order-by-disorder scenario to the columnar phase of the TFIM on the ruby lattice and potentially also the clock-ordered phase even in the presence of long-range interactions. The physics of Rydberg atom arrays can be modeled with the two-dimensional Fendley-Sengupta-Sachdev model \cite{Fendley2004, Verresen2021}
\begin{equation}
\mathcal{H}^\mathrm{FSS} = \frac{\Omega}{2} \sum_{i} \hspace{-0.1cm}\left(b_{i}^{\phantom\dagger} + b_{i}^{\dagger}\right) - \delta \sum_{i}n_{i}\ + \sum_{i<j}\hspace{-0.1cm}\frac{C_6}{|\bm{i}-\bm{j}|^{6}}n_{i} n_{j}	
\end{equation}
with hardcore bosonic operators $b_{i}^{\dagger}=\ket{e}_{i}\bra{g}_{i}$ (${b_{i}^{\phantom\dagger}=\ket{g}_{i}\bra{e}_{i}}$) (de)exciting a Rydberg atom and ${n_{i}=b_{i}^{\dagger}b_{i}^{\phantom\dagger}}$ counting the excitations at site $i$. Excited Rydberg atoms interact via an algebraically decaying long-range van-der-Waals interaction $\sim |\bm{i}-\bm{j}|^{-6}$ with van-der-Waals coefficient $C_6$. 
Identifying the Pauli matrices $\sigma_{i}^x=b_{i}^{\phantom\dagger}+b_{i}^{\dagger}$ and $\sigma_{i}^z=1-2n_{i}$, we arrive up to a constant at the spin Hamiltonian
\begin{equation}
\mathcal{H}^\mathrm{FSS} = h\sum_{i} \hspace{-0.1cm}\sigma_{i}^x +\hspace{-0.1cm} \left(\frac{\delta}{2}-\frac{C_6\mu}{4}\right)\hspace{-0.1cm}\sum_{i}\sigma_{i}^z + \frac{1}{4}\sum_{i<j}\hspace{-0.1cm}\frac{C_6}{|\bm{i}-\bm{j}|^{6}}\sigma_{i}^z \sigma_{j}^z
\end{equation}
with $\mu=\sum_{\bm{d}}|\bm{d}|^{-6}$, summing over all vectors $\bm{d}$ between interacting sites on the lattice, and $h=\Omega/2$. To recover the TFIM, the longitudinal field must be zero, yielding the condition $\delta=C_6\mu/2$ for the laser detuning. As the algebraically decaying interaction strength implies $J_1\geq J_2 > J_3$ for $\rho\geq 1$, from our previous analysis of the $J_1$-$J_2$-$J_3$ TFIM we expect an order-by-disorder mechanism into a columnar phase upon tuning the Rabi frequency $\Omega$ from zero to a finite value and eventually a phase transition to a clock-ordered phase which exhibits a 3d-XY quantum phase transition into the polarized phase. The case of $J_3 \geq J_2$ can not be realized on such a platform. 

% --- rho=sqrt(3) ---

The ruby lattice with aspect ratio $\rho=\sqrt{3}$ is equivalent to placing sites on the links of a Kagome lattice. In a recent work \cite{Semeghini2021}, Rydberg atoms on such a geometry were used to probe topological spin liquids in agreement with theoretical studies \cite{Verresen2021, Samajdar2023}. From this geometry we obtain $J_1=C_6/4$, $J_2=J_1/27$, $J_3=J_1/64$, truncating all $J_i\le J_4=J_1/343$ for the series expansion.
In this case the $J_1$ interactions dominate ($J_1\gg J_2 > J_3$) and the system consists of almost isolated triangles, resulting in an almost flat dispersion and no closing of the excitation gap at momentum $\bm{k} = (2\pi/3,-2\pi/3)$ can be detected within the perturbative series expansion (see Ref.~\cite{supplementary}). 

% --- rho=1 ---
Decreasing the aspect ratio $\rho$, which is a free tuning parameter in the experiment, decreases the ratio of $J_1/J_2$. Setting e.\,g. $\rho=1$, we have $J_1=J_2$ and $J_3=J_1/8$, again neglecting all $J_i\le J_4=J_1/27$. From pCUT calculations we predict a quantum phase transition at $C_6/\Omega\approx 1.202 \pm 0.012$ from the clock-ordered phase towards the trivial polarized phase. The associated critical exponent $z\nu = 0.801 \pm 0.046$ is in line with the predicted 3d-XY criticality within previously discussed limitations.

We further demonstrate the order-by-disorder mechanism in the presence of the full algebraically decaying long-range interaction taking into account $J_i\leq J_4$. Following the approach described in Ref.~\cite{Koziol2023} we search for the energetically most beneficial configuration in the limit of $h=0$, taking into account the entire, untruncated long-range interactions via resummed couplings. Coming from the degenerate zero-field ground state in the $J_1$-$J_2$-$J_3$ TFIM, we find that the same columnar order is selected by both: the long-range interaction and an infinitesimal transverse field. This is remarkable as in other well-known order-by-disorder scenarios, e.\,g., for the antiferromagnetic TFIM on the triangular lattice, the long-range interactions compete with the fluctuations induced by the transverse field \cite{Smerald2016,Smerald2018,Saadatmand2018,Fey2019,Koziol2019,Koziol2023}.\\

% Conclusion
%%%%%%%%%%%%%%%%%%%%%%%%%%%%%%%%%%%%%%%%%%%%%%%%%%%%%%%%%%%%%%%%%%%%%%%%%%%%%%%%%%%%%%%%%%%% 
%\noindent{\bf{Conclusions}}\\
\section{Conclusions}
In this article we investigated the antiferromagnetic $J_1$-$J_2$-$J_3$ TFIM on the ruby lattice employing perturbative approaches from both the high- and low-field limit. We found a rich quantum phase diagram as a function of the ratio $J_2/J_3$ including two distinct order-by-disorder scenarios.
In particular, for the $J_1$-$J_2$ TFIM we derived an effective low-field model analogous to the RK QDM where quantum fluctuations induce an order-by-disorder scenario giving rise to a columnar ordered phase. For intermediate field values we argued that a clock-ordered phase is present. 
We further showed that long-range interactions and the breakdown of geometric frustration by quantum fluctuations do not compete for different ground states but promote the same columnar state. Based on this we proposed an experimental implementation of the columnar and clock-ordered phase, and in general the RK QDM, in current set ups using Rydberg atom quantum simulators. Our results demonstrate that the interplay of long-range interactions and geometric frustration represents a rich playground for exotic quantum phenomena.\\

\section*{Code Availability.}
The code used to calculate the pCUT series expansions, as well as the code used for the unit-cell based optimization, is available from the corresponding author upon reasonable request. \\

%\noindent{\bf{Data availability}}\\
\section*{Data availability}
The data regarding this work is available from the corresponding author upon reasonable request and as a data repository on Zenodo URL (to be created).\\

%\noindent{\bf{Author contributions}}\\
\section*{Author contributions}
AD performed the series expansions. JAK calculated the zero-field ground state considering the long-range interactions. MM provided the code for the high-field series expansion and graph decomposition. All authors jointly derived the effective low-field model. KPS devised the subject of the work. All authors contributed to the interpretation of the results and the writing of the manuscript.\\ 

%\noindent{\bf{Competing interests}}\\
\section*{Competing interests}
The authors declare no competing interests. 

%\noindent{\bf{Additional Information}}\\
\section*{Additional Information}
Supplementary material accompanies this paper, see Ref.~\cite{supplementary}.

%\noindent{\bf{Acknowledgments}}\\
\section*{Acknowledgments} 
The authors gratefully acknowledge the support by the Deutsche Forschungsgemeinschaft (DFG, German Research Foundation) -- Project-ID 429529648—TRR 306 \mbox{QuCoLiMa} (``Quantum Cooperativity of Light and Matter'') and the Munich Quantum Valley, which is supported by the Bavarian state government with funds from the Hightech Agenda Bayern Plus. PA, AD, and KPS gratefully acknowledge the scientific support and HPC resources provided by the Erlangen National High Performance Computing Center (NHR@FAU) of the Friedrich-Alexander-Universität Erlangen-Nürnberg (FAU) under the NHR project b177dc (``SELRIQS''). NHR funding is provided by federal and Bavarian state authorities. NHR@FAU hardware is partially funded by the German Research Foundation (DFG) – 440719683.

\bibliographystyle{apsrev4-2}

\widetext
\clearpage 

\begin{center}
	\textbf{\large Supplemental Material for ``Order-by-disorder in the antiferromagnetic long-range transverse-field Ising model on the ruby lattice''}
	\\\vspace{0.5em}
	Antonia Duft$^{\rm 1}$, Jan A. Koziol$^{\rm 1}$, Patrick Adelhardt$^{\rm 1}$, Matthias Mühlhauser$^{\rm 1}$ and Kai P. Schmidt$^{\rm 1}$ \\
	\textit{$^{\text 1}$Friedrich-Alexander-Universit\"at
		Erlangen-N\"urnberg (FAU), \\Department Physik,
		Staudtstra{\ss}e 7, D-91058 Erlangen, Germany}
\end{center}

\setcounter{equation}{0}
\setcounter{figure}{0}
\setcounter{table}{0}
\setcounter{page}{1}
\makeatletter
\renewcommand{\theequation}{S\arabic{equation}}
\renewcommand{\thefigure}{S\arabic{figure}}
\renewcommand{\bibnumfmt}[1]{[S#1]}
\renewcommand{\citenumfont}[1]{S#1}
\setcounter{secnumdepth}{3}
\setcounter{section}{0}

In this Supplementary Material, we start by explaining the perturbative approaches used to derive series expansions about the low- and high-field limit of the investigated transverse-field Ising model on the ruby lattice. We then include additional insights into the analysis of the low-field limit, including discussions on the zero-field ground states. We offer further justifications for various statements made in the main text and present an outline of our perturbative calculations along with the results. Finally, we include comprehensive results from high-field series expansions that our reasoning in the main body of the work is based on.

\section{Series expansion methods}\label{sec:series-expansion}
In this section, we provide a description of the applied perturbative series expansion methods.
We start by shortly introducing Takahashi perturbation theory, with which we calculate ground-state energies in the low-field limit. We then focus on describing the method of perturbative continuous unitary transformations with which we calculate high-order series expansions in the high-field limit using linked-cluster expansions set up as a full graph decomposition. We use two different methods due to the distinct structures of the perturbative problem in the low- and high-field limit. Finally, we introduce the DlogPadé extrapolation technique which allows the quantitative extraction of quantum-critical properties.

\subsection{Takahashi perturbation theory}\label{sec:takahashi}
We calculate ground-state energy corrections to an unperturbed Hamiltonian $\mathcal{H}_0$ with unperturbed ground-state energy $E_0^{(0)}$ by a perturbation $\mathcal{V}$ with perturbation parameter $\lambda$,
\begin{align}
\mathcal{H} = \mathcal{H}_0 + \lambda \mathcal{V}\,,
\end{align}
using Takahashi perturbation theory \cite{Takahashi1977s}. In the scope of this work we calculate even order ground-state energy corrections up to and including order four in $\lambda$. These are given by the following expressions, where $P$ is a projector onto the unperturbed ground-state subspace, ${S = (1-P)/(E_0^{(0)}-\mathcal{H}_0)}$, and $\ket{\Psi_i}$ is one of the unperturbed ground states.
In second order, the diagonal correction to the ground-state energy is given by 
\begin{align}\label{eq:energy_correction_O2_takahashi}
E_0^{(2)}(\Psi_i) = \lambda^2 \bra{\Psi_i} P\mathcal{V}S\mathcal{V}P \ket{ \Psi_i} \,.
\end{align}
In fourth order, the diagonal correction to the ground-state energy is given by 
\begin{align}\label{eq:energy_correction_O4_takahashi}
E_0^{(4)}(\Psi_i) = \lambda^4  \bra{\Psi_i}  \left(P\mathcal{V}S\mathcal{V}S\mathcal{V}S\mathcal{V}P - \frac{1}{2} P\mathcal{V}S^2\mathcal{V}P\mathcal{V}S\mathcal{V}P - \frac{1}{2}P\mathcal{V}S\mathcal{V}P\mathcal{V}S^2\mathcal{V}P\right)\ket{\Psi_i} \,.
\end{align}
Note that the last two summands arise due to two consecutive second-order processes,  while the first summand describes a purely fourth-order process. In the problem we consider, the first off-diagonal matrix elements between unperturbed ground states appear in order six and can be evaluated using the operator sequence $P\mathcal{V}S\mathcal{V}S\mathcal{V}S\mathcal{V}S\mathcal{V}S\mathcal{V}P$.

\subsection{The pCUT method}
We use the method of perturbative continuous unitary transformations (pCUT) \cite{Knetter2000s,Knetter2003s} to calculate high-order series expansions of the elementary excitation gap in the high-field phase. 
In order to be able to apply the pCUT method to a system, it has to meet the following requirements. It must be possible to write the Hamiltonian as 
\begin{align}
\mathcal{H} = \mathcal{H}_0 + \lambda \mathcal{V}\,,
\end{align}
where the unperturbed Hamiltonian $\mathcal{H}_0$ is bounded from below and has an equidistant spectrum, i.\,e. $\mathcal{H}_0 = E_0 + \sum_i n_i$. This implies the interpretation of the elementary excitations of the model as quasi-particles (QPs) and the operator $Q=\sum_i n_i$ is defined to count the number of QPs in the system. The perturbation $\mathcal{V}$ similarly has to take the form 
\begin{align}
\mathcal{V} = \sum_{n=-N}^{N} T_n = \sum_{n=-N}^{N} \sum_l \tau_{n,l}
\end{align}
where $T_n$ changes the number of quasi-particles in the system by $n$, i.\,e., $[Q,T_n]=nT_n$, and can be further decomposed into operators $\tau_{n,l}$ acting on links $l$ between sites on the respective lattice. 

If those requirements are met, the pCUT method transforms the Hamiltonian $\mathcal{H}$ into an effective model $\mathcal{H}_\mathrm{eff}$ which conserves the QP number. As $\mathcal{H}_\mathrm{eff}$ is block-diagonal in the QP number, each QP block can be treated separately. The effective model can be written as 
\begin{align}\label{eq:pCUT_H_eff}
\mathcal{H}_\text{eff} &= \mathcal{H}_0 + \sum_k \lambda^k \sum_{\mathclap{\substack{|\underline{m}|=k \\ \sum_i m_i=0}}} C(\underline{m})T_{m_1} T_{m_2} \ldots T_{m_k}\,, 
\end{align}
where $C(\underline{m})\in \mathbb{Q}$ are model-independent coefficients and $\underline{m}=(m_1,m_2,\ldots, m_k)$ parameterizes all possible QP conserving processes in the respective order $k$.

\subsection{Linked-cluster expansion and graph decomposition}
In the evaluation of the effective Hamiltonian we exploit its cluster additive property as ensured by the pCUT method \cite{Knetter2003s}, i.\,e.,
\begin{align}\label{eq:def_cluster_additive}
\mathcal{H}_\mathrm{eff}^C = \mathcal{H}_\mathrm{eff}^A \otimes \mathrm{id}^B + \mathrm{id}^A \otimes \mathcal{H}_\mathrm{eff}^B\,,
\end{align}
where $A,B\subset C$ denote disjoint subsets of the full cluster $C$, which implies that only linked processes have a contribution (linked-cluster theorem \cite{Coester2015s}). The effective Hamiltonian can then be rewritten as
\begin{align}\label{eq:pCUT_H_eff_cluster_decomposition}
\mathcal{H}_\text{eff} = \mathcal{H}_0 + \sum_{k} \lambda^k \quad \sum_{\mathclap{\substack{|\underline{m}|=k \\ \sum_i m_i = 0}}} \quad \sum_{C_k} \, C(\underline{m})\quad \sum_{\mathclap{\substack{l_1,\ldots,l_k, \\ \cup_i l_i = C_k}}}\quad \tau_{m_1,l_1}\tau_{m_2,l_2}\ldots \tau_{m_k,l_k}\,,
\end{align}
where the sum runs over all possible linked clusters $C_k=\cup_i l_i$ in the respective perturbation order $k$ \cite{Coester2015s}. We set up the linked cluster expansion of $\mathcal{H}_\mathrm{eff}$ as a full graph decomposition where we evaluate the contributions of topologically distinct graphs in order to push the maximally achievable perturbation order \cite{Coester2015s,Muehlhauser2022s}. We use the contributions on the graphs and embed them on the lattice to determine the irreducible matrix elements of the effective one quasi-particle Hamiltonian in the thermodynamic limit \cite{Gelfand2000s, Oitmaa2006s}.

\subsection{DlogPadé extrapolations}\label{sec:extrapolation}
We use DlogPadé extrapolation techniques to quantitatively extract the critical point and exponent of a possible second-order quantum phase transition from the excitation gap $\Delta(\lambda)$  which we calculate as a power series in the perturbation parameter $\lambda$ with the pCUT method. For an in-depth discussion we refer to e.\,g. Refs.~\cite{Baker1996s, Guttmann1989s}. \\

We define the $[L,M]$ Padé extrapolant of the series $\Delta(\lambda)$ as 
\begin{align}\label{eq:pade}
P[L,M]_{\Delta}(\lambda):= \frac{P_L(\lambda)}{Q_M(\lambda)}=\frac{p_0+p_1 \lambda+...+p_L \lambda^L}{1+q_1\lambda+...+q_M\lambda^M}
\end{align}
with coefficients $p_i, q_i \in \mathbb{R}$ and $r=L+M$. The Taylor expansion of $P[L,M]$ up to order $r$ is required to recover the series $\Delta(\lambda)$. 
The $[L,M]$ DlogPadé extrapolant is based on the Padé approximant $P[L,M]_D(\lambda)$ of the logarithmic derivative of $\Delta(\lambda)$,
\begin{align}\label{eq:logarithmicDerivative}
D(\lambda) &:= \frac{\mathrm{d}}{\mathrm{d}\lambda}\ln \Delta(\lambda)\,, %= \frac{\Delta'(\lambda)}{\Delta(\lambda)}\,, 
\end{align}
and defined by 
\begin{align}
\mathrm{d}P[L,M]_\Delta(\lambda):=\exp\left(\int_0^\lambda\mathrm{d}\lambda'P[L,M]_D(\lambda')\right)\,.
\end{align} 

As the excitation gap is dominated by a power-law behavior around the critical point $\lambda_c$ of a second-order phase transition with $\Delta \propto |\lambda-\lambda_c|^{-z\nu}$, $\lambda_c$ can be extracted from the zero of $Q_M(\lambda)$ of the respective DlogPadé extrapolant. We exclude defective extrapolants that exhibit unphysical poles on the positive real axis up to the real critcial point $\lambda_c$. An estimate for the critical exponent is extracted from the residue
\begin{align}\label{eq:dlog_critical_exponent}
z\nu =\left. \frac{P_L(\lambda)}{\frac{\mathrm{d}}{\mathrm{d}\lambda}Q_M(\lambda)}\right\rvert_{\lambda=\lambda_c}\,.
\end{align}

\section{Low-field analysis}
In the following section we describe our strategy for investigating the low-field limit of the \jfull TFIM in detail and include illustrations of several statements made in the main text. We focus on the two limiting cases of the \jtwo and the \jthree TFIM, as introduced in the main text, and derive the \jfull case by combining the results of those two. For all cases we first reduce the complexity of the problem by regarding the ruby lattice as hexagonal plaquettes arranged on a triangular lattice. We understand the ground-state configurations by investigating those finite plaquettes first and then extending our findings to the full lattice. We note that for all illustrations of the ruby lattice in this section we chose the aspect ratio on the ruby lattice to be $\rho=r_2/r_1=\sqrt{3}$ (see Fig.~\ref{fig:j1j2-j1j3-gs_plaquettes}).

\subsection{Zero-field ground states}\label{sec:zero-field}
\begin{figure}[h]
	\centering
	\includegraphics[width=.7\textwidth]{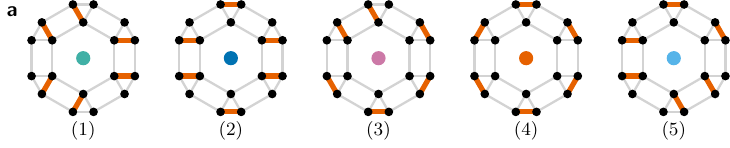}
	
	\vspace{.5cm}
	\includegraphics[width=.7\textwidth]{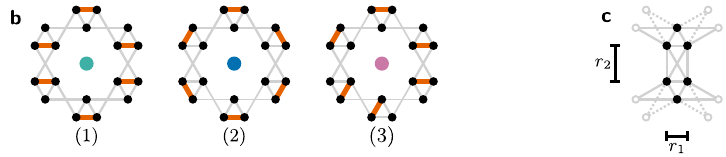}
	\caption{Depiction of the zero-field ground-state plaquette types in the \jtwo (\textbf{\textsf{a}}) and \jthree (\textbf{\textsf{b}}) TFIM and of a unit cell on the lattice (\textbf{\textsf{c}}). In (\textbf{\textsf{a}}) and (\textbf{\textsf{b}}), ferromagnetic bonds are drawn in orange, antiferromagnetic bonds in light gray. Each of the depicted plaquette types is a representative of all plaquettes related by rotations. In (\textbf{\textsf{c}}), the black dots and light gray lines show the sites and bonds on the ruby lattice which are parts of its unit cell. The dashed lines and gray circles in the unit cell indicate its integration into the lattice but are not part of the unit cell itself. }
	\label{fig:j1j2-j1j3-gs_plaquettes}
\end{figure}

We start by discussing ground-state configurations in the case of zero transverse magnetic field, $h=0$, where the model is reduced to only the antiferromagnetic Ising couplings $J_i\geq 0$ ($i=1,2,3$),
\begin{align}
\tilde{\mathcal{H}}_0 = J_1 \hspace{-0.1cm} \sum_{\langle i,j \rangle_1}\hspace{-0.1cm} \sigma_{i}^z\sigma_{j}^z + J_2 \hspace{-0.1cm} \sum_{\langle i,j \rangle_2} \hspace{-0.1cm} \sigma_{i}^z\sigma_{j}^z +J_3 \hspace{-0.1cm} \sum_{\langle i,j \rangle_3} \hspace{-0.1cm} \sigma_{i}^z\sigma_{j}^z \,.
\label{eq:H_zero-field}
\end{align}
Consequently, zero-field ground states are states with the smallest possible number of ferromagnetic frustrated bonds, each costing the energy penalty $J_i$. Due to geometrical frustration, all triangles on the lattice contain exactly one ferromagnetic bond. The \jfull model on the ruby lattice contains two types of triangles: $J_1$-triangles and $J_1$-$J_2$-$J_3$-triangles (see Fig.~\ref{fig:j1j2-j1j3-gs_plaquettes}c). If either the $J_2$ or the $J_3$ interaction is set to zero, only $J_1$ triangles remain. As argued, we analyze the zero-field ground-state configurations by regarding the ruby lattice as an arrangement of hexagonal plaquettes. 

\begin{figure}[h]
	\centering
	\begin{minipage}{.49\textwidth}
		\centering
		\includegraphics[width=.9\textwidth]{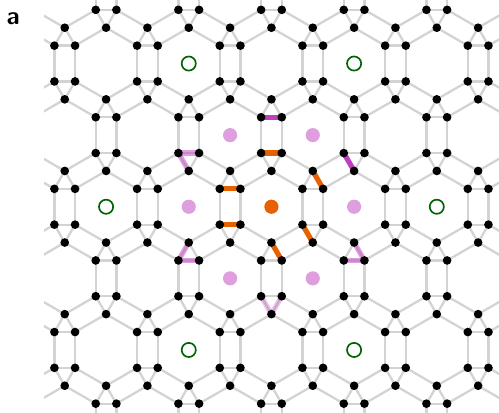}
	\end{minipage}
	\begin{minipage}{.49\textwidth}
		\centering
		\includegraphics[width=.9\textwidth]{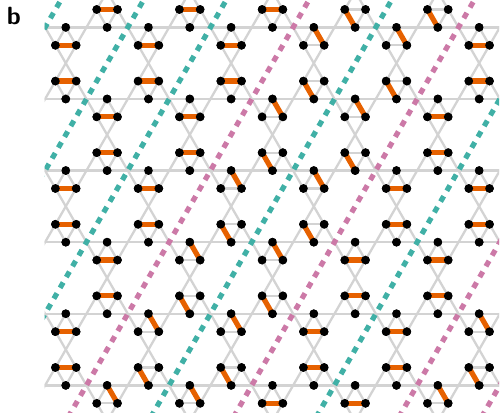}
	\end{minipage}
	\caption{Illustration of the extensive ground-state degeneracy in the zero-field limit of the \jtwo (\textbf{\textsf{a}}) and \jthree (\textbf{\textsf{b}}) case. \\
		(\textbf{\textsf{a}}) Choosing the configuration of a central plaquette (\protect\scircleorange), exemplarily type-(5) with ferromagnetic bonds in orange) places constraints on the six surrounding plaquettes (\protect\scirclelilaopa): dark purple lines resemble enforced ferromagnetic bonds, while light purple lines resemble the freedom of choosing one of the two respective bonds to be ferromagnetic. In the second-to-next surrounding circle, six plaquettes (\protect\scirclegreenborder) have no fixed ferromagnetic bonds and their configuration can be chosen from at least two plaquette types. \\
		(\textbf{\textsf{b}}) The exemplary ground-state configuration (ferromagnetic bonds in orange) consists of stripes of different plaquette types (indicated by dashed lines). Next to a type-(1) stripe (\protect\sdashedpink) one can always place either another type-(1) stripe or a type-(3) stripe (\protect\sdashedbluegreen) and vice versa. }
	\label{fig:j1j2-j1j3-gs_degeneracy}
\end{figure}

For the \jtwo TFIM with $J_3=0$ and $J_{1,2}>0$ we find five types of ground-state plaquettes as depicted in Fig.~\ref{fig:j1j2-j1j3-gs_plaquettes}a. Note that the depiction in terms of ferromagnetic and antiferromagnetic bonds does not specify the orientation of the respective spins, giving rise to a global spin flip symmetry. We define each plaquette-type to include all plaquettes related to the one depicted by rotations. On the full lattice we find an extensive number of possibilities to combine those five types of plaquette configurations as illustrated in Fig.~\ref{fig:j1j2-j1j3-gs_degeneracy}a, resulting in an extensively degenerate ground-state manifold. Note, an extensive ground-state degeneracy means that for large systems, the logarithm of the number of ground states scales with the system size.

For the \jthree TFIM with $J_2=0$ and $J_{1,3}>0$ we find three types of ground-state plaquettes as depicted in Fig.~\ref{fig:j1j2-j1j3-gs_plaquettes}b. Again, each of those plaquette types yields multiple explicit spin configurations. Note, however, that none of the spin configurations which are a ground state for the \jthree case are a ground state for the \jtwo case and vice versa, even though the configurations of ferromagnetic bonds look the same. On the full lattice we find an extensive number of possibilities to combine the three types of plaquette configurations as illustrated in Fig.~\ref{fig:j1j2-j1j3-gs_degeneracy}b, resulting again in an extensively degenerate ground-state manifold. 

For the full \jfull TFIM with $J_{1,2,3}>0$ and $J_1>J_{2,3}$, the lattice contains additional $J_1$-$J_2$-$J_3$-triangles (see Fig.~\ref{fig:j1j2-j1j3-gs_plaquettes}c). Since all $J_1$-triangles have exactly one ferromagnetic $J_1$ bond, in 1/3 of those triangles the $J_1$ bond is ferromagnetic. For the remaining 2/3 of those triangles, this results in one additional ferromagnetic $J_2$- or $J_3$-bond per triangle. The energy per site is given by $-\frac{1}{3}J_1+ ( \frac{4}{3}\chi-1)J_2 + ( \frac{4}{3}(1-\chi)-1)J_3$ with $\chi\in[0,1]$ depending on the ratio $J_2/J_3$. We distinguish the following three cases. If $J_2>J_3$, it is energetically beneficial for all $J_2$ bonds to be antiferromagnetic and we find $2/3$ ferromagnetic $J_3$ bonds. With $\chi=0$ the energy per site is $-\frac{1}{3}J_1-J_2+\frac{1}{3}J_3$. Configurations with zero ferromagnetic $J_2$ bonds are exactly the \jtwo ground states, which thus realize the ground states of the \jfull model with $J_2>J_3$. There exist no configurations with a smaller number of ferromagnetic bonds. The same holds for $J_3>J_2$ with $J_2$ and $J_3$ interchanged and $\chi=1$. 

If $J_2=J_3$, additional frustration is present as the two bonds can not be distinguished and the energy per site is $-\frac{1}{3}J_1-\frac{2}{3}J_2$ independent of $\chi$. In this case, all zero-field \jtwo and \jthree ground states are ground states of the full $J_1$-$J_2$-$J_3$ model. Simultaneously, we find multiple further ground states: the ground-state plaquettes of the \jtwo and \jthree model can be mixed and we also find further possible plaquette configurations. Thus, an enhanced ground-state degeneracy is present. 

\subsection{Leading order low-field effects}
In the following, we illustrate our analysis of the leading order processes in the limit of a small transverse magnetic field $h$ in the limiting cases of the \jtwo and \jthree TFIM, for which we perform degenerate perturbation theory in the transverse field around $h=0$ using Takahashi's method \cite{Takahashi1977s} (see Sec.~\ref{sec:takahashi}). 

\subsubsection{$J_1$-$J_2$ case}\label{sec:j1j2_lowfield}
\begin{figure}[h]
	\centering
	\includegraphics[width=.3\textwidth]{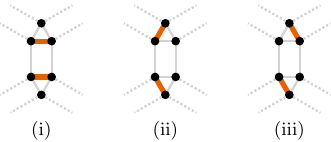}
	\caption{Unit cell configurations appearing in zero-field ground states in the \jtwo TFIM. Ferromagnetic bonds are drawn in orange, antiferromagnetic bonds in light gray. Antiferromagnetic $J_2$ bonds leaving the unit cell are drawn as dotted lines and have to be taken into account for the calculation of the corrections of the ground-state energy.}
	\label{fig:j1j2-unit_cells}
\end{figure}

\textbf{Diagonal corrections -} We calculate diagonal corrections to the zero-field ground-state energy of the \jtwo TFIM up to order four in the field $h$. For this we note that every zero-field plaquette in Fig.~\ref{fig:j1j2-j1j3-gs_plaquettes} and thus every zero-field ground state can be constructed from the three unit-cell configurations depicted in Fig.~\ref{fig:j1j2-unit_cells}. As all processes relevant up to fourth order are contained within such a unit cell, the respective ground-state energy correction of any zero-field state can be calculated from the corrections of those three configurations which we do in the following. 

The unperturbed Hamiltonian is given by $\tilde{\mathcal{H}}_0$ in Eq.~\eqref{eq:H_zero-field} and yields the unperturbed ground-state energy
\begin{align}
E_0^{(0)} = -2 J_1 - 6J_2 \,
\end{align}
for all three configurations. The perturbation $\mathcal{V}=\sum_i \sigma_i^x$ with perturbation parameter $\lambda=h\ll J$ only yields nonzero corrections for even orders in $h$. For the calculations it is illustrative to distinguish the sites in a zero-field configuration by whether they are connected to one or zero ferromagnetic $J_1$ bonds and we introduce the notion of (anti)ferromagnetic sites, implying that the site has (zero) one ferromagnetic $J_1$ bond(s). 

The second-order ground-state energy correction includes all processes acting twice on the same site. The contribution of a process depends only on whether the site is ferro- or antiferromagnetic, yielding the same contribution for all three configurations:
\begin{align}
E_0^{(2)} = -h^2 \left(2\cdot\frac{1}{4J_1 + 4J_2} + 4\cdot\frac{1}{4J_2}\right)\,.
\end{align}
The fourth-order ground-state energy correction as given in Eq.~\eqref{eq:energy_correction_O4_takahashi} contains summands consisting of two consecutive second-order processes, which are the same for all configurations, and a purely fourth-order process acting on two interacting spins which distinguishes the configurations energetically. The result of a process depends on whether the two involved sites are both ferromagnetic, both antiferromagnetic or one of each. For processes acting on two spins interacting via a $J_1$ bond, this implies that the result depends only on whether the bond itself is ferro- or antiferromagnetic, yielding the same contribution for all configurations. The antiferromagnetic $J_2$ bonds however connect different types of sites for the three configurations and respective fourth-order processes distinguish the configurations energetically as follows:
\begin{align}
E_{J_2}^{(4)}(\mathrm{i}) &= -h^4\cdot 8\cdot \left( \frac{1}{4J_2}\right)^3 \nonumber\,,\\
E_{J_2}^{(4)}(\mathrm{ii}) &=	-h^4 \cdot \left(4\cdot \left(\frac{1}{4J_2}\right) ^3 + 4\cdot \frac{1}{8J_1 + 4J_2} \cdot \left(\frac{1}{4J_1 + 4J_2}\right)^2\right)\,, \label{eq:j1j2_o4_corr}\\
E_{J_2}^{(4)}(\mathrm{iii}) &= -h^4  \cdot 2 \cdot \frac{1}{4J_1 + 4J_2}\cdot \left(\frac{1}{4J_2} + \frac{1}{4J_1 + 4J_2} \right)^2\nonumber \,.
\end{align}
From this we find the following hierarchy of energy lowering corrections $E_{J_2}^{(4)}$ for arbitrary ${J_1,J_2>0}$:
\begin{align}
|E_{J_2}^{(4)}(\mathrm{i})| > |E_{J_2}^{(4)}(\mathrm{ii})| > |E_{J_2}^{(4)}(\mathrm{iii})|\,.
\end{align}
As all other contributions to the fourth-order corrections yield the same result for all three unit cell configurations this defines the overall energetic hierarchy of the configurations up to fourth order, $(\mathrm{i})\to (\mathrm{ii}) \to (\mathrm{iii})$. We confirm this analytically derived result by a numeric evaluation of the respective ground-state energy correction of the configurations for $J_1 = J_2$.

\begin{figure}[h]
	\begin{minipage}{.49\textwidth}
		\centering
		\includegraphics[width=.9\textwidth]{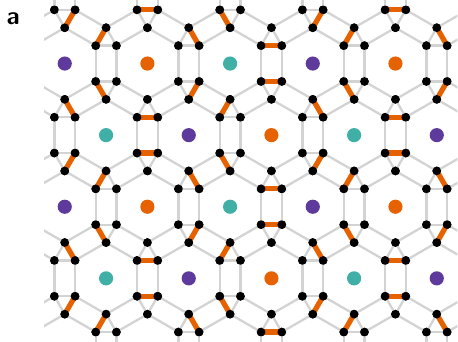}
	\end{minipage}
	\begin{minipage}{.49\textwidth}
		\centering
		\includegraphics[width=.9\textwidth]{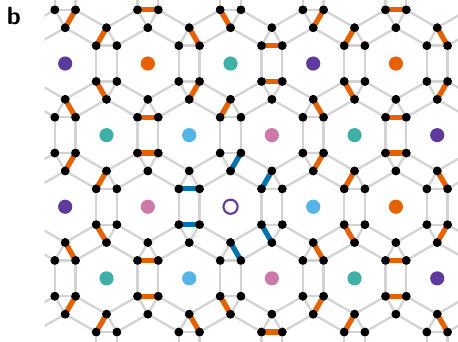}
	\end{minipage}
	\caption{Two zero-field ground state configurations of the \jtwo TFIM. The  colored dots indicate the plaquette types with the following color code: \protect\scirclebluegreen \, type (1), \protect\scirclepurple\, and \protect\scirclepurpleborder\, type ($1^\prime$), \protect\scirclepink\, type (3), \protect\scircleorange\, type (4), \protect\scircleskyblue\, type (5). \\
		(\textbf{\textsf{a}}) The zero-field ground-state configuration containing only type-(1) and type-(4) plaquettes (columnar state). Here we explicitly distinguish the rotated type-(1) \protect\scirclebluegreen \, plaquettes as type-($1^\prime$)  \protect\scirclepurple\, in order to illustrate the $\bm{k}=(2\pi/3,-2\pi/3)$ order. We find that states which exhibit this order are selected from the zero-field ground-state manifold by a small transverse magnetic field in fourth order.  \\
		(\textbf{\textsf{b}}) The configuration resulting from the one in the left panel by a resonance of the type-(1)  plaquette marked with \protect\scirclepurpleborder\, (ferromagnetic bonds in dark blue) to type ($1^\prime$) (compare Fig.~\ref{fig:j1j2-resonating_plaquette}), showing the presence of the resonating sixth-order process on the full lattice.}
	\label{fig:j1j2-full_lattice_resonance}
\end{figure}

We continue by discussing the implications regarding the energy splitting of zero-field ground-states on the full lattice. All zero-field plaquettes in Fig.~\ref{fig:j1j2-j1j3-gs_plaquettes} contain the same fraction of 1/3 of the energetically most beneficial configuration (i). For this statement, we also consider that a ferromagnetic bond along the outer hexagon of a plaquette enforces the continuation to a configuration of type (i) in order to yield a valid zero-field ground state. The most beneficial plaquettes are thus the ones maximizing the number of configuration (ii), i.\,e., plaquettes (1) and (4). The respective zero-field ground states containing only those plaquettes realize structures with ordering vector $\bm{k}=(2\pi/3,-2\pi/3)$ as depicted in in Fig.~\ref{fig:j1j2-full_lattice_resonance}\,a. We also refer to this as columnar order, motivated by the analogy to the Rokhsar Kivelson quantum dimer model (RK QDM) \cite{Rokhsar1988s,Moessner2001news} (see main text and following paragraphs). Thus, the zero-field ground-state degeneracy is lifted in a diagonal order-by-disorder scenario in fourth order in the magnetic field. \\

\begin{figure}[h]
	\centering
	\includegraphics[width=.4\textwidth]{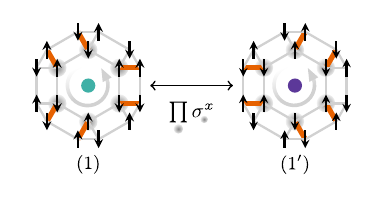}
	\caption{Illustration of the lowest-order off-diagonal process contributing to the effective low-field description of the \jtwo TFIM. Applying a spin-flip to the six spins along the inner hexagon of a type-(1) plaquette, highlighted by gray shadows, switches the position of the ferromagnetic $J_1$ bond at each of those spins as shown. The resulting configuration is equivalent to the initial configuration rotated clockwise by $60^\circ$ and thus also a zero-field ground state we term type-($1^\prime$). Applying the same spin flip sequence for a second time results in the initial ground state.}
	\label{fig:j1j2-resonating_plaquette}
\end{figure}

\textbf{Off-diagonal corrections -} On the level of single plaquettes we find the lowest-order off-diagonal process within the zero-field ground-state manifold to be a sixth-order process acting on the six sites of the inner hexagon a type-(1) plaquette. This is equivalent to a rotation of the plaquette by $60^\circ$ into a type-($1^\prime$) plaquette as shown in Fig.~\ref{fig:j1j2-resonating_plaquette}. 
We find that mapping between plaquettes of type (1) and ($1^\prime$) within an arbitrary zero-field ground state changes each neighboring plaquette into another valid zero-field ground-state plaquette. Thus, the sixth-order process on type-(1) plaquettes induces resonances between different zero-field ground states. We illustrate the new ground-state configuration in Fig.~\ref{fig:j1j2-full_lattice_resonance}\,b after such a resonating process acted on the ground state in Fig.~\ref{fig:j1j2-full_lattice_resonance}\,a.
The respective ground states with maximal density of type-(1) and ($1^\prime$) plaquettes are the same $\bm{k}=(2\pi/3,-2\pi/3)$ ordered structures that are selected by diagonal energy corrections in order four in the magnetic field (see Fig.~\ref{fig:j1j2-full_lattice_resonance}a). \\

Summarizing, fourth-order diagonal energy corrections lift the zero-field ground-state degeneracy and benefit the subspace of columnar states with $\bm{k}=(2\pi/3,-2\pi/3)$ with maximal density of type-(1) and ($1^\prime$) plaquettes. The leading-order off-diagonal process in sixth order induces resonances between those type-(1) and ($1^\prime$) plaquettes which are thus maximized within the subspace selected by the fourth-order diagonal corrections. In analogy to the RK QDM on the honeycomb lattice \cite{Rokhsar1988s,Moessner2001news}, we write down an effective low-field description containing the leading-order diagonal and off-diagonal contribution as a sum over all plaquettes on the lattice:
\begin{align}\label{eq:j1j2_effective}
\mathcal{H}_\mathrm{eff} = \bar{E}_0 &+ \frac{h^4}{2} \sum_{\plaquette[9pt]} \bigg(\big( \underbrace{3 E_{J_2}^{(4)}(\mathrm{i}) + 3 E_{J_2}^{(4)}(\mathrm{ii})}_{=:E_{J_2}^{(4)}(\protect\circlebluegreen)}\big)\ket{\resonatingplaquette}\bra{\resonatingplaquette} + \big(\underbrace{2 E_{J_2}^{(4)}(\mathrm{i}) + 4 E_{J_2}^{(4)}(\mathrm{iii})}_{=:E_{J_2}^{(4)}(\protect\circleblue)}\big)\ket{\plaquettetwo}\bra{\plaquettetwo}   \\[1.5ex]
&+ \big(\underbrace{1 E_{J_2}^{(4)}(\mathrm{i}) + 3 E_{J_2}^{(4)}(\mathrm{ii}) + 2 E_{J_2}^{(4)}(\mathrm{iii})}_{=:E_{J_2}^{(4)}(\protect\circlepink)}\big)\ket{\plaquettethree}\bra{\plaquettethree} + \underbrace{6 E_{J_2}^{(4)}(\mathrm{ii})}_{=:E_{J_2}^{(4)}(\protect\circleorange)} \ket{\plaquettefour}\bra{\plaquettefour} + \nonumber\\  &+  \big(\underbrace{2 E_{J_2}^{(4)}(\mathrm{i}) + 2 E_{J_2}^{(4)}(\mathrm{ii}) + 2 E_{J_2}^{(4)}(\mathrm{iii})}_{=:E_{J_2}^{(4)}(\protect\circleskyblue)}\big)\ket{\plaquettefive}\bra{\plaquettefive} \bigg)	+  h^6 \sum_{\plaquette[9pt]}  \frac{63}{16 J_2^5 }  \left(\ket{\resonatingplaquette}\bra{\resonatingplaquettetwo} +\mathrm{h.c.} \right)  \nonumber \\
=: \bar{E}_0 &+\frac{ h^4}{2} \sum_{\plaquette[8pt]} \quad \sum_{\mathclap{\substack{\protect\circlelightgray=\protect\circlebluegreen,\protect\circleblue,\protect\circlepink, \\ \quad \protect\circleorange,\protect\circleskyblue}}} \quad
\left( E_{J_2}^{(4)}({\adjustbox{valign=c}{\protect\scirclelightgray}})\ket{\plaquetteplain}\bra{\plaquetteplain} \right)	+  h^6 \sum_{\plaquette[8pt]}  E_\mathrm{res}^{(6)} \left(\ket{\resonatingplaquette}\bra{\resonatingplaquettetwo} +\mathrm{h.c.} \right) \nonumber \,,
\end{align} 
with constant $\bar{E}_0$ containing perturbative contributions which are equal for all ground states. In the diagonal part each of the five plaquette types is defined to include plaquettes related to the depicted one by symmetry operations like rotations like before. In the off-diagonal part we explicitly write down the plaquette types (1) and ($1^\prime$) related by rotation as it is physically relevant for the resonance. Each relevant diagonal fourth-order process acting on the $J_2$ bonds is part of two plaquettes and thus considered twice in the sum over plaquettes which is accounted for by a factor 1/2.

The effective model in Eq.~\eqref{eq:j1j2_effective} is closely related to the RK QDM on the honeycomb lattice discussed in Ref.~\cite{Moessner2001news}. This allows to make estimates regarding the phase diagram, in particular for the extent of the columnar phase. As discussed in Sec.~\ref{sec:j1j2-j1j3-high-field}, the findings resulting from these estimates are consistent with our high-field high-order series expansions. The RK QDM contains a kinetic term that we directly identify with the resonating sixth-order process, $-t=h^6E_\mathrm{res}^{(6)}$. The RK QDM further contains a single diagonal term giving the respective resonating plaquette, in our case the plaquette of type-(1), an energy offset $v$. In contrast, the effective model in Eq.~\eqref{eq:j1j2_effective} contains five different plaquette configurations, each associated with a distinct diagonal energy correction $E_{J_2}^{(4)}$. Thus, to associate our low-field description exactly with the RK QDM, we need the condition $E_{J_2}^{(4)}({\adjustbox{valign=c}{\circleblue}})=E_{J_2}^{(4)}({\adjustbox{valign=c}{\circlepink}})=E_{J_2}^{(4)}({\adjustbox{valign=c}{\circleorange}})=E_{J_2}^{(4)}({\adjustbox{valign=c}{\circleskyblue}})$, which we will assume for our estimation.%, which is not the case here (c.\,f. Eqs.~\ref{eq:j1j2_o4_corr}). 
	
The RK QDM on the honeycomb lattice exhibits a columnar phase for $v/t<-0.2$, which we associate with the low-field order selected by the fourth-order diagonal energy corrections (see Fig.~\ref{fig:j1j2-full_lattice_resonance}a). At $v/t=-0.2\pm0.05$ the RK QDM undergoes a quantum phase transition to a resonating plaquette phase \cite{Moessner2001news}. In order to estimate the point $\lambda_\mathrm{clock}$ of the respective phase transition from the columnar ordered to the resonating (clock-ordered) phase in the \jtwo TFIM on the ruby lattice as a function of $h$, $J_1$ and $J_2$, we %assume that plaquettes (2) to (5) all have the same energy correction . We can then 
identify $v=h^4(E_{J_2}^{(4)}({\adjustbox{valign=c}{\protect\circlebluegreen}})-E_{J_2}^{(4)}({\adjustbox{valign=c}{\protect\circlelightgray}}))/2$ and calculate $v/t=-0.2$ to determine $\lambda_\mathrm{clock}$ for ${\adjustbox{valign=c}{\circlelightgray}}={\adjustbox{valign=c}{\circleblue}},{\adjustbox{valign=c}{\circlepink}},{\adjustbox{valign=c}{\circleorange}},{\adjustbox{valign=c}{\circleskyblue}}$. For a plot of the results, see Fig.~\ref{fig:j1j2-j1j3-ratio}. 

We note that the resonance of a type-(1) plaquette within a columnar state as illustrated in Fig.~\ref{fig:j1j2-full_lattice_resonance} leaves the subspace of zero-field ground states benefited by diagonal corrections in fourth order, as six configurations (ii) are turned into (iii). The subspaces of one resonated plaquette and zero resonated plaquettes have an energy splitting of ${6 |E_{J_2}^{(4)}(\mathrm{ii}) - E_{J_2}^{(4)}(\mathrm{iii})|}$. This can be extended to multiple resonated plaquettes. Note that resonating a second plaquette does not necessarily turn six configurations from (ii) to (iii): if the two resonated plaquettes are connected by a unit cell, this unit cell undergoes the process $(\mathrm{ii}) \to (\mathrm{iii}) \to(\mathrm{ii})$ upon the two resonating processes, i.\,e., it is of type (ii) in the end and hence only 11 unit cells are turned from (ii) to (iii). From this we conclude that it is energetically beneficial for resonances to occur on plaquettes connected by a unit cell, i.e., plaquettes close to each other. 

\subsubsection{$J_1$-$J_3$ case}
\begin{figure}[h]
	\centering
	\includegraphics[width=.3\textwidth]{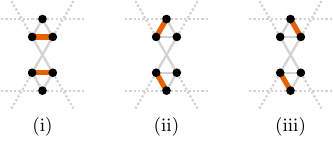}
	\caption{Unit-cell configurations appearing in zero-field ground states in the \jthree TFIM. Ferromagnetic bonds are drawn in orange, antiferromagnetic bonds in light gray. Antiferromagnetic $J_3$ bonds leaving the unit cell are drawn as dotted lines and have to be taken into account for the calculation of the right corrections of the ground-state energy. }
	\label{fig:j1j3-unit_cells}
\end{figure}

\textbf{Diagonal corrections -} We calculate diagonal corrections to the zero-field ground-state energy of the \jthree TFIM up to order four in $h$ analogous to the procedure for the \jtwo TFIM described in Sec.~\ref{sec:j1j2_lowfield}. As each zero-field ground-state configuration is built from different mixtures of the three unit-cell configurations illustrated in Fig.~\ref{fig:j1j3-unit_cells}, we determine the most beneficial ground state by ranking those three unit cell configurations by their energy corrections. 

We note that the three configurations of ferromagnetic $J_1$ bonds look like the ones in the unit cells for the \jtwo case (compare Fig.~\ref{fig:j1j2-unit_cells}) and the unperturbed ground-state energy of the three configurations is equally given by $E_0^{(0)}=-2J_1-6J_3$. However, the $J_2$ and $J_3$ bonds connect different sites within the unit cells such that the spin configurations may differ. We find that configuration (i) from the two models are equivalent regarding the explicit spin configuration and geometry of (anti)ferromagnetic bonds. Further, configuration (ii) in the \jthree case is equivalent to configuration (iii) in the \jtwo case and vice versa. Consequently, we find that the energy corrections and the resulting energetic hierarchy of unit cell configurations are equivalent to the \jtwo case for $J_2 \to J_3$ and upon swapping (ii) and (iii). 
In second order, the correction to the ground-state energy is given by 
\begin{align}
E_0^{(2)} = -h^2 \left(4\cdot\frac{1}{4J_3} + 2\cdot \frac{1}{4J_1 + 4J_3}\right)\,.
\end{align} 
for all three configurations. In fourth order, the three configurations are distinguished by purely fourth-order processes acting on two spins connected by a $J_3$ bond,
\begin{align}
E_{J_3}^{(4)}(\mathrm{i}) &= -h^4\cdot 8\cdot \left( \frac{1}{4J_3}\right)^3 \,,\\
E_{J_3}^{(4)}(\mathrm{ii}) &= -h^4  \cdot 2 \cdot \frac{1}{4J_1 + 4J_3}\cdot \left(\frac{1}{4J_3} + \frac{1}{4J_1 + 4J_3} \right)^2 \,\\
E_{J_3}^{(4)}(\mathrm{iii}) &=	-h^4 \cdot \left(4\cdot \left(\frac{1}{4J_3}\right) ^3 + 4\cdot \frac{1}{8J_1 + 4J_3} \cdot \left(\frac{1}{4J_1 + 4J_3}\right)^2\right)\,,
\end{align}
resulting in the overall energetic hierarchy of the configurations up to fourth order, $(\mathrm{i}) \to (\mathrm{iii}) \to (\mathrm{ii})$, for arbitrary $J_1,J_3>0$.

\begin{figure}[h]
	\centering
	\includegraphics[width=.49\textwidth]{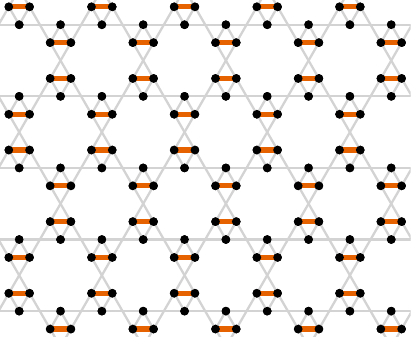}
	\caption{Zero-field ground state of the \jthree TFIM containing only type-(1) plaquettes which is favored by the a small transverse magnetic field in fourth order. Orange lines resemble ferromagnetic bonds, light gray lines antiferromagnetic ones.}
	\label{fig:j1j3-gs_configuration}
\end{figure}

Similar to the \jtwo case, all \jthree zero-field plaquettes in Fig.~\ref{fig:j1j2-j1j3-gs_plaquettes} enforce the same fraction of 1/3 of the energetically most beneficial configuration (i). Plaquettes of type-(1) maximize the number of the second-most-beneficial configuration (iii). The configuration on the full lattice which maximizes the number of such type-(1) plaquettes is the one containing only type-(1) plaquettes exhibiting $\bm{k}=(0,0)$ order, as illustrated in Fig.~\ref{fig:j1j3-gs_configuration}, and the ground-state degeneracy at zero magnetic field is lifted in a diagonal order-by-disorder scenario in fourth order in the magnetic field.

\textbf{Off-diagonal corrections -} In contrast to the \jtwo TFIM we find no resonating process between different zero-field ground states in any finite order in $h$, as can be understood from the geometry of $J_3$ bonds which form decoupled chains over the full lattice (compare straight gray lines in Fig.~\ref{fig:j1j3-gs_configuration}) with each site being contained in exactly one chain. In a zero-field ground state, all $J_3$ bonds are antiferromagnetic and the spins are aligned antiparallel along all chains. Mapping between different zero-field ground states requires inverting an entire chain of spins and thus an infinite order in $h$.

In summary, we express the effective low-field model in the \jthree model as 
\begin{align}
\mathcal{H}_\mathrm{eff} &= \bar{E}_0 +  h^4 \sum_{\schiffchen[5pt]} \left( E_{J_3}^{(4)}(\mathrm{i})\ket{\schiffchena}\bra{\schiffchena} + E_{J_3}^{(4)}(\mathrm{ii}) \ket{\schiffchenb}\bra{\schiffchenb} + E_{J_3}^{(4)}(\mathrm{iii}) \ket{\schiffchenc}\bra{\schiffchenc} \right) \,,
\end{align}
with constant $\bar{E}_0$ containing perturbative contributions which are equal for all ground states. 

\section{High-field series expansions}
In this section we discuss our results in the limit of high magnetic fields, $h\gg J$, of the \jfull TFIM. Applying the pCUT method in combination with a graph decomposition (see Sec.~\ref{sec:series-expansion}) we obtain the one quasi-particle effective Hamiltonian $\mathcal{H}_\mathrm{eff}$ in order 9 in $\lambda=J/2h$ for the general \jfull TFIM, in order 10 for the reduced \jtwo and \jthree TFIM, and in order 11 for the respective limiting cases $J_1=J_2$ and $J_1=J_3$. Note that we define the three interaction strengths $J_{1,2,3}$ to be related to each other, with the explicit definitions stated in the respective sections, and thus only consider a single perturbation parameter $\lambda$.

\subsection{Dispersion}\label{sec:dispersion}
As the ruby lattice has a unit cell containing six sites the one quasi-particle effective Hamiltonian reduces to a $6\times 6$ matrix $h_{lm}(\bm{k})$ with $l,m\in {1,2,\ldots,6}$ in momentum space using a Fourier transformation. Due to the lattice symmetries, this matrix exhibits the following relations:
\begin{align}
h_{11} (k_1,k_2) = h_{66} (k_1,k_2) =	h_{22} (k_1,k_1-k_2) &= h_{55} (k_1,k_1-k_2) =	h_{33} (k_1-k_2,k_1) = h_{44} (k_1-k_2,k_1) \\	%checked
h_{12} (k_1,k_2) = h_{56} (k_1,k_2) &= h_{13} (k_2,k_1) = h_{46} (k_2,k_1) \nonumber \\ %checked
h_{14} (k_1,k_2) = h_{36} (k_1,k_2) &= h_{15} (k_2,k_1) = h_{26} (k_2,k_1) \nonumber\\ %checked
h_{23} (k_1,k_2) &= h_{45} (k_1,k_2) \nonumber\\ %checked
h_{24} (k_1,k_2) &= h_{35} (k_1,k_2) \nonumber\\ %checked
h_{25} (k_1,k_2) &= h_{34} (k_2,k_1) \nonumber\\ %checked
h_{\alpha\beta} &= (h_{\beta\alpha})^*\nonumber
\end{align}
and only the seven elements $h_{11}, h_{12}, h_{14}, h_{16}, h_{23}, h_{24}, h_{25}$ have to be calculated. 
We obtain the dispersion containing six energy bands by diagonalizing $h(\bm{k})$, where we find a qualitative difference between the \jtwo and the \jthree TFIM which extends to the full \jfull TFIM.  

\begin{figure}[h]
	\centering
	\includegraphics[width=.325\textwidth]{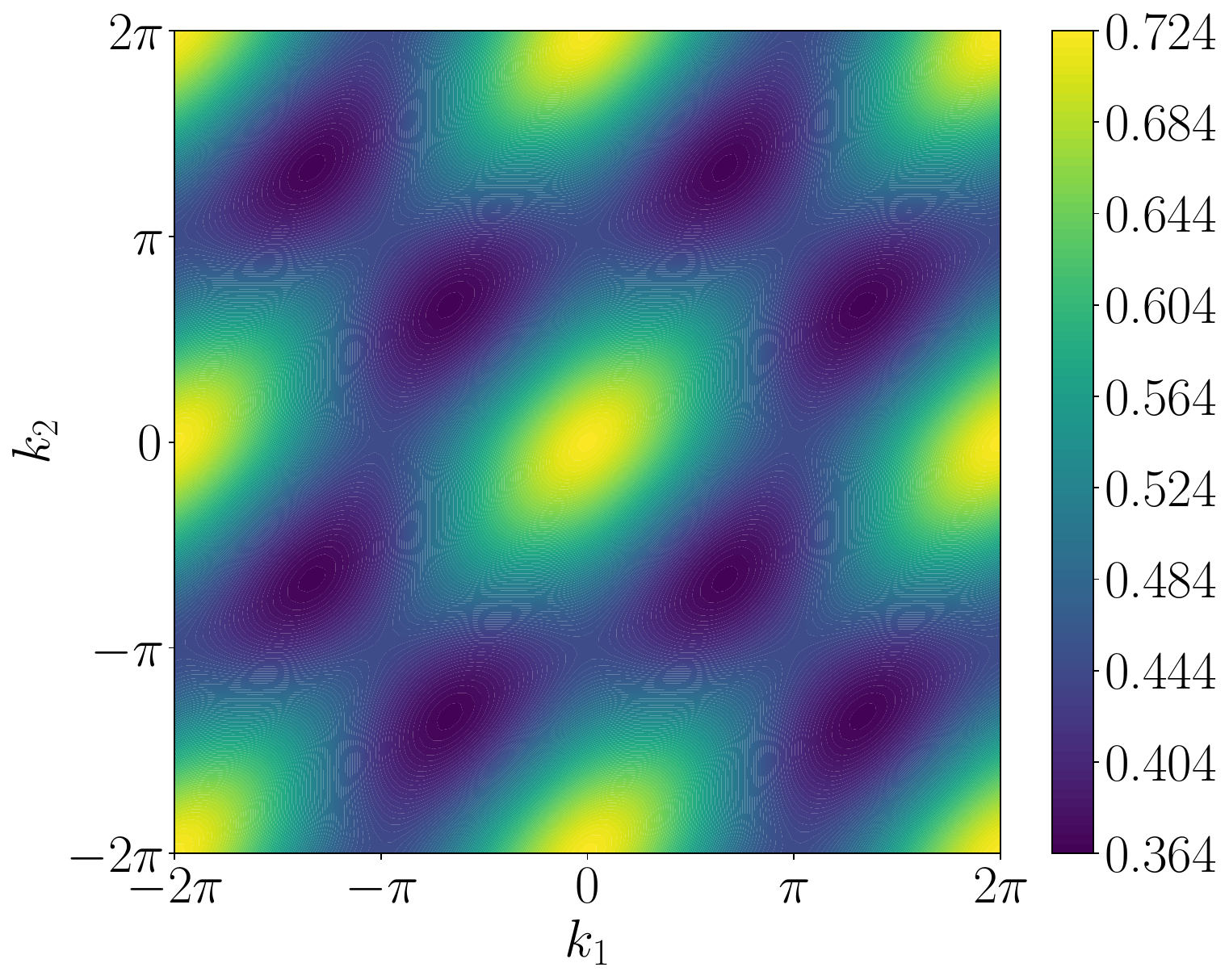}
	\includegraphics[width=.325\textwidth]{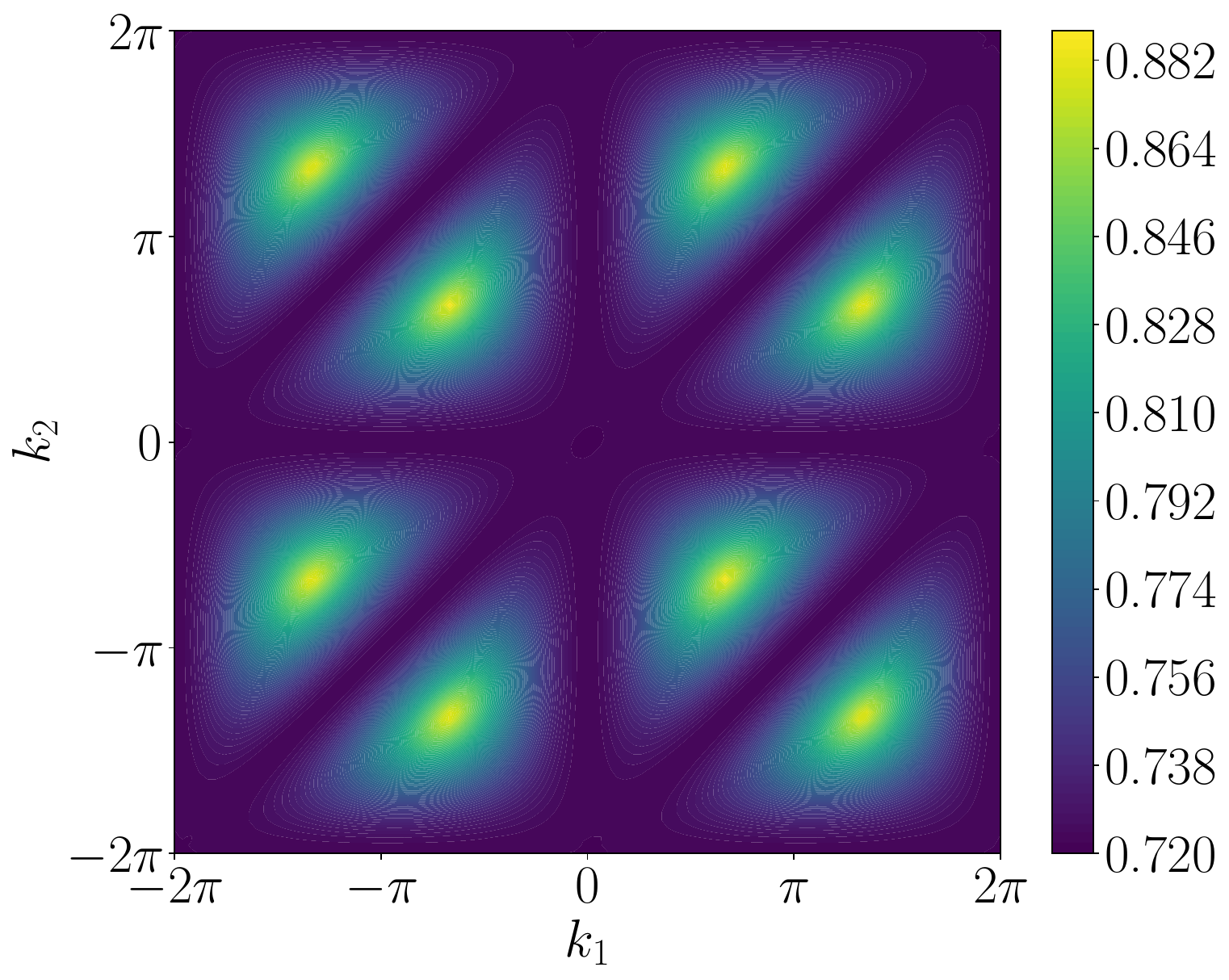}
	\includegraphics[width=.325\textwidth]{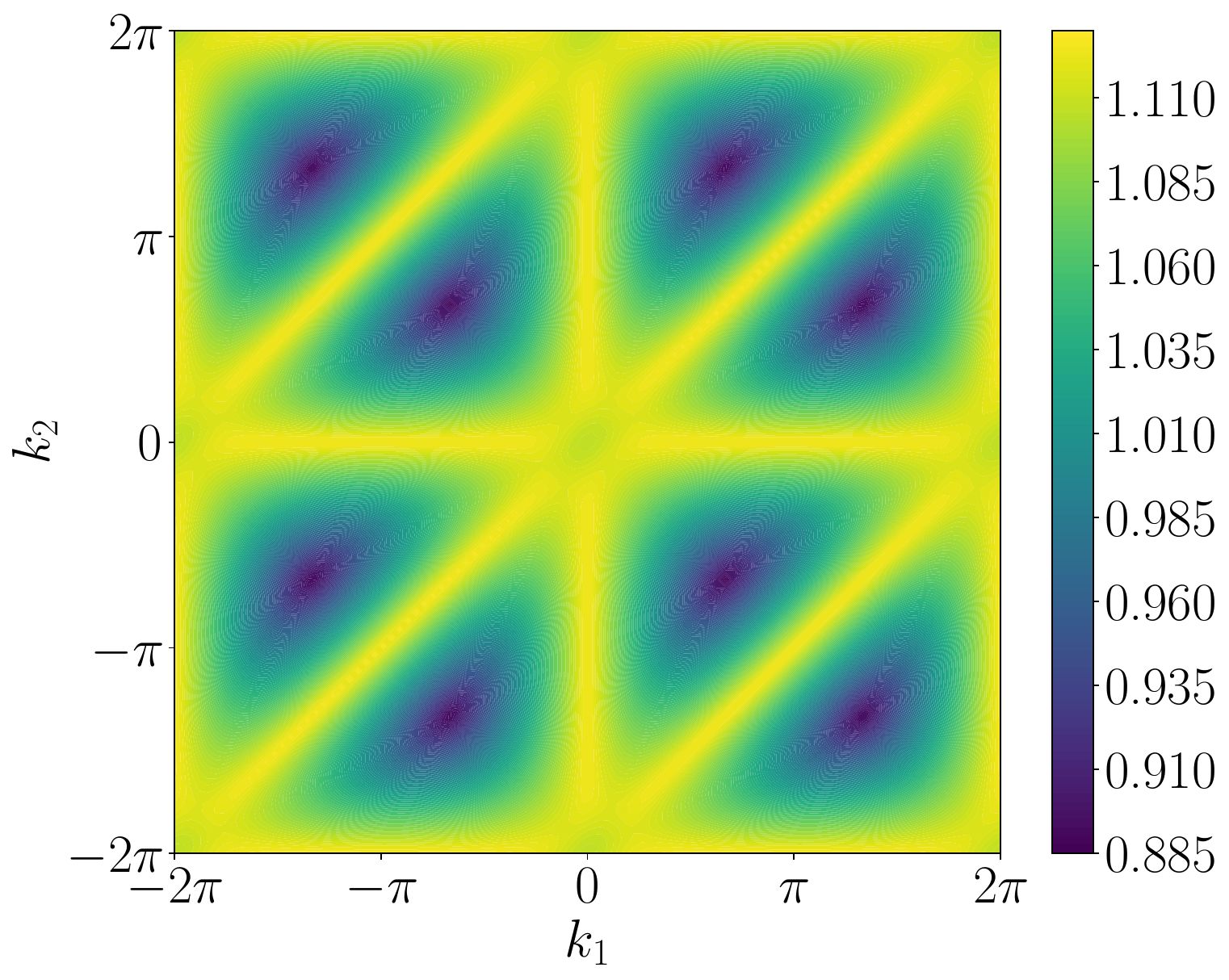}
	\includegraphics[width=.325\textwidth]{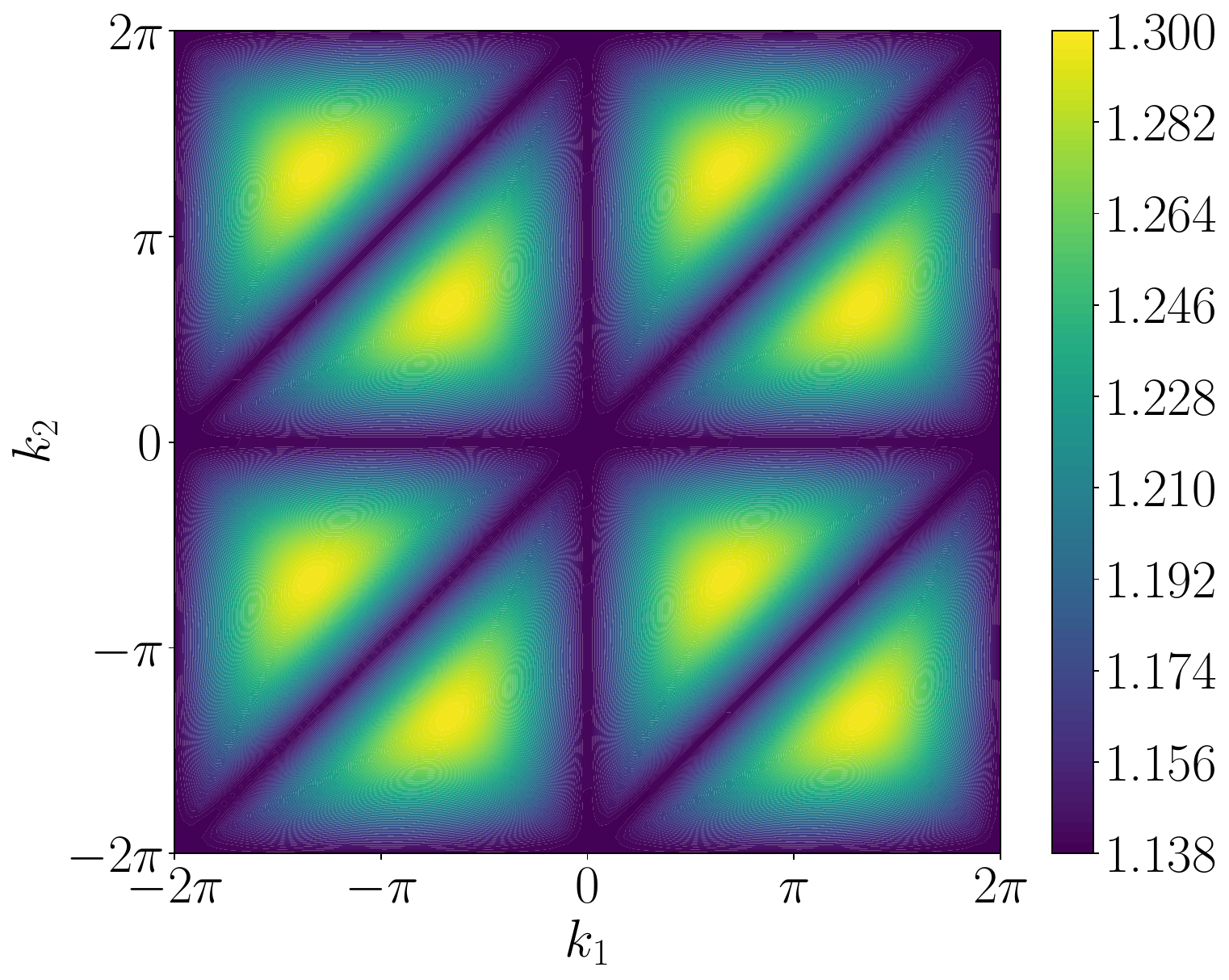}
	\includegraphics[width=.325\textwidth]{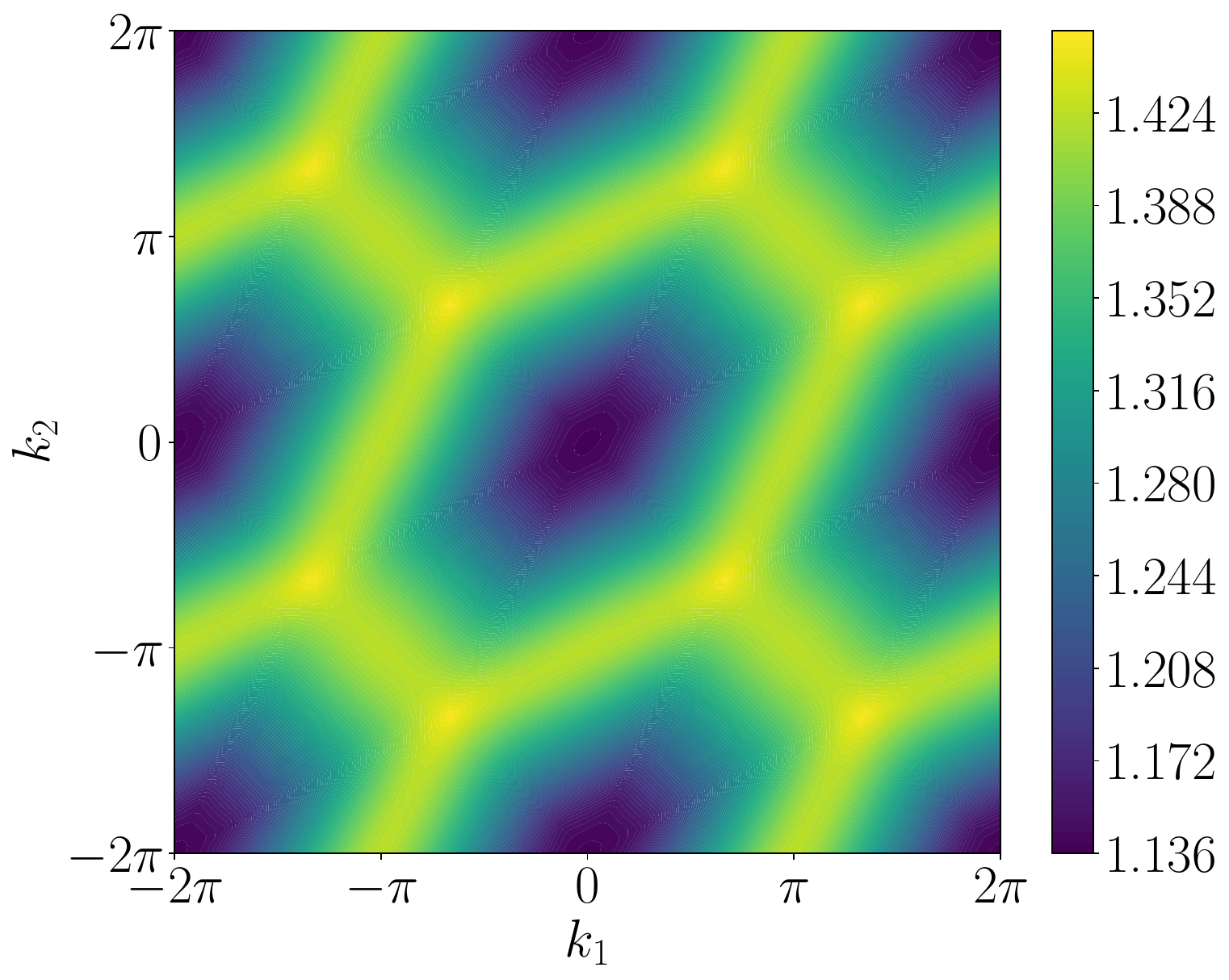}
	\includegraphics[width=.325\textwidth]{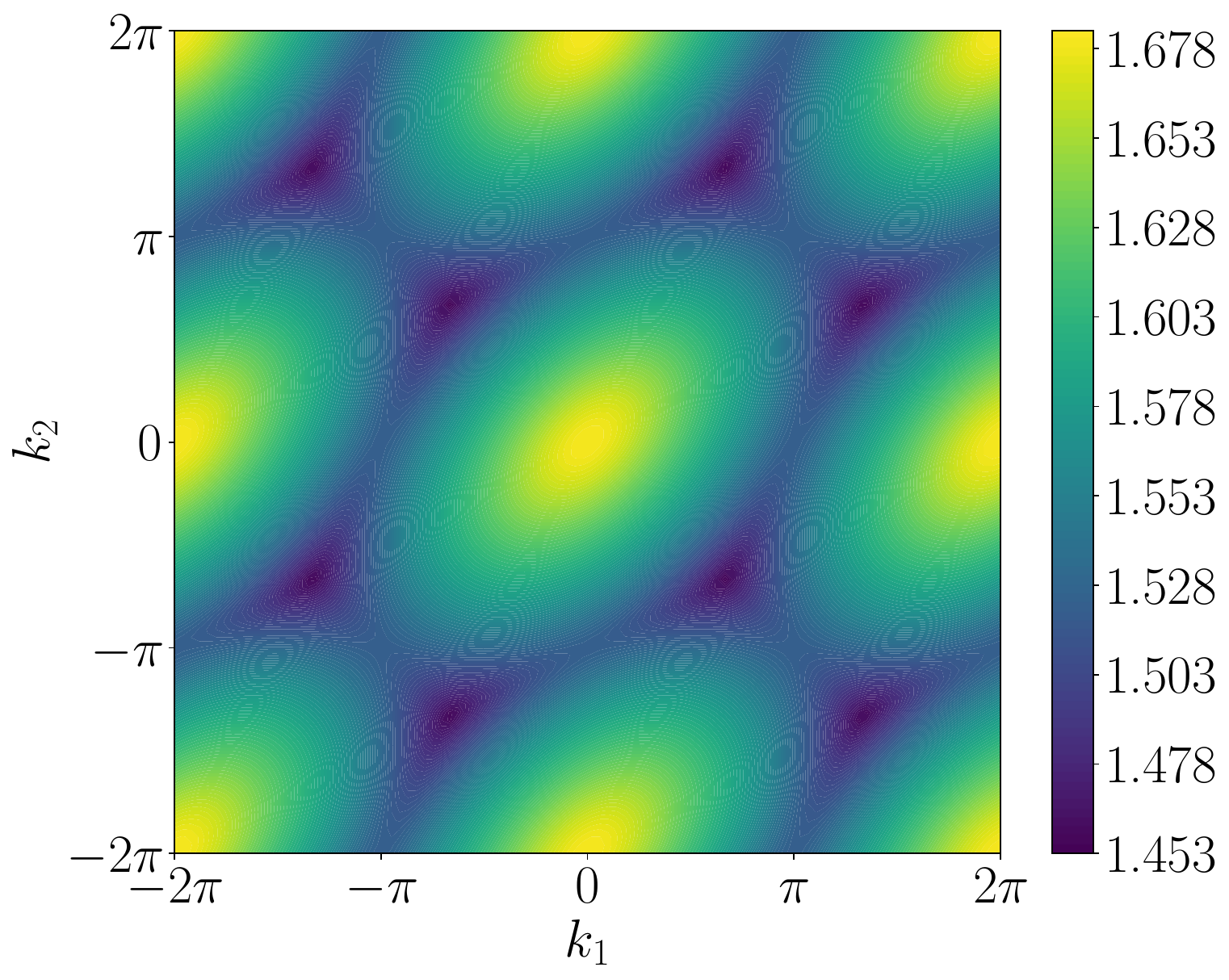}
	\caption{One quasi-particle dispersion in the \jtwo TFIM for $J_{1,2}/2h = 0.2$ in units of $2h$. The momenta $k_1, k_2$ are defined with respect to the lattice vectors $\bm{t}_1,\bm{t}_2$.}
	\label{fig:j1j2-dispersion}
\end{figure}

For the $J_1$-$J_2$ case the six dispersive energy bands qualitatively take the shape depicted in Fig.~\ref{fig:j1j2-dispersion} exemplarily for the case $J_1 = J_2$ with $J_1/2h = 0.2$. We locate the minima of the dispersion, corresponding to the one quasi-particle excitation gap $\Delta$, at ${\bm{k} = (2\pi/3,-2\pi/3)}$ for all finite ratios $J_1/J_2>0$.

\begin{figure}[h]
	\centering
	\includegraphics[width=.325\textwidth]{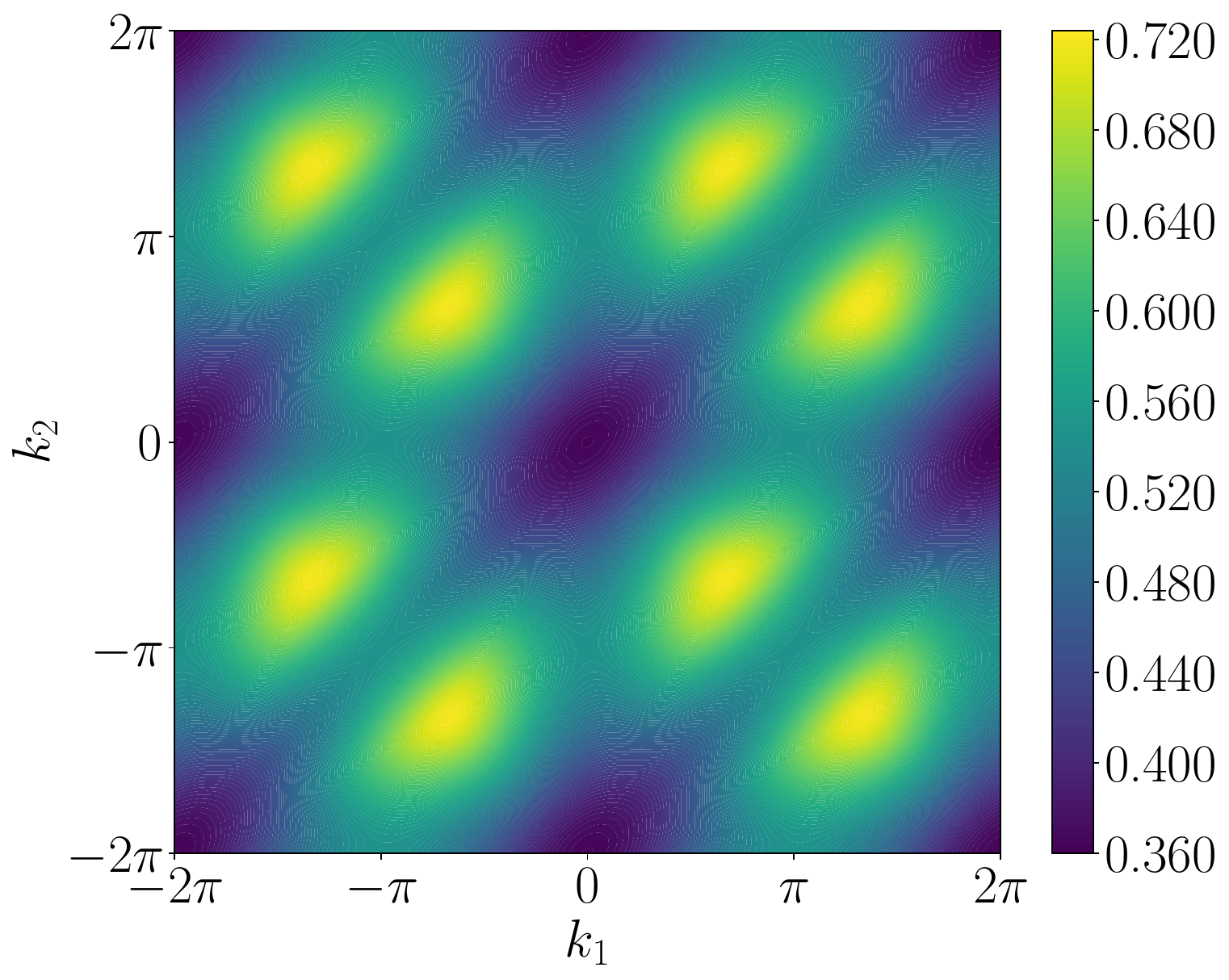}
	\includegraphics[width=.325\textwidth]{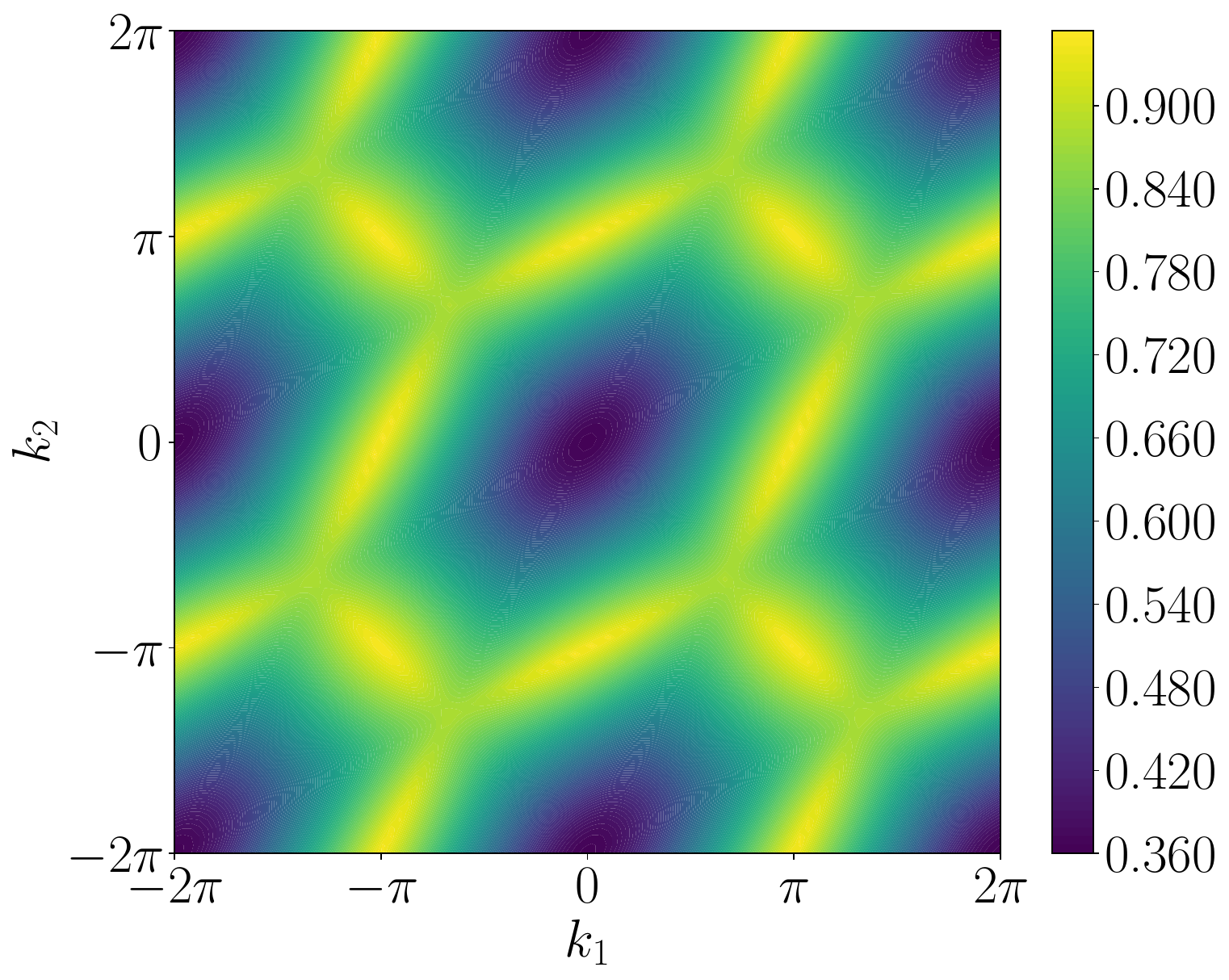}
	\includegraphics[width=.325\textwidth]{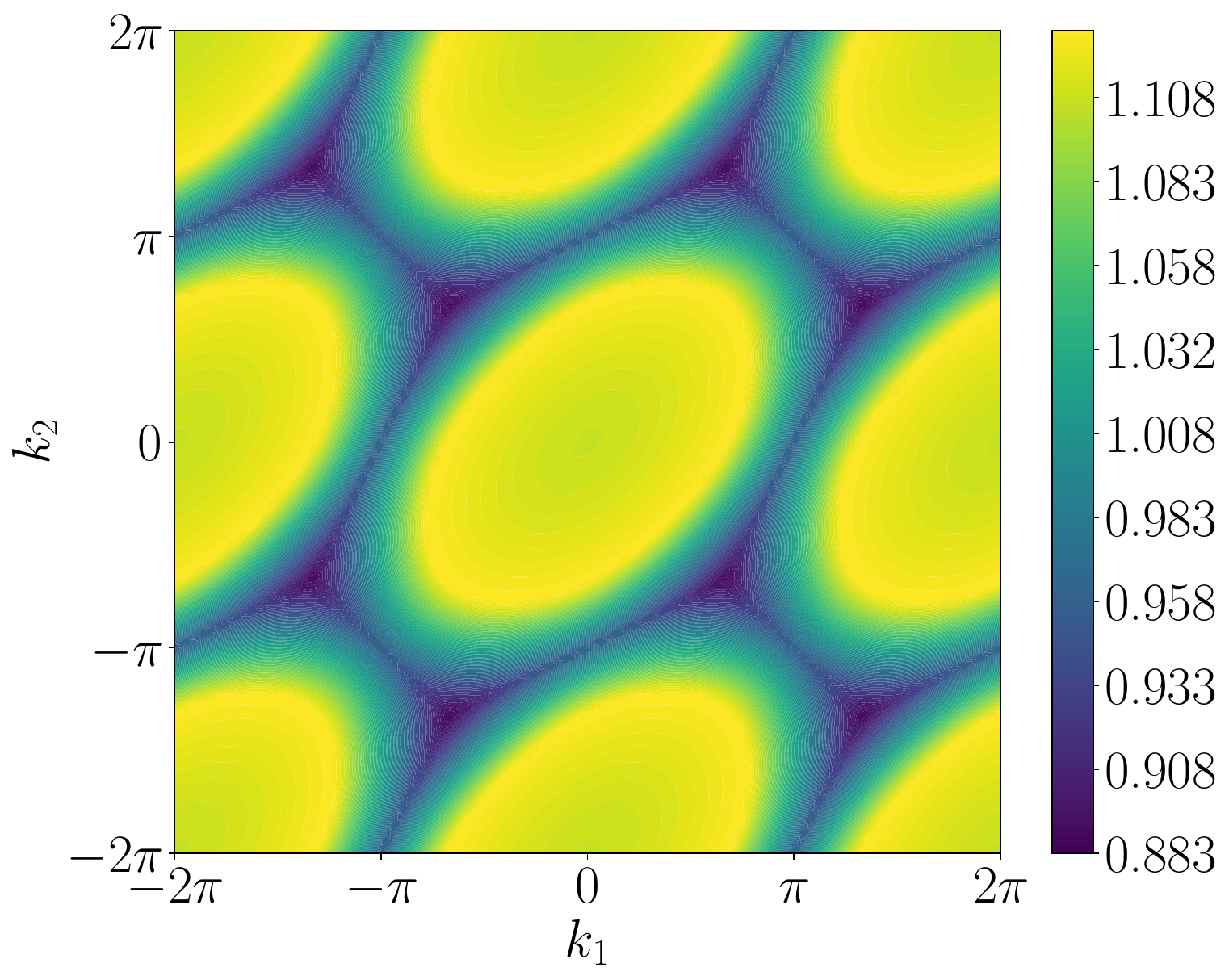}
	\includegraphics[width=.325\textwidth]{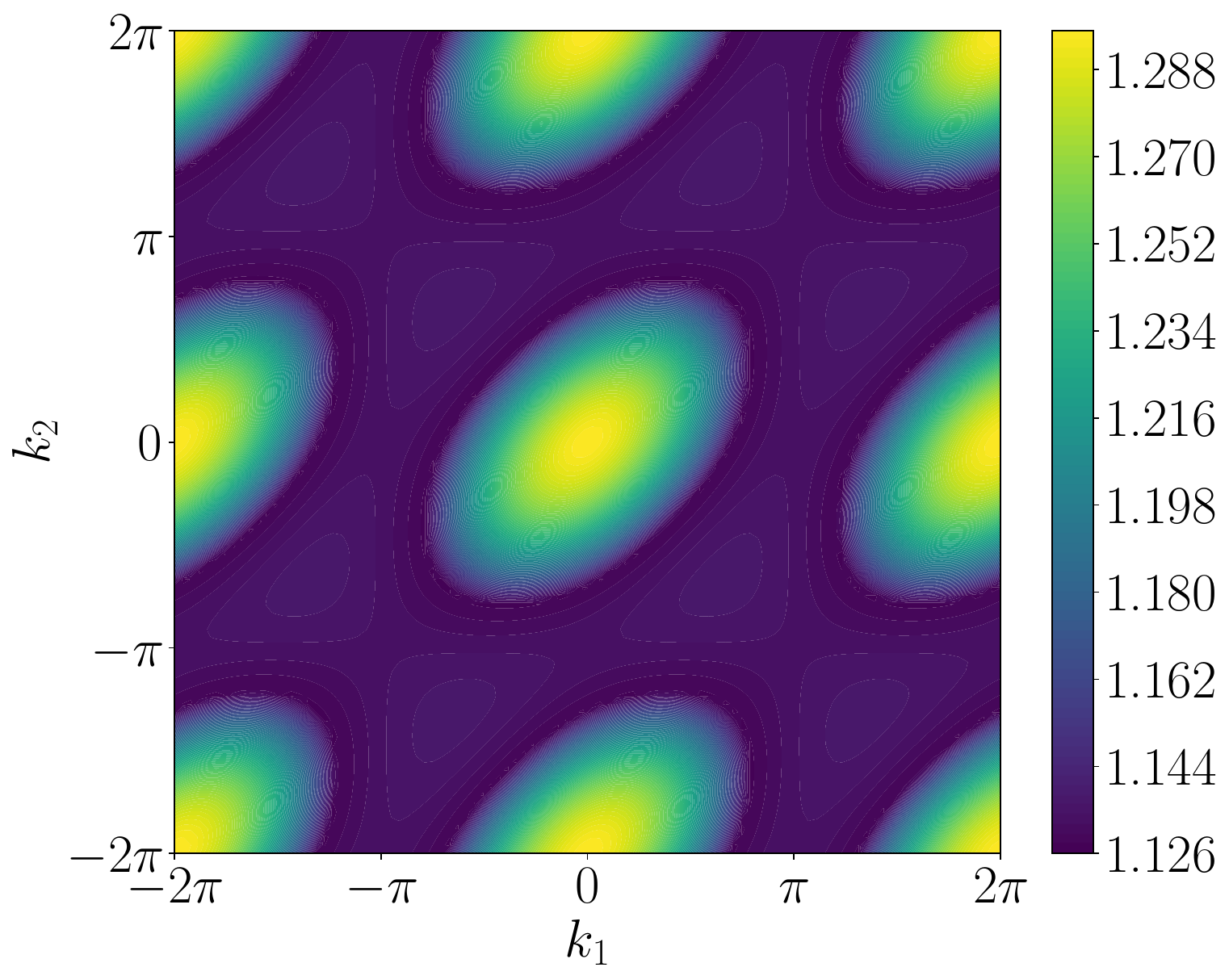}
	\includegraphics[width=.325\textwidth]{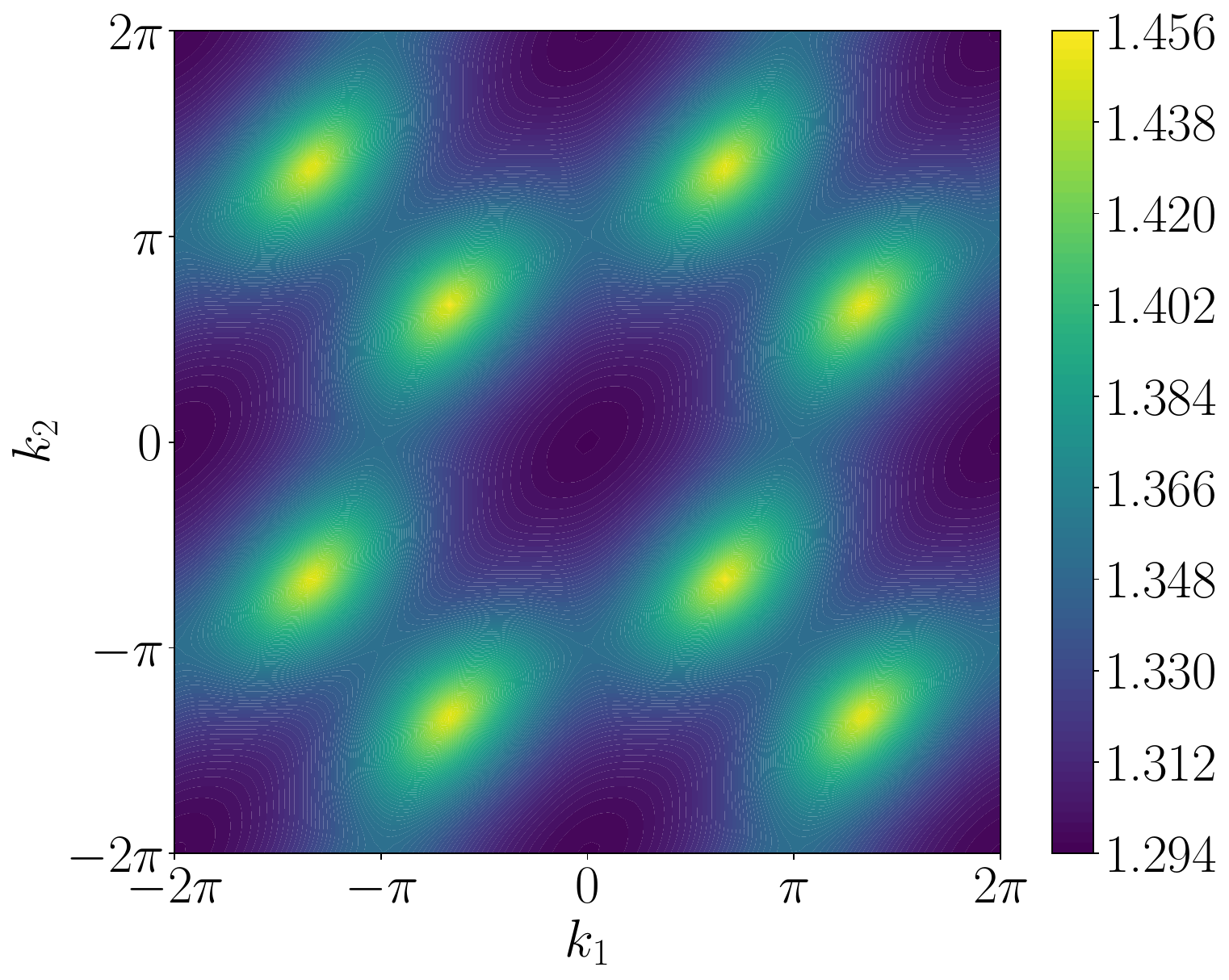}
	\includegraphics[width=.325\textwidth]{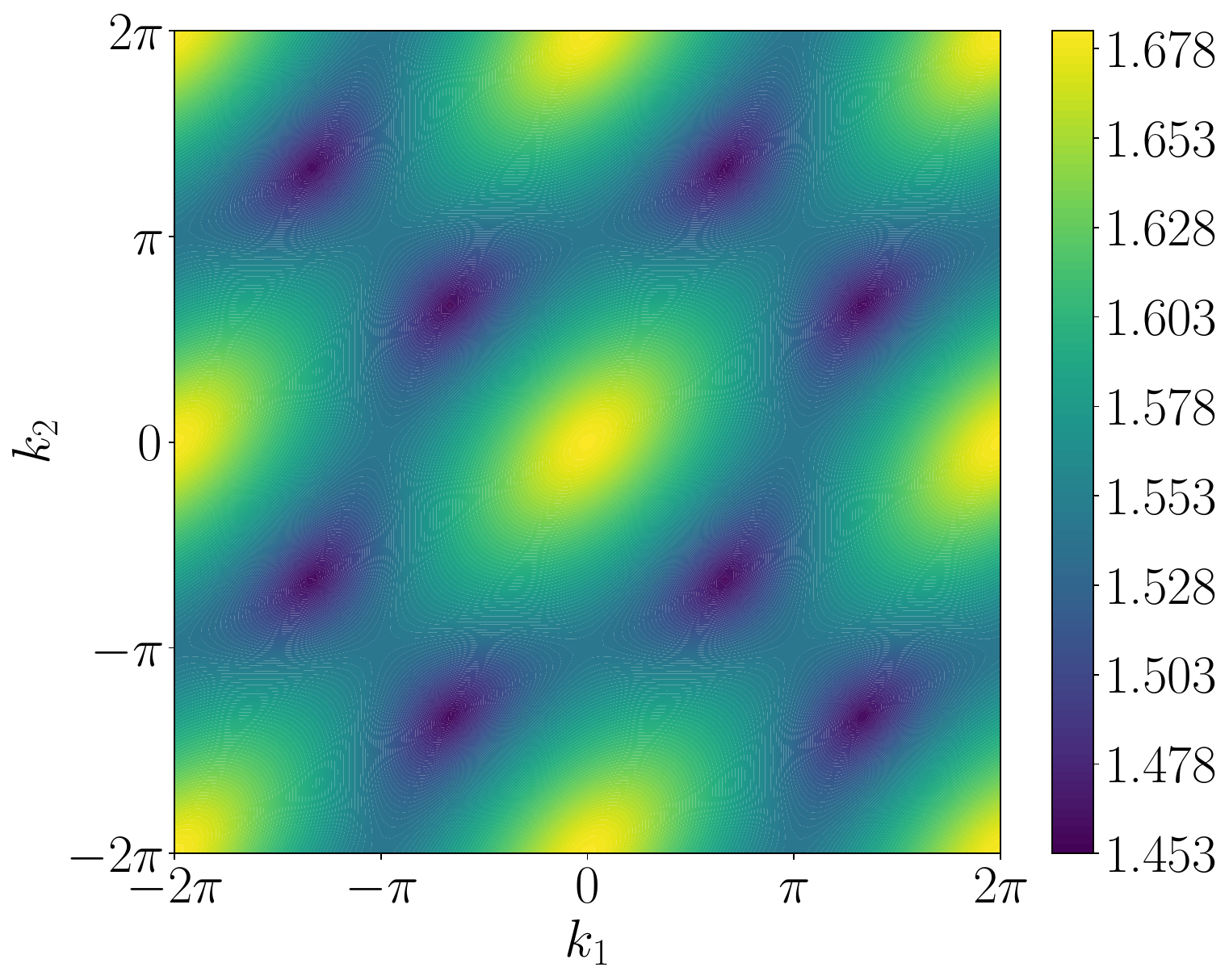}
	\caption{One quasi-particle dispersion in the \jthree TFIM for $J_{1,3}/2h = 0.2$ in units of $2h$. The momenta $k_1, k_2$ are defined with respect to the lattice vectors $\bm{t}_1,\bm{t}_2$.}
	\label{fig:j1j3-dispersion}
\end{figure}

For the $J_1$-$J_3$ case the six dispersive energy bands qualitatively take the shape depicted in Fig.~\ref{fig:j1j3-dispersion} exemplarily for the case $J_1 = J_3$ with $J_1/2h = 0.2$. We locate the minima of the dispersion at $\bm{k} = (0,0)$ for all finite ratios $J_1/J_3>0$. 

For the full \jfull case we find that the dispersion qualitatively depends on the ratio $J_2/J_3$. For $J_2>J_3$ we find the same qualitative dispersion as for the pure \jtwo TFIM, while for $J_3>J_2$ the dispersion resembles the one of the pure \jthree TFIM.


\subsection{Elementary excitation gap}
The elementary excitation gap $\Delta(\lambda)$ is given by the one quasi-particle energy at the minimum of the dispersion and obtained as a series in $\lambda$ by explicit diagonalization of the effective Hamiltonian at the respective gap momentum. 
We perform DlogPadé extrapolations (compare Sec.~\ref{sec:extrapolation}) on $\Delta(\lambda)$ to approximate the assumed underlying power-law behavior $\Delta\propto|\lambda-\lambda_c|^{z\nu}$ of the elementary excitation gap and extract the critical point $\lambda_c$ and the critical exponent $z\nu$.

\subsubsection{$J_1$-$J_2$ and $J_1$-$J_3$ case}\label{sec:j1j2-j1j3-high-field}
\begin{figure}[h]
	\centering
	\includegraphics[width=\textwidth]{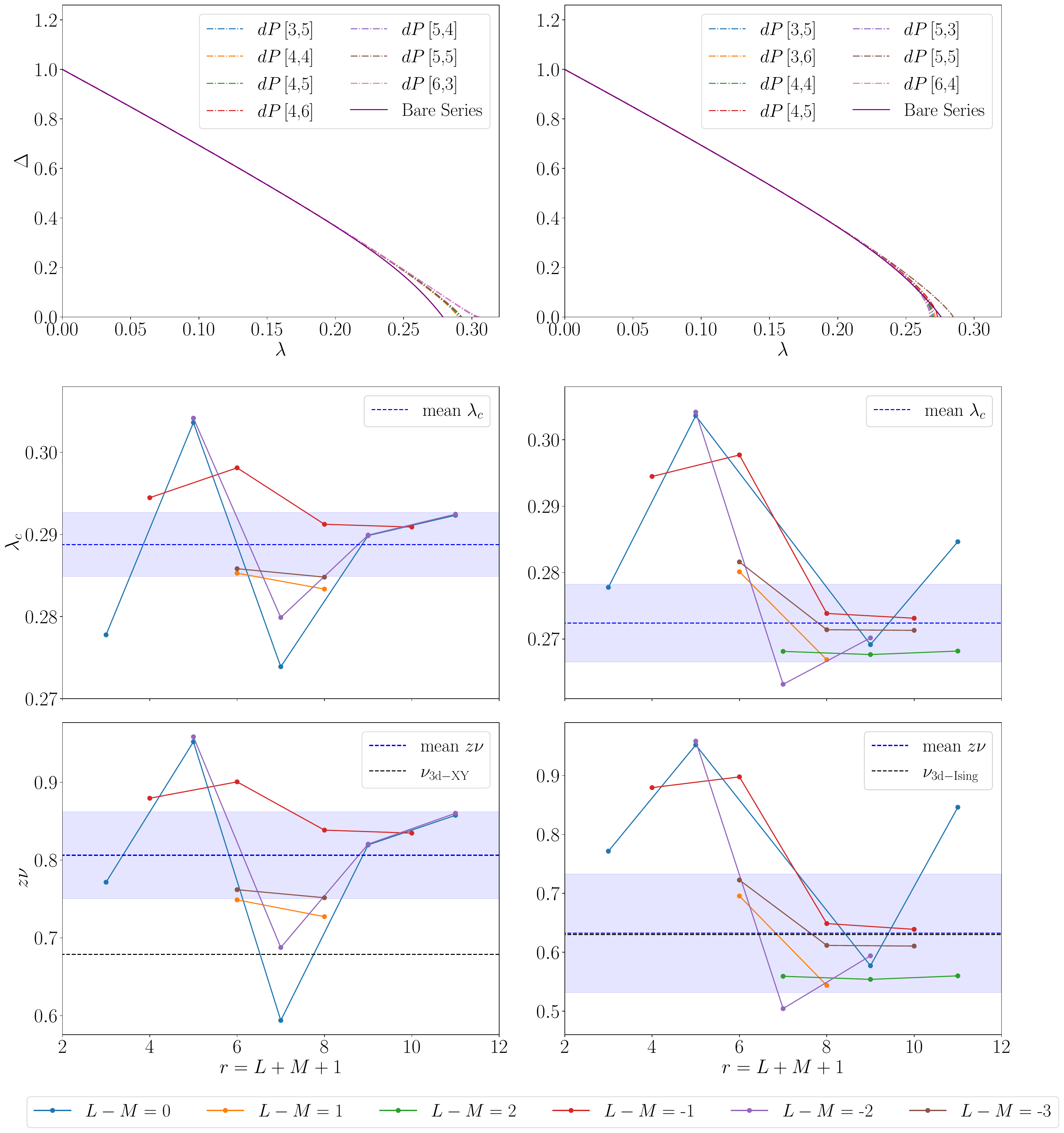}
	\caption{Results from high-field series expansions in the \jtwo TFIM with $J_1=J_2$ (left panels) and the \jthree TFIM with $J_1=J_3$ (right panels). \\
		\textit{First row:} Elementary excitation gap $\Delta(\lambda)$ in units of $2h$ as a function of $\lambda=J_1/2h$. The plots show the bare series in order $r=11$ alongside DlogPadé extrapolants obtained from orders $r\geq 9$ within families $|L-M|\leq 3$. \\
		\textit{Second and third row:} Convergence behavior of the critical points $\lambda_c$ (second row) and critical exponents $z\nu$ (third row) extracted from the DlogPadé extrapolants $dP[L,M]$ in the considered order $r=L+M+1$. The extrapolants are structured into families by connecting extrapolants with the same $d=L-M$. We show only extrapolants with $|L-M|\leq 3$. To obtain a mean value for $\lambda_c$ and $z\nu$ we average over the extrapolants of highest order for each shown family. The calculated means $\lambda_c = 0.289\pm  0.004$ and $z\nu = 0.806 \pm 0.056$ for $J_1=J_2$ and $\lambda_c = 0.272\pm  0.006$ and $z\nu = 0.632 \pm 0.101$ for $J_1=J_3$ are drawn as a dashed blue lines, with the highlighted areas indicating the standard deviations of the individual extrapolants. The black dashed lines represent the literature values for $\nu_\mathrm{3d-XY}=0.67169(7)$ \cite{Hasenbusch2019s,Chester2020s} and $\nu_\mathrm{3d-Ising}=0.629971(4)$ \cite{Kos2016s} respectively.}
	\label{fig:j1j2-j1j3-all}
\end{figure}

% --- case j1 = j2/j3 ---
We first consider the cases $J_1=J_2$ and $J_1=J_3$ in the reduced \jtwo and \jthree models in detail. The  bare series of the excitation gap $\Delta(\lambda)$ in order $r=11$ in $\lambda=J_1/2h$ are shown in Fig.~\ref{fig:j1j2-j1j3-all} (first row) alongside extrapolants obtained from orders $r\geq 9$ and with $|L-M| \leq 3$. For both cases we find a critical point $\lambda_c$ where the gap closes and extract the critical exponent $z\nu$. We structure the extrapolants $dP[L,M]$ into families with the same $d=L-M$ and plot the critical point and exponent as a function of the order $r=L+M+1$ in Fig.~\ref{fig:j1j2-j1j3-all}, taking only extrapolants with $|L-M|\leq 3$ into account. To obtain an estimate for $\lambda_c$ and $z\nu$ we average over the extrapolants of the highest order of all considered families. We obtain $\lambda_c = 0.289\pm 0.004$ and $z\nu=0.806\pm0.056$ for the $J_1=J_2$ case and $\lambda_c = 0.272\pm 0.006$ and $z\nu=0.632\pm0.101$ for the $J_1=J_3$ case. \\

\begin{figure}[h]
	\centering
	\includegraphics[width=\textwidth]{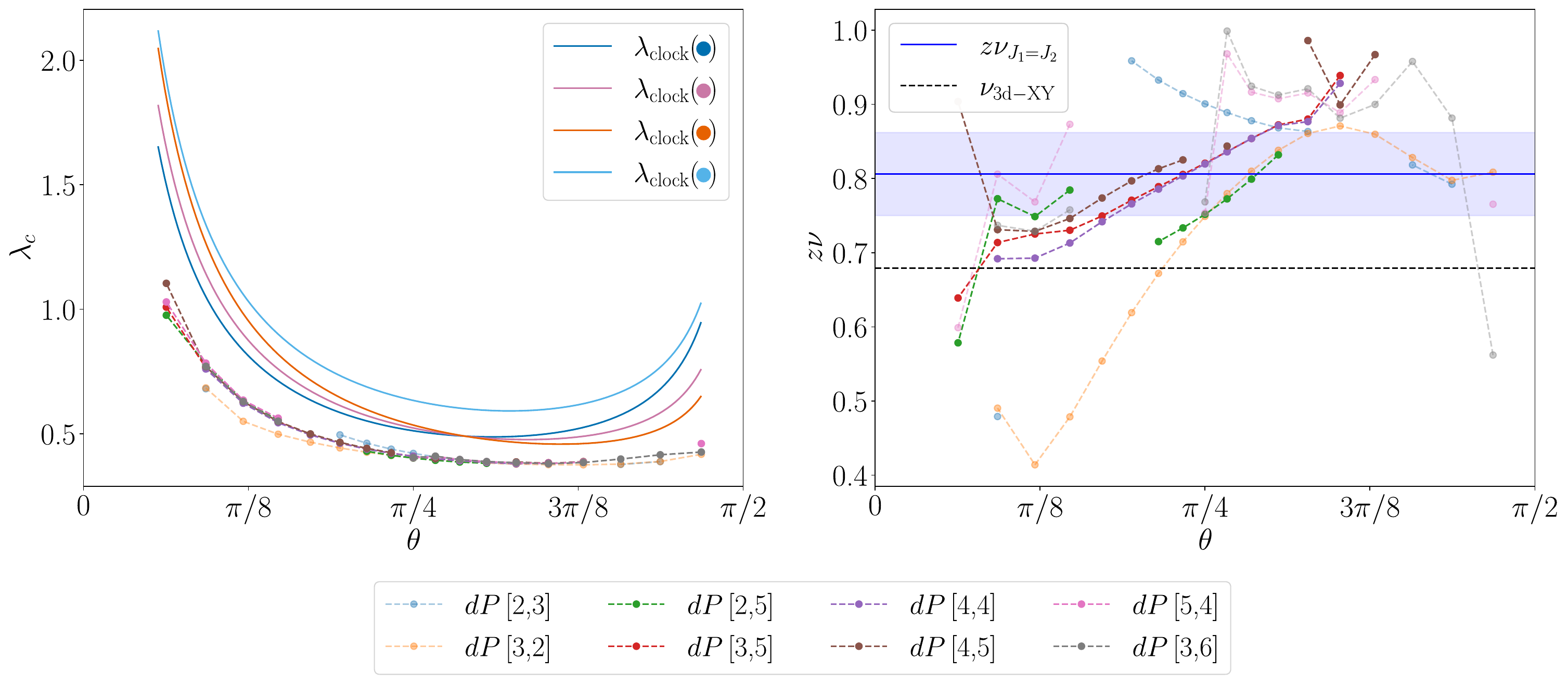}
	
	\vspace{0.5cm}
	
	\includegraphics[width=\textwidth]{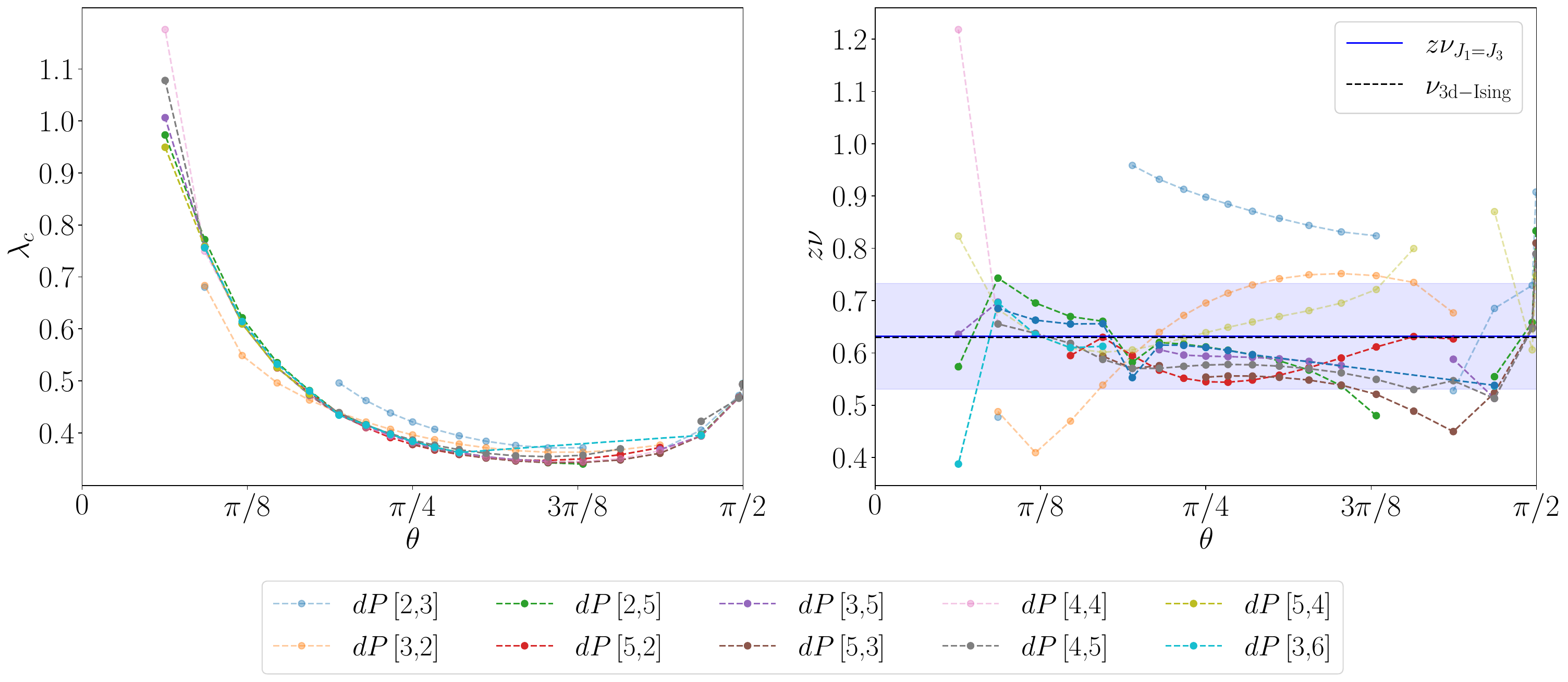}
	\caption{Critical points $\lambda_c$ (left panels) and critical exponents $z\nu$ (right panels) for various ratios of $J_2/J_1$ (upper panels) and $J_3/J_1$ (lower panels) parameterized by $\theta$ as given in the main text. The plots show the results obtained from individual selected DlogPadé extrapolants in different orders. Extrapolants which show deviations from the (qualitative) behavior of the bulk are shown in lighter opacities. The blue lines and shaded regions show the values $z\nu_{J_1=J_{2,3}}$ with their standard deviations as calculated in Sec.~\ref{sec:j1j2-j1j3-high-field}, respectively. The black dashed lines in the right panels show the literature value for the assumed 3d-XY and 3d-Ising criticalities of the phase transition, $\nu_\mathrm{3d-XY}=0.67169(7)$ \cite{Hasenbusch2019s,Chester2020s} and $\nu_\mathrm{3d-Ising}=0.629971(4)$ \cite{Kos2016s}. The upper left plot further shows an estimate for the phase transition from the columnar to the clock-ordered phase ($\lambda_\text{clock}$) derived from the analogy to the RK QDM on the honeycomb lattice \cite{Moessner2001news}.  }
	\label{fig:j1j2-j1j3-ratio}
\end{figure}

% --- arbitrary ratios --- 
We continue by discussing the \jtwo and \jthree TFIM for arbitrary ratios $J_1/J_2$ and $J_1/J_3$ respectively, where we calculate the dispersion up to order $r=10$ in $\lambda=J/2h$. We define the two nonzero interaction strengths as 
\begin{align}
J_1 &= \cos(\theta)\, J\,, \\
J_{2,3} &= \sin(\theta)\,J\,.
\end{align}
As noted in Sec.~\ref{sec:dispersion}, the gap momentum is independent from the ratio $J_1/J_{2,3}$ within the respective reduced model. 
In Fig.~\ref{fig:j1j2-j1j3-ratio} we show the critical points $\lambda_c$ and the critical exponents $z\nu$ determined from selected (non-defective) DlogPadé extrapolants as a function of $\theta$ for both models. Note that the previously discussed cases $J_1=J_2$ and $J_1 = J_3$ are recovered for $\theta=\pi/4$, respectively. However, $\theta=\pi/4$ corresponds to $J_1 = J_{2,3} = J/\sqrt{2}$ in this parametrization which results in a rescaling of $\lambda_c$. 

The selected extrapolants show the same qualitative behavior regarding the extracted critical points $\lambda_c$ and we find a closing of the excitation gap for $\theta\in(0.1,1.47)$ in the \jtwo TFIM, which corresponds to $J_1/J_2 = \cot\theta \in (0.1,10)$, and for $\theta\in(0.1,1.57)$ in the \jthree TFIM, which corresponds to ${J_1/J_2 \in (0.1,100)}$.
Coming from the symmetric case $\theta=\pi/4$ and decreasing $\theta$, the critical point gets pushed to larger $\lambda$ as the limit of decoupled $J_1$-triangles is approached in both cases. For $\theta = \pi/2$ in the \jtwo TFIM, the system consists of isolated hexagons and there is no phase transition, while in the \jthree TFIM, the system consists of decoupled $J_3$-chains with the critical point $\lambda_c = 0.5$ from the one-dimensional TFIM. Regarding the critical exponents we find some more variety in the qualitative behavior. Nevertheless we can identify a consistent tendency among extrapolants of high orders. Within the \jtwo TFIM, the critical exponent increases with $\theta$ while within the \jthree TFIM, the critical exponent decreases slightly with $\theta$. In both cases we argue that the critical exponent remains roughly constant, considering the range of deviation indicated by the results we obtain for $\theta=\pi/4$ respectively, as indicated by the blue shaded regions in Fig.~\ref{fig:j1j2-j1j3-ratio}.

As argued in the main text and Sec.~\ref{sec:j1j2_lowfield}, the effective low-field description of the \jtwo TFIM is analogous to the RK QDM and there is a phase transition from the columnar low-field phase to a clock-ordered phase at intermediate field strengths. In order to estimate where this phase transition occurs in comparison to the phase transition to the $x$-polarized high-field phase, we calculate the ratio $v/t = - (E_{J_2}^{(4)}({\adjustbox{valign=c}{\protect\circlebluegreen}})-E_{J_2}^{(4)}({\adjustbox{valign=c}{\protect\circlelightgray}}))/2h^2E_\mathrm{res}^{(6)}$ for ${\adjustbox{valign=c}{\protect\circlelightgray}}={\adjustbox{valign=c}{\protect\circleblue}},{\adjustbox{valign=c}{\protect\circlepink}},{\adjustbox{valign=c}{\protect\circleorange}},{\adjustbox{valign=c}{\protect\circleskyblue}}$ and plot the point $\lambda_\text{clock}({\adjustbox{valign=c}{\circlelightgray}})$ (in units $J/2h$) of the columnar to clock-order phase transition at $v/t=-0.2$ \cite{Moessner2001news} as a function of $\theta$ in the respective upper left panel in Fig.~\ref{fig:j1j2-j1j3-ratio}. While these values are only an estimate, it is clear that the phase transition to the polarized phase occurs at larger $h$-values and therefore we argue that an intermediate clock-ordered phase is present for all $\theta$. 

\subsubsection{Experimental realization of the $J_1$-$J_2$-$J_3$ TFIM}
In the following we highlight two experimentally relevant cases of the \jfull TFIM which can be implemented on a Rydberg atom quantum simulator with appropriately fixed laser detuning. Interactions between excited Rydberg atoms are of van-der-Waals type and decay algebraically with the distance $r$ between atoms like $\sim r^{-6}$, which defines the relations between the $J_i$. 
The distances $r_i$ between the three nearest neighbors we consider in the series expansion depend on the aspect ratio $\rho$ of the ruby lattice as follows: setting $r_1=1$, $r_2=\rho$ and $r_3= (1+\rho^2)^{1/2}$. Note that which site is the fourth-nearest neighbor depends on $\rho$, for $\rho>5/4$ we find $r_4 = (\rho(\rho+\sqrt{3}) +1)^{1/2}$ while for $\rho<5/4$ we find $r_4 = \sqrt{3}\rho$. For all $\rho\geq 1$ we have $J_i \geq J_{i+1}$ for $i\in\mathbb{N}$. 

\begin{figure}[h]
	\centering
	\includegraphics[width=\textwidth]{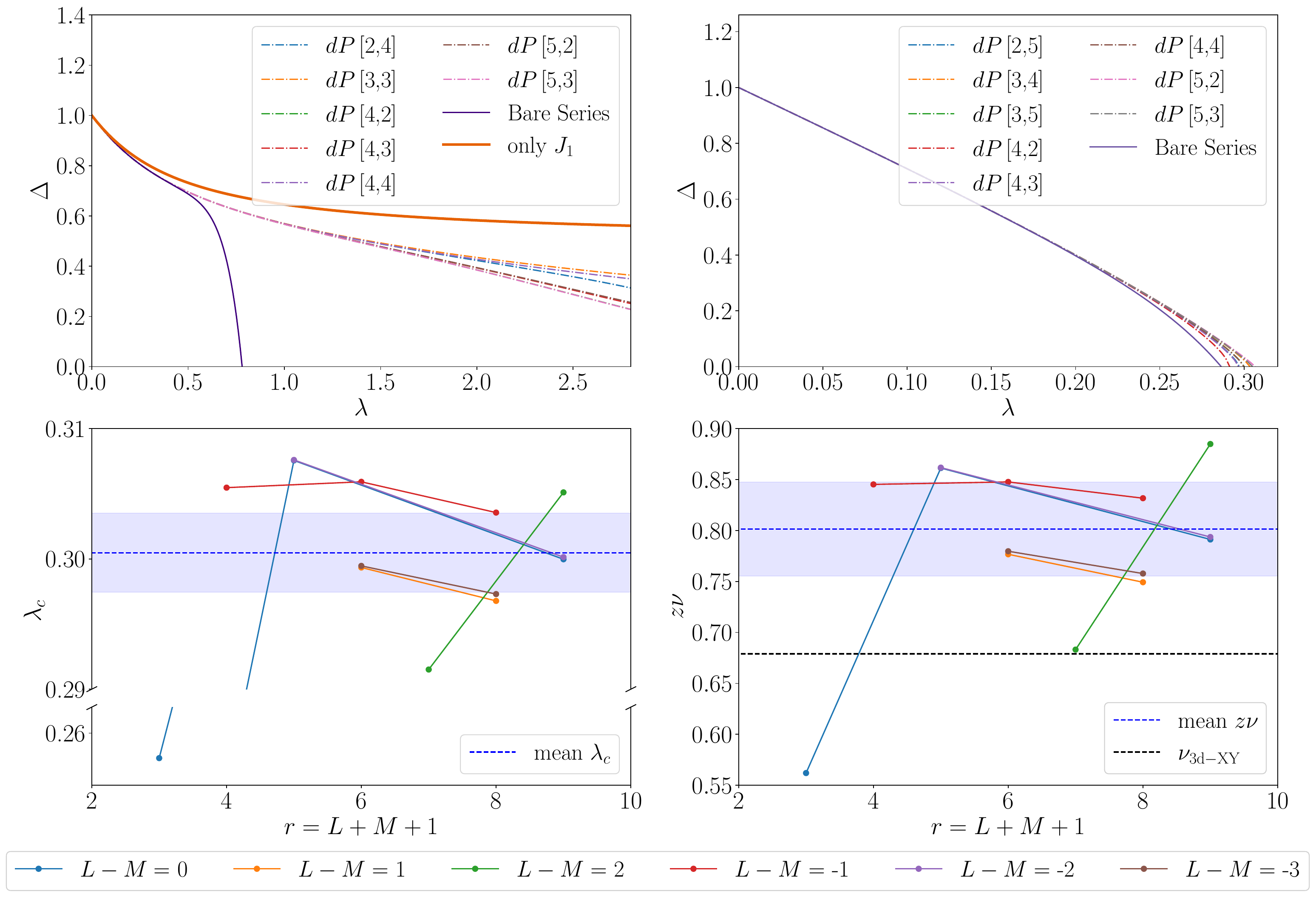}
	\caption{High-field results for two experimentally relevant cases. Upper panels: One quasi-particle excitation gap $\Delta$ in the $J_1$-$J_2$-$J_3$ TFIM with $\rho=\sqrt{3}$ (left) and $\rho=1$ (right) in units of $2h$ as a function of $\lambda=J_1/2h$. The bare series in the maximal order $r=9$ is shown along lower orders in lower opacities and the obtained DlogPadé extrapolants $dP[L,M]$ in orders $r\geq7$ with in families $|L-M|\leq 3$ as a function of $\lambda=J_1/2h$. The thick orange line in the left panel shows the gap in the limit of $J_2 = J_3 =0$, where the lattice is decomposed into isolated triangles. Lower panels: Convergence behavior of the critical point $\lambda_c$ (left) and critical exponent $z\nu$ (right) for $\rho=1$ in the $J_1$-$J_2$-$J_3$ TFIM extracted from the DlogPadé extrapolants $dP[L,M]$ in the considered order $r=L+M+1$. The extrapolants are structured into families by connecting extrapolants with the same $d=L-M$. We show only extrapolants with $|L-M|\leq 3$. To obtain a mean value for $\lambda_c$ and $z\nu$ we average over the extrapolants of highest order for each shown family. The calculated means $\lambda_c = 0.301\pm  0.003$ and $z\nu = 0.801 \pm 0.046$ for are drawn as a dashed blue lines, with the highlighted areas indicating the standard deviations of the individual extrapolants. The black dashed line represents the literature value for $\nu_\mathrm{3d-XY}=0.67169(7)$ \cite{Hasenbusch2019s,Chester2020s}.}
	\label{fig:j1j2j3-experimental}
\end{figure}

The first case we highlight is the structure with the sites located on the links of a Kagome lattice, realized by the ruby lattice for an aspect ratio of $\rho=\sqrt{3}$. Here the interaction strengths as defined by the algebraic decay, $J_i =J_1 r_i^{-6}$, are given by $J_2=J_1/27$ and $J_3=J_1/64$, neglecting all $J_i \leq J_4 = J_1/364$. We depict our results for the excitation gap obtained by pCUT high-order high-field series expansions in $\lambda=J_1/2h$ and subsequent DlogPadé extrapolations in Fig.~\ref{fig:j1j2j3-experimental} (upper left panel) in analogous fashion to Fig.~\ref{fig:j1j2-j1j3-all}. The gap momentum is $\bm{k} = (2\pi/3,-2\pi/3)$ as $J_1 > J_2 > J_3$. The system is dominated by the $J_1$ interaction which defines isolated triangles with no phase transition. Regarding only $J_1$ results in the energy gap as depicted in orange, which is very close to the results obtained from considering all three interactions which explains why no phase transition can be detected within our perturbative approach. 

The second case is the one of $\rho=1$, which results in $J_1=J_2$ and $J_3=J_1/8$, again neglecting all $J_i \le J_4 = J_1/27$. Our high-field results for the excitation gap are depicted in Fig.~\ref{fig:j1j2j3-experimental} (upper right panel). We find a gap momentum $\bm{k} = (2\pi/3,-2\pi/3)$ in line with previous results for $J_1>J_2>J_3$ with a quantum phase transition at $\lambda_c=0.301\pm 0.003$ ($C_6/\Omega= 1.202 \pm 0.012$) with exponent $z\nu=0.801\pm 0.046$ (values obtained by averaging over extrapolants of highest order for all families depicted in Fig.~\ref{fig:j1j2j3-experimental}, lower panels).

\bibliographystyle{apsrev4-2}

\begin{thebibliography}{81}%
\makeatletter
\providecommand \@ifxundefined [1]{%
 \@ifx{#1\undefined}
}%
\providecommand \@ifnum [1]{%
 \ifnum #1\expandafter \@firstoftwo
 \else \expandafter \@secondoftwo
 \fi
}%
\providecommand \@ifx [1]{%
 \ifx #1\expandafter \@firstoftwo
 \else \expandafter \@secondoftwo
 \fi
}%
\providecommand \natexlab [1]{#1}%
\providecommand \enquote  [1]{``#1''}%
\providecommand \bibnamefont  [1]{#1}%
\providecommand \bibfnamefont [1]{#1}%
\providecommand \citenamefont [1]{#1}%
\providecommand \href@noop [0]{\@secondoftwo}%
\providecommand \href [0]{\begingroup \@sanitize@url \@href}%
\providecommand \@href[1]{\@@startlink{#1}\@@href}%
\providecommand \@@href[1]{\endgroup#1\@@endlink}%
\providecommand \@sanitize@url [0]{\catcode `\\12\catcode `\$12\catcode
  `\&12\catcode `\#12\catcode `\^12\catcode `\_12\catcode `\%12\relax}%
\providecommand \@@startlink[1]{}%
\providecommand \@@endlink[0]{}%
\providecommand \url  [0]{\begingroup\@sanitize@url \@url }%
\providecommand \@url [1]{\endgroup\@href {#1}{\urlprefix }}%
\providecommand \urlprefix  [0]{URL }%
\providecommand \Eprint [0]{\href }%
\providecommand \doibase [0]{https://doi.org/}%
\providecommand \selectlanguage [0]{\@gobble}%
\providecommand \bibinfo  [0]{\@secondoftwo}%
\providecommand \bibfield  [0]{\@secondoftwo}%
\providecommand \translation [1]{[#1]}%
\providecommand \BibitemOpen [0]{}%
\providecommand \bibitemStop [0]{}%
\providecommand \bibitemNoStop [0]{.\EOS\space}%
\providecommand \EOS [0]{\spacefactor3000\relax}%
\providecommand \BibitemShut  [1]{\csname bibitem#1\endcsname}%
\let\auto@bib@innerbib\@empty
%</preamble>
\bibitem [{\citenamefont {Kitaev}(2006)}]{Kitaev2006}%
  \BibitemOpen
  \bibfield  {author} {\bibinfo {author} {\bibfnamefont {A.}~\bibnamefont
  {Kitaev}},\ }\href {https://doi.org/10.1016/j.aop.2005.10.005} {\bibfield
  {journal} {\bibinfo  {journal} {Annals of Physics}\ }\textbf {\bibinfo
  {volume} {321}},\ \bibinfo {pages} {2} (\bibinfo {year} {2006})}\BibitemShut
  {NoStop}%
\bibitem [{\citenamefont {Wannier}(1950)}]{Wannier1950}%
  \BibitemOpen
  \bibfield  {author} {\bibinfo {author} {\bibfnamefont {G.~H.}\ \bibnamefont
  {Wannier}},\ }\href {https://doi.org/10.1103/PhysRev.79.357} {\bibfield
  {journal} {\bibinfo  {journal} {Phys. Rev.}\ }\textbf {\bibinfo {volume}
  {79}},\ \bibinfo {pages} {357} (\bibinfo {year} {1950})}\BibitemShut
  {NoStop}%
\bibitem [{\citenamefont {Wannier}(1973)}]{Wannier1973}%
  \BibitemOpen
  \bibfield  {author} {\bibinfo {author} {\bibfnamefont {G.~H.}\ \bibnamefont
  {Wannier}},\ }\href {https://doi.org/10.1103/PhysRevB.7.5017} {\bibfield
  {journal} {\bibinfo  {journal} {Phys. Rev. B}\ }\textbf {\bibinfo {volume}
  {7}},\ \bibinfo {pages} {5017} (\bibinfo {year} {1973})}\BibitemShut
  {NoStop}%
\bibitem [{\citenamefont {Moessner}\ \emph {et~al.}(2000)\citenamefont
  {Moessner}, \citenamefont {Sondhi},\ and\ \citenamefont
  {Chandra}}]{Moessner2000}%
  \BibitemOpen
  \bibfield  {author} {\bibinfo {author} {\bibfnamefont {R.}~\bibnamefont
  {Moessner}}, \bibinfo {author} {\bibfnamefont {S.~L.}\ \bibnamefont
  {Sondhi}},\ and\ \bibinfo {author} {\bibfnamefont {P.}~\bibnamefont
  {Chandra}},\ }\href {https://doi.org/10.1103/PhysRevLett.84.4457} {\bibfield
  {journal} {\bibinfo  {journal} {Phys. Rev. Lett.}\ }\textbf {\bibinfo
  {volume} {84}},\ \bibinfo {pages} {4457} (\bibinfo {year}
  {2000})}\BibitemShut {NoStop}%
\bibitem [{\citenamefont {Moessner}\ and\ \citenamefont
  {Sondhi}(2001)}]{Moessner2001}%
  \BibitemOpen
  \bibfield  {author} {\bibinfo {author} {\bibfnamefont {R.}~\bibnamefont
  {Moessner}}\ and\ \bibinfo {author} {\bibfnamefont {S.~L.}\ \bibnamefont
  {Sondhi}},\ }\href {https://doi.org/10.1103/PhysRevB.63.224401} {\bibfield
  {journal} {\bibinfo  {journal} {Phys. Rev. B}\ }\textbf {\bibinfo {volume}
  {63}},\ \bibinfo {pages} {224401} (\bibinfo {year} {2001})}\BibitemShut
  {NoStop}%
\bibitem [{\citenamefont {Isakov}\ and\ \citenamefont
  {Moessner}(2003)}]{Isakov2003}%
  \BibitemOpen
  \bibfield  {author} {\bibinfo {author} {\bibfnamefont {S.~V.}\ \bibnamefont
  {Isakov}}\ and\ \bibinfo {author} {\bibfnamefont {R.}~\bibnamefont
  {Moessner}},\ }\href {https://doi.org/10.1103/PhysRevB.68.104409} {\bibfield
  {journal} {\bibinfo  {journal} {Phys. Rev. B}\ }\textbf {\bibinfo {volume}
  {68}},\ \bibinfo {pages} {104409} (\bibinfo {year} {2003})}\BibitemShut
  {NoStop}%
\bibitem [{\citenamefont {Powalski}\ \emph {et~al.}(2013)\citenamefont
  {Powalski}, \citenamefont {Coester}, \citenamefont {Moessner},\ and\
  \citenamefont {Schmidt}}]{Powalski2013}%
  \BibitemOpen
  \bibfield  {author} {\bibinfo {author} {\bibfnamefont {M.}~\bibnamefont
  {Powalski}}, \bibinfo {author} {\bibfnamefont {K.}~\bibnamefont {Coester}},
  \bibinfo {author} {\bibfnamefont {R.}~\bibnamefont {Moessner}},\ and\
  \bibinfo {author} {\bibfnamefont {K.~P.}\ \bibnamefont {Schmidt}},\ }\href
  {https://doi.org/10.1103/PhysRevB.87.054404} {\bibfield  {journal} {\bibinfo
  {journal} {Phys. Rev. B}\ }\textbf {\bibinfo {volume} {87}},\ \bibinfo
  {pages} {054404} (\bibinfo {year} {2013})}\BibitemShut {NoStop}%
\bibitem [{\citenamefont {Kano}\ and\ \citenamefont {Naya}(1953)}]{Kano1953}%
  \BibitemOpen
  \bibfield  {author} {\bibinfo {author} {\bibfnamefont {K.}~\bibnamefont
  {Kano}}\ and\ \bibinfo {author} {\bibfnamefont {S.}~\bibnamefont {Naya}},\
  }\href {https://doi.org/10.1143/ptp/10.2.158} {\bibfield  {journal} {\bibinfo
   {journal} {Progress of Theoretical Physics}\ }\textbf {\bibinfo {volume}
  {10}},\ \bibinfo {pages} {158} (\bibinfo {year} {1953})}\BibitemShut
  {NoStop}%
\bibitem [{\citenamefont {Nikoli\ifmmode~\acute{c}\else \'{c}\fi{}}\ and\
  \citenamefont {Senthil}(2005)}]{Nikolic2005}%
  \BibitemOpen
  \bibfield  {author} {\bibinfo {author} {\bibfnamefont {P.}~\bibnamefont
  {Nikoli\ifmmode~\acute{c}\else \'{c}\fi{}}}\ and\ \bibinfo {author}
  {\bibfnamefont {T.}~\bibnamefont {Senthil}},\ }\href
  {https://doi.org/10.1103/PhysRevB.71.024401} {\bibfield  {journal} {\bibinfo
  {journal} {Phys. Rev. B}\ }\textbf {\bibinfo {volume} {71}},\ \bibinfo
  {pages} {024401} (\bibinfo {year} {2005})}\BibitemShut {NoStop}%
\bibitem [{\citenamefont {Balents}(2010)}]{Balents2010}%
  \BibitemOpen
  \bibfield  {author} {\bibinfo {author} {\bibfnamefont {L.}~\bibnamefont
  {Balents}},\ }\href {https://doi.org/10.1038/nature08917} {\bibfield
  {journal} {\bibinfo  {journal} {Nature}\ }\textbf {\bibinfo {volume} {464}},\
  \bibinfo {pages} {199} (\bibinfo {year} {2010})}\BibitemShut {NoStop}%
\bibitem [{\citenamefont {R\"ochner}\ \emph {et~al.}(2016)\citenamefont
  {R\"ochner}, \citenamefont {Balents},\ and\ \citenamefont
  {Schmidt}}]{Roechner2016}%
  \BibitemOpen
  \bibfield  {author} {\bibinfo {author} {\bibfnamefont {J.}~\bibnamefont
  {R\"ochner}}, \bibinfo {author} {\bibfnamefont {L.}~\bibnamefont {Balents}},\
  and\ \bibinfo {author} {\bibfnamefont {K.~P.}\ \bibnamefont {Schmidt}},\
  }\href {https://doi.org/10.1103/PhysRevB.94.201111} {\bibfield  {journal}
  {\bibinfo  {journal} {Phys. Rev. B}\ }\textbf {\bibinfo {volume} {94}},\
  \bibinfo {pages} {201111} (\bibinfo {year} {2016})}\BibitemShut {NoStop}%
\bibitem [{\citenamefont {Villain}\ \emph {et~al.}(1980)\citenamefont
  {Villain}, \citenamefont {Bidaux}, \citenamefont {Carton},\ and\
  \citenamefont {Conte}}]{Villain1980}%
  \BibitemOpen
  \bibfield  {author} {\bibinfo {author} {\bibfnamefont {J.}~\bibnamefont
  {Villain}}, \bibinfo {author} {\bibfnamefont {R.}~\bibnamefont {Bidaux}},
  \bibinfo {author} {\bibfnamefont {J.-P.}\ \bibnamefont {Carton}},\ and\
  \bibinfo {author} {\bibfnamefont {R.}~\bibnamefont {Conte}},\ }\href
  {https://doi.org/10.1051/jphys:0198000410110126300} {\bibfield  {journal}
  {\bibinfo  {journal} {Journal de Physique}\ }\textbf {\bibinfo {volume}
  {41}},\ \bibinfo {pages} {1263} (\bibinfo {year} {1980})}\BibitemShut
  {NoStop}%
\bibitem [{\citenamefont {Priour}\ \emph {et~al.}(2001)\citenamefont {Priour},
  \citenamefont {Gelfand},\ and\ \citenamefont {Sondhi}}]{Priour2001}%
  \BibitemOpen
  \bibfield  {author} {\bibinfo {author} {\bibfnamefont {D.~J.}\ \bibnamefont
  {Priour}}, \bibinfo {author} {\bibfnamefont {M.~P.}\ \bibnamefont
  {Gelfand}},\ and\ \bibinfo {author} {\bibfnamefont {S.~L.}\ \bibnamefont
  {Sondhi}},\ }\bibfield  {journal} {\bibinfo  {journal} {Physical Review B}\
  }\textbf {\bibinfo {volume} {64}},\ \href
  {https://doi.org/10.1103/physrevb.64.134424} {10.1103/physrevb.64.134424}
  (\bibinfo {year} {2001})\BibitemShut {NoStop}%
\bibitem [{\citenamefont {Coester}\ \emph {et~al.}(2013)\citenamefont
  {Coester}, \citenamefont {Malitz}, \citenamefont {Fey},\ and\ \citenamefont
  {Schmidt}}]{Malitz2013}%
  \BibitemOpen
  \bibfield  {author} {\bibinfo {author} {\bibfnamefont {K.}~\bibnamefont
  {Coester}}, \bibinfo {author} {\bibfnamefont {W.}~\bibnamefont {Malitz}},
  \bibinfo {author} {\bibfnamefont {S.}~\bibnamefont {Fey}},\ and\ \bibinfo
  {author} {\bibfnamefont {K.~P.}\ \bibnamefont {Schmidt}},\ }\href
  {https://doi.org/10.1103/PhysRevB.88.184402} {\bibfield  {journal} {\bibinfo
  {journal} {Phys. Rev. B}\ }\textbf {\bibinfo {volume} {88}},\ \bibinfo
  {pages} {184402} (\bibinfo {year} {2013})}\BibitemShut {NoStop}%
\bibitem [{\citenamefont {Castelnovo}\ \emph {et~al.}(2008)\citenamefont
  {Castelnovo}, \citenamefont {Moessner},\ and\ \citenamefont
  {Sondhi}}]{Castelnovo2008}%
  \BibitemOpen
  \bibfield  {author} {\bibinfo {author} {\bibfnamefont {C.}~\bibnamefont
  {Castelnovo}}, \bibinfo {author} {\bibfnamefont {R.}~\bibnamefont
  {Moessner}},\ and\ \bibinfo {author} {\bibfnamefont {S.~L.}\ \bibnamefont
  {Sondhi}},\ }\href {https://doi.org/10.1038/nature06433} {\bibfield
  {journal} {\bibinfo  {journal} {Nature}\ }\textbf {\bibinfo {volume} {451}},\
  \bibinfo {pages} {42} (\bibinfo {year} {2008})}\BibitemShut {NoStop}%
\bibitem [{\citenamefont {Jaubert}\ and\ \citenamefont
  {Holdsworth}(2009)}]{Jaubert2009}%
  \BibitemOpen
  \bibfield  {author} {\bibinfo {author} {\bibfnamefont {L.~D.~C.}\
  \bibnamefont {Jaubert}}\ and\ \bibinfo {author} {\bibfnamefont {P.~C.~W.}\
  \bibnamefont {Holdsworth}},\ }\href {https://doi.org/10.1038/nphys1227}
  {\bibfield  {journal} {\bibinfo  {journal} {Nature Physics}\ }\textbf
  {\bibinfo {volume} {5}},\ \bibinfo {pages} {258} (\bibinfo {year}
  {2009})}\BibitemShut {NoStop}%
\bibitem [{\citenamefont {Fennell}\ \emph {et~al.}(2009)\citenamefont
  {Fennell}, \citenamefont {Deen}, \citenamefont {Wildes}, \citenamefont
  {Schmalzl}, \citenamefont {Prabhakaran}, \citenamefont {Boothroyd},
  \citenamefont {Aldus}, \citenamefont {McMorrow},\ and\ \citenamefont
  {Bramwell}}]{Fennell2009}%
  \BibitemOpen
  \bibfield  {author} {\bibinfo {author} {\bibfnamefont {T.}~\bibnamefont
  {Fennell}}, \bibinfo {author} {\bibfnamefont {P.~P.}\ \bibnamefont {Deen}},
  \bibinfo {author} {\bibfnamefont {A.~R.}\ \bibnamefont {Wildes}}, \bibinfo
  {author} {\bibfnamefont {K.}~\bibnamefont {Schmalzl}}, \bibinfo {author}
  {\bibfnamefont {D.}~\bibnamefont {Prabhakaran}}, \bibinfo {author}
  {\bibfnamefont {A.~T.}\ \bibnamefont {Boothroyd}}, \bibinfo {author}
  {\bibfnamefont {R.~J.}\ \bibnamefont {Aldus}}, \bibinfo {author}
  {\bibfnamefont {D.~F.}\ \bibnamefont {McMorrow}},\ and\ \bibinfo {author}
  {\bibfnamefont {S.~T.}\ \bibnamefont {Bramwell}},\ }\href
  {https://doi.org/10.1126/science.1177582} {\bibfield  {journal} {\bibinfo
  {journal} {Science}\ }\textbf {\bibinfo {volume} {326}},\ \bibinfo {pages}
  {415} (\bibinfo {year} {2009})}\BibitemShut {NoStop}%
\bibitem [{\citenamefont {Savary}\ and\ \citenamefont
  {Balents}(2017)}]{Savary2017}%
  \BibitemOpen
  \bibfield  {author} {\bibinfo {author} {\bibfnamefont {L.}~\bibnamefont
  {Savary}}\ and\ \bibinfo {author} {\bibfnamefont {L.}~\bibnamefont
  {Balents}},\ }\href {https://doi.org/10.1103/PhysRevLett.118.087203}
  {\bibfield  {journal} {\bibinfo  {journal} {Phys. Rev. Lett.}\ }\textbf
  {\bibinfo {volume} {118}},\ \bibinfo {pages} {087203} (\bibinfo {year}
  {2017})}\BibitemShut {NoStop}%
\bibitem [{\citenamefont {Defenu}\ \emph {et~al.}(2023)\citenamefont {Defenu},
  \citenamefont {Donner}, \citenamefont {Macr\`{\i}}, \citenamefont {Pagano},
  \citenamefont {Ruffo},\ and\ \citenamefont {Trombettoni}}]{Defenu2023}%
  \BibitemOpen
  \bibfield  {author} {\bibinfo {author} {\bibfnamefont {N.}~\bibnamefont
  {Defenu}}, \bibinfo {author} {\bibfnamefont {T.}~\bibnamefont {Donner}},
  \bibinfo {author} {\bibfnamefont {T.}~\bibnamefont {Macr\`{\i}}}, \bibinfo
  {author} {\bibfnamefont {G.}~\bibnamefont {Pagano}}, \bibinfo {author}
  {\bibfnamefont {S.}~\bibnamefont {Ruffo}},\ and\ \bibinfo {author}
  {\bibfnamefont {A.}~\bibnamefont {Trombettoni}},\ }\href
  {https://doi.org/10.1103/RevModPhys.95.035002} {\bibfield  {journal}
  {\bibinfo  {journal} {Rev. Mod. Phys.}\ }\textbf {\bibinfo {volume} {95}},\
  \bibinfo {pages} {035002} (\bibinfo {year} {2023})}\BibitemShut {NoStop}%
\bibitem [{\citenamefont {Gross}\ and\ \citenamefont
  {Bloch}(2017)}]{Gross2017}%
  \BibitemOpen
  \bibfield  {author} {\bibinfo {author} {\bibfnamefont {C.}~\bibnamefont
  {Gross}}\ and\ \bibinfo {author} {\bibfnamefont {I.}~\bibnamefont {Bloch}},\
  }\href {https://doi.org/10.1126/science.aal3837} {\bibfield  {journal}
  {\bibinfo  {journal} {Science}\ }\textbf {\bibinfo {volume} {357}},\ \bibinfo
  {pages} {995–1001} (\bibinfo {year} {2017})}\BibitemShut {NoStop}%
\bibitem [{\citenamefont {Browaeys}\ and\ \citenamefont
  {Lahaye}(2020)}]{Browaeys2020}%
  \BibitemOpen
  \bibfield  {author} {\bibinfo {author} {\bibfnamefont {A.}~\bibnamefont
  {Browaeys}}\ and\ \bibinfo {author} {\bibfnamefont {T.}~\bibnamefont
  {Lahaye}},\ }\href {https://doi.org/10.1038/s41567-019-0733-z} {\bibfield
  {journal} {\bibinfo  {journal} {Nature Physics}\ }\textbf {\bibinfo {volume}
  {16}},\ \bibinfo {pages} {132} (\bibinfo {year} {2020})}\BibitemShut
  {NoStop}%
\bibitem [{\citenamefont {Chomaz}\ \emph {et~al.}(2022)\citenamefont {Chomaz},
  \citenamefont {Ferrier-Barbut}, \citenamefont {Ferlaino}, \citenamefont
  {Laburthe-Tolra}, \citenamefont {Lev},\ and\ \citenamefont
  {Pfau}}]{Chomaz2022}%
  \BibitemOpen
  \bibfield  {author} {\bibinfo {author} {\bibfnamefont {L.}~\bibnamefont
  {Chomaz}}, \bibinfo {author} {\bibfnamefont {I.}~\bibnamefont
  {Ferrier-Barbut}}, \bibinfo {author} {\bibfnamefont {F.}~\bibnamefont
  {Ferlaino}}, \bibinfo {author} {\bibfnamefont {B.}~\bibnamefont
  {Laburthe-Tolra}}, \bibinfo {author} {\bibfnamefont {B.~L.}\ \bibnamefont
  {Lev}},\ and\ \bibinfo {author} {\bibfnamefont {T.}~\bibnamefont {Pfau}},\
  }\href {https://doi.org/10.1088/1361-6633/aca814} {\bibfield  {journal}
  {\bibinfo  {journal} {Reports on Progress in Physics}\ }\textbf {\bibinfo
  {volume} {86}},\ \bibinfo {pages} {026401} (\bibinfo {year}
  {2022})}\BibitemShut {NoStop}%
\bibitem [{\citenamefont {Friedenauer}\ \emph {et~al.}(2008)\citenamefont
  {Friedenauer}, \citenamefont {Schmitz}, \citenamefont {Glueckert},
  \citenamefont {Porras},\ and\ \citenamefont {Schaetz}}]{Friedenauer2008}%
  \BibitemOpen
  \bibfield  {author} {\bibinfo {author} {\bibfnamefont {A.}~\bibnamefont
  {Friedenauer}}, \bibinfo {author} {\bibfnamefont {H.}~\bibnamefont
  {Schmitz}}, \bibinfo {author} {\bibfnamefont {J.~T.}\ \bibnamefont
  {Glueckert}}, \bibinfo {author} {\bibfnamefont {D.}~\bibnamefont {Porras}},\
  and\ \bibinfo {author} {\bibfnamefont {T.}~\bibnamefont {Schaetz}},\ }\href
  {https://doi.org/10.1038/nphys1032} {\bibfield  {journal} {\bibinfo
  {journal} {Nature Physics}\ }\textbf {\bibinfo {volume} {4}},\ \bibinfo
  {pages} {757–761} (\bibinfo {year} {2008})}\BibitemShut {NoStop}%
\bibitem [{\citenamefont {Kim}\ \emph {et~al.}(2009)\citenamefont {Kim},
  \citenamefont {Chang}, \citenamefont {Islam}, \citenamefont {Korenblit},
  \citenamefont {Duan},\ and\ \citenamefont {Monroe}}]{Kim2009}%
  \BibitemOpen
  \bibfield  {author} {\bibinfo {author} {\bibfnamefont {K.}~\bibnamefont
  {Kim}}, \bibinfo {author} {\bibfnamefont {M.-S.}\ \bibnamefont {Chang}},
  \bibinfo {author} {\bibfnamefont {R.}~\bibnamefont {Islam}}, \bibinfo
  {author} {\bibfnamefont {S.}~\bibnamefont {Korenblit}}, \bibinfo {author}
  {\bibfnamefont {L.-M.}\ \bibnamefont {Duan}},\ and\ \bibinfo {author}
  {\bibfnamefont {C.}~\bibnamefont {Monroe}},\ }\href
  {https://doi.org/10.1103/PhysRevLett.103.120502} {\bibfield  {journal}
  {\bibinfo  {journal} {Phys. Rev. Lett.}\ }\textbf {\bibinfo {volume} {103}},\
  \bibinfo {pages} {120502} (\bibinfo {year} {2009})}\BibitemShut {NoStop}%
\bibitem [{\citenamefont {Kim}\ \emph {et~al.}(2010)\citenamefont {Kim},
  \citenamefont {Chang}, \citenamefont {Korenblit}, \citenamefont {Islam},
  \citenamefont {Edwards}, \citenamefont {Freericks}, \citenamefont {Lin},
  \citenamefont {Duan},\ and\ \citenamefont {Monroe}}]{Kim2010}%
  \BibitemOpen
  \bibfield  {author} {\bibinfo {author} {\bibfnamefont {K.}~\bibnamefont
  {Kim}}, \bibinfo {author} {\bibfnamefont {M.-S.}\ \bibnamefont {Chang}},
  \bibinfo {author} {\bibfnamefont {S.}~\bibnamefont {Korenblit}}, \bibinfo
  {author} {\bibfnamefont {R.}~\bibnamefont {Islam}}, \bibinfo {author}
  {\bibfnamefont {E.~E.}\ \bibnamefont {Edwards}}, \bibinfo {author}
  {\bibfnamefont {J.~K.}\ \bibnamefont {Freericks}}, \bibinfo {author}
  {\bibfnamefont {G.-D.}\ \bibnamefont {Lin}}, \bibinfo {author} {\bibfnamefont
  {L.-M.}\ \bibnamefont {Duan}},\ and\ \bibinfo {author} {\bibfnamefont
  {C.}~\bibnamefont {Monroe}},\ }\href {https://doi.org/10.1038/nature09071}
  {\bibfield  {journal} {\bibinfo  {journal} {Nature}\ }\textbf {\bibinfo
  {volume} {465}},\ \bibinfo {pages} {590–593} (\bibinfo {year}
  {2010})}\BibitemShut {NoStop}%
\bibitem [{\citenamefont {Islam}\ \emph {et~al.}(2011)\citenamefont {Islam},
  \citenamefont {Edwards}, \citenamefont {Kim}, \citenamefont {Korenblit},
  \citenamefont {Noh}, \citenamefont {Carmichael}, \citenamefont {Lin},
  \citenamefont {Duan}, \citenamefont {Joseph~Wang}, \citenamefont
  {Freericks},\ and\ \citenamefont {Monroe}}]{Islam2011}%
  \BibitemOpen
  \bibfield  {author} {\bibinfo {author} {\bibfnamefont {R.}~\bibnamefont
  {Islam}}, \bibinfo {author} {\bibfnamefont {E.~E.}\ \bibnamefont {Edwards}},
  \bibinfo {author} {\bibfnamefont {K.}~\bibnamefont {Kim}}, \bibinfo {author}
  {\bibfnamefont {S.}~\bibnamefont {Korenblit}}, \bibinfo {author}
  {\bibfnamefont {C.}~\bibnamefont {Noh}}, \bibinfo {author} {\bibfnamefont
  {H.}~\bibnamefont {Carmichael}}, \bibinfo {author} {\bibfnamefont {G.-D.}\
  \bibnamefont {Lin}}, \bibinfo {author} {\bibfnamefont {L.-M.}\ \bibnamefont
  {Duan}}, \bibinfo {author} {\bibfnamefont {C.-C.}\ \bibnamefont
  {Joseph~Wang}}, \bibinfo {author} {\bibfnamefont {J.~K.}\ \bibnamefont
  {Freericks}},\ and\ \bibinfo {author} {\bibfnamefont {C.}~\bibnamefont
  {Monroe}},\ }\href {https://doi.org/10.1038/ncomms1374} {\bibfield  {journal}
  {\bibinfo  {journal} {Nature Communications}\ }\textbf {\bibinfo {volume}
  {2}},\ \bibinfo {pages} {377} (\bibinfo {year} {2011})}\BibitemShut {NoStop}%
\bibitem [{\citenamefont {Schneider}\ \emph {et~al.}(2012)\citenamefont
  {Schneider}, \citenamefont {Porras},\ and\ \citenamefont
  {Schaetz}}]{Schneider2012}%
  \BibitemOpen
  \bibfield  {author} {\bibinfo {author} {\bibfnamefont {C.}~\bibnamefont
  {Schneider}}, \bibinfo {author} {\bibfnamefont {D.}~\bibnamefont {Porras}},\
  and\ \bibinfo {author} {\bibfnamefont {T.}~\bibnamefont {Schaetz}},\ }\href
  {https://doi.org/10.1088/0034-4885/75/2/024401} {\bibfield  {journal}
  {\bibinfo  {journal} {Reports on Progress in Physics}\ }\textbf {\bibinfo
  {volume} {75}},\ \bibinfo {pages} {024401} (\bibinfo {year}
  {2012})}\BibitemShut {NoStop}%
\bibitem [{\citenamefont {Britton}\ \emph {et~al.}(2012)\citenamefont
  {Britton}, \citenamefont {Sawyer}, \citenamefont {Keith}, \citenamefont
  {Wang}, \citenamefont {Freericks}, \citenamefont {Uys}, \citenamefont
  {Biercuk},\ and\ \citenamefont {Bollinger}}]{Britton2012}%
  \BibitemOpen
  \bibfield  {author} {\bibinfo {author} {\bibfnamefont {J.~W.}\ \bibnamefont
  {Britton}}, \bibinfo {author} {\bibfnamefont {B.~C.}\ \bibnamefont {Sawyer}},
  \bibinfo {author} {\bibfnamefont {A.~C.}\ \bibnamefont {Keith}}, \bibinfo
  {author} {\bibfnamefont {C.-C.~J.}\ \bibnamefont {Wang}}, \bibinfo {author}
  {\bibfnamefont {J.~K.}\ \bibnamefont {Freericks}}, \bibinfo {author}
  {\bibfnamefont {H.}~\bibnamefont {Uys}}, \bibinfo {author} {\bibfnamefont
  {M.~J.}\ \bibnamefont {Biercuk}},\ and\ \bibinfo {author} {\bibfnamefont
  {J.~J.}\ \bibnamefont {Bollinger}},\ }\href
  {https://doi.org/10.1038/nature10981} {\bibfield  {journal} {\bibinfo
  {journal} {Nature}\ }\textbf {\bibinfo {volume} {484}},\ \bibinfo {pages}
  {489–492} (\bibinfo {year} {2012})}\BibitemShut {NoStop}%
\bibitem [{\citenamefont {Islam}\ \emph {et~al.}(2013)\citenamefont {Islam},
  \citenamefont {Senko}, \citenamefont {Campbell}, \citenamefont {Korenblit},
  \citenamefont {Smith}, \citenamefont {Lee}, \citenamefont {Edwards},
  \citenamefont {Wang}, \citenamefont {Freericks},\ and\ \citenamefont
  {Monroe}}]{Islam2013}%
  \BibitemOpen
  \bibfield  {author} {\bibinfo {author} {\bibfnamefont {R.}~\bibnamefont
  {Islam}}, \bibinfo {author} {\bibfnamefont {C.}~\bibnamefont {Senko}},
  \bibinfo {author} {\bibfnamefont {W.~C.}\ \bibnamefont {Campbell}}, \bibinfo
  {author} {\bibfnamefont {S.}~\bibnamefont {Korenblit}}, \bibinfo {author}
  {\bibfnamefont {J.}~\bibnamefont {Smith}}, \bibinfo {author} {\bibfnamefont
  {A.}~\bibnamefont {Lee}}, \bibinfo {author} {\bibfnamefont {E.~E.}\
  \bibnamefont {Edwards}}, \bibinfo {author} {\bibfnamefont {C.-C.~J.}\
  \bibnamefont {Wang}}, \bibinfo {author} {\bibfnamefont {J.~K.}\ \bibnamefont
  {Freericks}},\ and\ \bibinfo {author} {\bibfnamefont {C.}~\bibnamefont
  {Monroe}},\ }\href {https://doi.org/10.1126/science.1232296} {\bibfield
  {journal} {\bibinfo  {journal} {Science}\ }\textbf {\bibinfo {volume}
  {340}},\ \bibinfo {pages} {583–587} (\bibinfo {year} {2013})}\BibitemShut
  {NoStop}%
\bibitem [{\citenamefont {Jurcevic}\ \emph {et~al.}(2014)\citenamefont
  {Jurcevic}, \citenamefont {Lanyon}, \citenamefont {Hauke}, \citenamefont
  {Hempel}, \citenamefont {Zoller}, \citenamefont {Blatt},\ and\ \citenamefont
  {Roos}}]{Jurcevic2014}%
  \BibitemOpen
  \bibfield  {author} {\bibinfo {author} {\bibfnamefont {P.}~\bibnamefont
  {Jurcevic}}, \bibinfo {author} {\bibfnamefont {B.~P.}\ \bibnamefont
  {Lanyon}}, \bibinfo {author} {\bibfnamefont {P.}~\bibnamefont {Hauke}},
  \bibinfo {author} {\bibfnamefont {C.}~\bibnamefont {Hempel}}, \bibinfo
  {author} {\bibfnamefont {P.}~\bibnamefont {Zoller}}, \bibinfo {author}
  {\bibfnamefont {R.}~\bibnamefont {Blatt}},\ and\ \bibinfo {author}
  {\bibfnamefont {C.~F.}\ \bibnamefont {Roos}},\ }\href
  {https://doi.org/10.1038/nature13461} {\bibfield  {journal} {\bibinfo
  {journal} {Nature}\ }\textbf {\bibinfo {volume} {511}},\ \bibinfo {pages}
  {202–205} (\bibinfo {year} {2014})}\BibitemShut {NoStop}%
\bibitem [{\citenamefont {Bohnet}\ \emph {et~al.}(2016)\citenamefont {Bohnet},
  \citenamefont {Sawyer}, \citenamefont {Britton}, \citenamefont {Wall},
  \citenamefont {Rey}, \citenamefont {Foss-Feig},\ and\ \citenamefont
  {Bollinger}}]{Bohnet2016}%
  \BibitemOpen
  \bibfield  {author} {\bibinfo {author} {\bibfnamefont {J.~G.}\ \bibnamefont
  {Bohnet}}, \bibinfo {author} {\bibfnamefont {B.~C.}\ \bibnamefont {Sawyer}},
  \bibinfo {author} {\bibfnamefont {J.~W.}\ \bibnamefont {Britton}}, \bibinfo
  {author} {\bibfnamefont {M.~L.}\ \bibnamefont {Wall}}, \bibinfo {author}
  {\bibfnamefont {A.~M.}\ \bibnamefont {Rey}}, \bibinfo {author} {\bibfnamefont
  {M.}~\bibnamefont {Foss-Feig}},\ and\ \bibinfo {author} {\bibfnamefont
  {J.~J.}\ \bibnamefont {Bollinger}},\ }\href
  {https://doi.org/10.1126/science.aad9958} {\bibfield  {journal} {\bibinfo
  {journal} {Science}\ }\textbf {\bibinfo {volume} {352}},\ \bibinfo {pages}
  {1297–1301} (\bibinfo {year} {2016})}\BibitemShut {NoStop}%
\bibitem [{\citenamefont {Lukin}\ \emph {et~al.}(2001)\citenamefont {Lukin},
  \citenamefont {Fleischhauer}, \citenamefont {Cote}, \citenamefont {Duan},
  \citenamefont {Jaksch}, \citenamefont {Cirac},\ and\ \citenamefont
  {Zoller}}]{Lukin2001}%
  \BibitemOpen
  \bibfield  {author} {\bibinfo {author} {\bibfnamefont {M.~D.}\ \bibnamefont
  {Lukin}}, \bibinfo {author} {\bibfnamefont {M.}~\bibnamefont {Fleischhauer}},
  \bibinfo {author} {\bibfnamefont {R.}~\bibnamefont {Cote}}, \bibinfo {author}
  {\bibfnamefont {L.~M.}\ \bibnamefont {Duan}}, \bibinfo {author}
  {\bibfnamefont {D.}~\bibnamefont {Jaksch}}, \bibinfo {author} {\bibfnamefont
  {J.~I.}\ \bibnamefont {Cirac}},\ and\ \bibinfo {author} {\bibfnamefont
  {P.}~\bibnamefont {Zoller}},\ }\href
  {https://doi.org/10.1103/PhysRevLett.87.037901} {\bibfield  {journal}
  {\bibinfo  {journal} {Phys. Rev. Lett.}\ }\textbf {\bibinfo {volume} {87}},\
  \bibinfo {pages} {037901} (\bibinfo {year} {2001})}\BibitemShut {NoStop}%
\bibitem [{\citenamefont {Jaksch}\ \emph {et~al.}(2000)\citenamefont {Jaksch},
  \citenamefont {Cirac}, \citenamefont {Zoller}, \citenamefont {Rolston},
  \citenamefont {C\^ot\'e},\ and\ \citenamefont {Lukin}}]{Jaksch2000}%
  \BibitemOpen
  \bibfield  {author} {\bibinfo {author} {\bibfnamefont {D.}~\bibnamefont
  {Jaksch}}, \bibinfo {author} {\bibfnamefont {J.~I.}\ \bibnamefont {Cirac}},
  \bibinfo {author} {\bibfnamefont {P.}~\bibnamefont {Zoller}}, \bibinfo
  {author} {\bibfnamefont {S.~L.}\ \bibnamefont {Rolston}}, \bibinfo {author}
  {\bibfnamefont {R.}~\bibnamefont {C\^ot\'e}},\ and\ \bibinfo {author}
  {\bibfnamefont {M.~D.}\ \bibnamefont {Lukin}},\ }\href
  {https://doi.org/10.1103/PhysRevLett.85.2208} {\bibfield  {journal} {\bibinfo
   {journal} {Phys. Rev. Lett.}\ }\textbf {\bibinfo {volume} {85}},\ \bibinfo
  {pages} {2208} (\bibinfo {year} {2000})}\BibitemShut {NoStop}%
\bibitem [{\citenamefont {Sachdev}\ \emph {et~al.}(2002)\citenamefont
  {Sachdev}, \citenamefont {Sengupta},\ and\ \citenamefont
  {Girvin}}]{Sachdev2002}%
  \BibitemOpen
  \bibfield  {author} {\bibinfo {author} {\bibfnamefont {S.}~\bibnamefont
  {Sachdev}}, \bibinfo {author} {\bibfnamefont {K.}~\bibnamefont {Sengupta}},\
  and\ \bibinfo {author} {\bibfnamefont {S.~M.}\ \bibnamefont {Girvin}},\
  }\href {https://doi.org/10.1103/PhysRevB.66.075128} {\bibfield  {journal}
  {\bibinfo  {journal} {Phys. Rev. B}\ }\textbf {\bibinfo {volume} {66}},\
  \bibinfo {pages} {075128} (\bibinfo {year} {2002})}\BibitemShut {NoStop}%
\bibitem [{\citenamefont {Fendley}\ \emph {et~al.}(2004)\citenamefont
  {Fendley}, \citenamefont {Sengupta},\ and\ \citenamefont
  {Sachdev}}]{Fendley2004}%
  \BibitemOpen
  \bibfield  {author} {\bibinfo {author} {\bibfnamefont {P.}~\bibnamefont
  {Fendley}}, \bibinfo {author} {\bibfnamefont {K.}~\bibnamefont {Sengupta}},\
  and\ \bibinfo {author} {\bibfnamefont {S.}~\bibnamefont {Sachdev}},\ }\href
  {https://doi.org/10.1103/PhysRevB.69.075106} {\bibfield  {journal} {\bibinfo
  {journal} {Phys. Rev. B}\ }\textbf {\bibinfo {volume} {69}},\ \bibinfo
  {pages} {075106} (\bibinfo {year} {2004})}\BibitemShut {NoStop}%
\bibitem [{\citenamefont {Labuhn}\ \emph {et~al.}(2016)\citenamefont {Labuhn},
  \citenamefont {Barredo}, \citenamefont {Ravets}, \citenamefont
  {de~L{\'{e}}s{\'{e}}leuc}, \citenamefont {Macr{\`{\i}}}, \citenamefont
  {Lahaye},\ and\ \citenamefont {Browaeys}}]{Labuhn2016}%
  \BibitemOpen
  \bibfield  {author} {\bibinfo {author} {\bibfnamefont {H.}~\bibnamefont
  {Labuhn}}, \bibinfo {author} {\bibfnamefont {D.}~\bibnamefont {Barredo}},
  \bibinfo {author} {\bibfnamefont {S.}~\bibnamefont {Ravets}}, \bibinfo
  {author} {\bibfnamefont {S.}~\bibnamefont {de~L{\'{e}}s{\'{e}}leuc}},
  \bibinfo {author} {\bibfnamefont {T.}~\bibnamefont {Macr{\`{\i}}}}, \bibinfo
  {author} {\bibfnamefont {T.}~\bibnamefont {Lahaye}},\ and\ \bibinfo {author}
  {\bibfnamefont {A.}~\bibnamefont {Browaeys}},\ }\href
  {https://doi.org/10.1038/nature18274} {\bibfield  {journal} {\bibinfo
  {journal} {Nature}\ }\textbf {\bibinfo {volume} {534}},\ \bibinfo {pages}
  {667} (\bibinfo {year} {2016})}\BibitemShut {NoStop}%
\bibitem [{\citenamefont {Schauss}(2018)}]{Schauss2018}%
  \BibitemOpen
  \bibfield  {author} {\bibinfo {author} {\bibfnamefont {P.}~\bibnamefont
  {Schauss}},\ }\href {https://doi.org/10.1088/2058-9565/aa9c59} {\bibfield
  {journal} {\bibinfo  {journal} {Quantum Science and Technology}\ }\textbf
  {\bibinfo {volume} {3}},\ \bibinfo {pages} {023001} (\bibinfo {year}
  {2018})}\BibitemShut {NoStop}%
\bibitem [{\citenamefont {Scholl}\ \emph {et~al.}(2021)\citenamefont {Scholl},
  \citenamefont {Schuler}, \citenamefont {Williams}, \citenamefont
  {Eberharter}, \citenamefont {Barredo}, \citenamefont {Schymik}, \citenamefont
  {Lienhard}, \citenamefont {Henry}, \citenamefont {Lang}, \citenamefont
  {Lahaye}, \citenamefont {L\"{a}uchli},\ and\ \citenamefont
  {Browaeys}}]{Scholl2021}%
  \BibitemOpen
  \bibfield  {author} {\bibinfo {author} {\bibfnamefont {P.}~\bibnamefont
  {Scholl}}, \bibinfo {author} {\bibfnamefont {M.}~\bibnamefont {Schuler}},
  \bibinfo {author} {\bibfnamefont {H.~J.}\ \bibnamefont {Williams}}, \bibinfo
  {author} {\bibfnamefont {A.~A.}\ \bibnamefont {Eberharter}}, \bibinfo
  {author} {\bibfnamefont {D.}~\bibnamefont {Barredo}}, \bibinfo {author}
  {\bibfnamefont {K.-N.}\ \bibnamefont {Schymik}}, \bibinfo {author}
  {\bibfnamefont {V.}~\bibnamefont {Lienhard}}, \bibinfo {author}
  {\bibfnamefont {L.-P.}\ \bibnamefont {Henry}}, \bibinfo {author}
  {\bibfnamefont {T.~C.}\ \bibnamefont {Lang}}, \bibinfo {author}
  {\bibfnamefont {T.}~\bibnamefont {Lahaye}}, \bibinfo {author} {\bibfnamefont
  {A.~M.}\ \bibnamefont {L\"{a}uchli}},\ and\ \bibinfo {author} {\bibfnamefont
  {A.}~\bibnamefont {Browaeys}},\ }\href
  {https://doi.org/10.1038/s41586-021-03585-1} {\bibfield  {journal} {\bibinfo
  {journal} {Nature}\ }\textbf {\bibinfo {volume} {595}},\ \bibinfo {pages}
  {233} (\bibinfo {year} {2021})}\BibitemShut {NoStop}%
\bibitem [{\citenamefont {Semeghini}\ \emph {et~al.}(2021)\citenamefont
  {Semeghini}, \citenamefont {Levine}, \citenamefont {Keesling}, \citenamefont
  {Ebadi}, \citenamefont {Wang}, \citenamefont {Bluvstein}, \citenamefont
  {Verresen}, \citenamefont {Pichler}, \citenamefont {Kalinowski},
  \citenamefont {Samajdar}, \citenamefont {Omran}, \citenamefont {Sachdev},
  \citenamefont {Vishwanath}, \citenamefont {Greiner}, \citenamefont
  {Vuleti{\'{c}}},\ and\ \citenamefont {Lukin}}]{Semeghini2021}%
  \BibitemOpen
  \bibfield  {author} {\bibinfo {author} {\bibfnamefont {G.}~\bibnamefont
  {Semeghini}}, \bibinfo {author} {\bibfnamefont {H.}~\bibnamefont {Levine}},
  \bibinfo {author} {\bibfnamefont {A.}~\bibnamefont {Keesling}}, \bibinfo
  {author} {\bibfnamefont {S.}~\bibnamefont {Ebadi}}, \bibinfo {author}
  {\bibfnamefont {T.~T.}\ \bibnamefont {Wang}}, \bibinfo {author}
  {\bibfnamefont {D.}~\bibnamefont {Bluvstein}}, \bibinfo {author}
  {\bibfnamefont {R.}~\bibnamefont {Verresen}}, \bibinfo {author}
  {\bibfnamefont {H.}~\bibnamefont {Pichler}}, \bibinfo {author} {\bibfnamefont
  {M.}~\bibnamefont {Kalinowski}}, \bibinfo {author} {\bibfnamefont
  {R.}~\bibnamefont {Samajdar}}, \bibinfo {author} {\bibfnamefont
  {A.}~\bibnamefont {Omran}}, \bibinfo {author} {\bibfnamefont
  {S.}~\bibnamefont {Sachdev}}, \bibinfo {author} {\bibfnamefont
  {A.}~\bibnamefont {Vishwanath}}, \bibinfo {author} {\bibfnamefont
  {M.}~\bibnamefont {Greiner}}, \bibinfo {author} {\bibfnamefont
  {V.}~\bibnamefont {Vuleti{\'{c}}}},\ and\ \bibinfo {author} {\bibfnamefont
  {M.~D.}\ \bibnamefont {Lukin}},\ }\href
  {https://doi.org/10.1126/science.abi8794} {\bibfield  {journal} {\bibinfo
  {journal} {Science}\ }\textbf {\bibinfo {volume} {374}},\ \bibinfo {pages}
  {1242} (\bibinfo {year} {2021})}\BibitemShut {NoStop}%
\bibitem [{\citenamefont {Verresen}\ \emph {et~al.}(2021)\citenamefont
  {Verresen}, \citenamefont {Lukin},\ and\ \citenamefont
  {Vishwanath}}]{Verresen2021}%
  \BibitemOpen
  \bibfield  {author} {\bibinfo {author} {\bibfnamefont {R.}~\bibnamefont
  {Verresen}}, \bibinfo {author} {\bibfnamefont {M.~D.}\ \bibnamefont
  {Lukin}},\ and\ \bibinfo {author} {\bibfnamefont {A.}~\bibnamefont
  {Vishwanath}},\ }\href {https://doi.org/10.1103/PhysRevX.11.031005}
  {\bibfield  {journal} {\bibinfo  {journal} {Phys. Rev. X}\ }\textbf {\bibinfo
  {volume} {11}},\ \bibinfo {pages} {031005} (\bibinfo {year}
  {2021})}\BibitemShut {NoStop}%
\bibitem [{\citenamefont {Samajdar}\ \emph {et~al.}(2023)\citenamefont
  {Samajdar}, \citenamefont {Joshi}, \citenamefont {Teng},\ and\ \citenamefont
  {Sachdev}}]{Samajdar2023}%
  \BibitemOpen
  \bibfield  {author} {\bibinfo {author} {\bibfnamefont {R.}~\bibnamefont
  {Samajdar}}, \bibinfo {author} {\bibfnamefont {D.~G.}\ \bibnamefont {Joshi}},
  \bibinfo {author} {\bibfnamefont {Y.}~\bibnamefont {Teng}},\ and\ \bibinfo
  {author} {\bibfnamefont {S.}~\bibnamefont {Sachdev}},\ }\href
  {https://doi.org/10.1103/PhysRevLett.130.043601} {\bibfield  {journal}
  {\bibinfo  {journal} {Phys. Rev. Lett.}\ }\textbf {\bibinfo {volume} {130}},\
  \bibinfo {pages} {043601} (\bibinfo {year} {2023})}\BibitemShut {NoStop}%
\bibitem [{\citenamefont {Dutta}\ and\ \citenamefont
  {Bhattacharjee}(2001)}]{Dutta2001}%
  \BibitemOpen
  \bibfield  {author} {\bibinfo {author} {\bibfnamefont {A.}~\bibnamefont
  {Dutta}}\ and\ \bibinfo {author} {\bibfnamefont {J.~K.}\ \bibnamefont
  {Bhattacharjee}},\ }\href {https://doi.org/10.1103/PhysRevB.64.184106}
  {\bibfield  {journal} {\bibinfo  {journal} {Phys. Rev. B}\ }\textbf {\bibinfo
  {volume} {64}},\ \bibinfo {pages} {184106} (\bibinfo {year}
  {2001})}\BibitemShut {NoStop}%
\bibitem [{\citenamefont {Fey}\ and\ \citenamefont {Schmidt}(2016)}]{Fey2016}%
  \BibitemOpen
  \bibfield  {author} {\bibinfo {author} {\bibfnamefont {S.}~\bibnamefont
  {Fey}}\ and\ \bibinfo {author} {\bibfnamefont {K.~P.}\ \bibnamefont
  {Schmidt}},\ }\href {https://doi.org/10.1103/PhysRevB.94.075156} {\bibfield
  {journal} {\bibinfo  {journal} {Phys. Rev. B}\ }\textbf {\bibinfo {volume}
  {94}},\ \bibinfo {pages} {075156} (\bibinfo {year} {2016})}\BibitemShut
  {NoStop}%
\bibitem [{\citenamefont {Defenu}\ \emph {et~al.}(2017)\citenamefont {Defenu},
  \citenamefont {Trombettoni},\ and\ \citenamefont {Ruffo}}]{Defenu2017}%
  \BibitemOpen
  \bibfield  {author} {\bibinfo {author} {\bibfnamefont {N.}~\bibnamefont
  {Defenu}}, \bibinfo {author} {\bibfnamefont {A.}~\bibnamefont
  {Trombettoni}},\ and\ \bibinfo {author} {\bibfnamefont {S.}~\bibnamefont
  {Ruffo}},\ }\href {https://doi.org/10.1103/PhysRevB.96.104432} {\bibfield
  {journal} {\bibinfo  {journal} {Phys. Rev. B}\ }\textbf {\bibinfo {volume}
  {96}},\ \bibinfo {pages} {104432} (\bibinfo {year} {2017})}\BibitemShut
  {NoStop}%
\bibitem [{\citenamefont {Fey}\ \emph {et~al.}(2019)\citenamefont {Fey},
  \citenamefont {Kapfer},\ and\ \citenamefont {Schmidt}}]{Fey2019}%
  \BibitemOpen
  \bibfield  {author} {\bibinfo {author} {\bibfnamefont {S.}~\bibnamefont
  {Fey}}, \bibinfo {author} {\bibfnamefont {S.~C.}\ \bibnamefont {Kapfer}},\
  and\ \bibinfo {author} {\bibfnamefont {K.~P.}\ \bibnamefont {Schmidt}},\
  }\href {https://doi.org/10.1103/PhysRevLett.122.017203} {\bibfield  {journal}
  {\bibinfo  {journal} {Phys. Rev. Lett.}\ }\textbf {\bibinfo {volume} {122}},\
  \bibinfo {pages} {017203} (\bibinfo {year} {2019})}\BibitemShut {NoStop}%
\bibitem [{\citenamefont {Zhu}\ \emph {et~al.}(2018)\citenamefont {Zhu},
  \citenamefont {Sun}, \citenamefont {You},\ and\ \citenamefont
  {Shi}}]{Zhu2019}%
  \BibitemOpen
  \bibfield  {author} {\bibinfo {author} {\bibfnamefont {Z.}~\bibnamefont
  {Zhu}}, \bibinfo {author} {\bibfnamefont {G.}~\bibnamefont {Sun}}, \bibinfo
  {author} {\bibfnamefont {W.-L.}\ \bibnamefont {You}},\ and\ \bibinfo {author}
  {\bibfnamefont {D.-N.}\ \bibnamefont {Shi}},\ }\href
  {https://doi.org/10.1103/PhysRevA.98.023607} {\bibfield  {journal} {\bibinfo
  {journal} {Phys. Rev. A}\ }\textbf {\bibinfo {volume} {98}},\ \bibinfo
  {pages} {023607} (\bibinfo {year} {2018})}\BibitemShut {NoStop}%
\bibitem [{\citenamefont {Adelhardt}\ \emph {et~al.}(2020)\citenamefont
  {Adelhardt}, \citenamefont {Koziol}, \citenamefont {Schellenberger},\ and\
  \citenamefont {Schmidt}}]{Adelhardt2020}%
  \BibitemOpen
  \bibfield  {author} {\bibinfo {author} {\bibfnamefont {P.}~\bibnamefont
  {Adelhardt}}, \bibinfo {author} {\bibfnamefont {J.~A.}\ \bibnamefont
  {Koziol}}, \bibinfo {author} {\bibfnamefont {A.}~\bibnamefont
  {Schellenberger}},\ and\ \bibinfo {author} {\bibfnamefont {K.~P.}\
  \bibnamefont {Schmidt}},\ }\href
  {https://doi.org/10.1103/PhysRevB.102.174424} {\bibfield  {journal} {\bibinfo
   {journal} {Phys. Rev. B}\ }\textbf {\bibinfo {volume} {102}},\ \bibinfo
  {pages} {174424} (\bibinfo {year} {2020})}\BibitemShut {NoStop}%
\bibitem [{\citenamefont {Koziol}\ \emph {et~al.}(2021)\citenamefont {Koziol},
  \citenamefont {Langheld}, \citenamefont {Kapfer},\ and\ \citenamefont
  {Schmidt}}]{Koziol2021}%
  \BibitemOpen
  \bibfield  {author} {\bibinfo {author} {\bibfnamefont {J.~A.}\ \bibnamefont
  {Koziol}}, \bibinfo {author} {\bibfnamefont {A.}~\bibnamefont {Langheld}},
  \bibinfo {author} {\bibfnamefont {S.~C.}\ \bibnamefont {Kapfer}},\ and\
  \bibinfo {author} {\bibfnamefont {K.~P.}\ \bibnamefont {Schmidt}},\ }\href
  {https://doi.org/10.1103/PhysRevB.103.245135} {\bibfield  {journal} {\bibinfo
   {journal} {Phys. Rev. B}\ }\textbf {\bibinfo {volume} {103}},\ \bibinfo
  {pages} {245135} (\bibinfo {year} {2021})}\BibitemShut {NoStop}%
\bibitem [{\citenamefont {Langheld}\ \emph {et~al.}(2022)\citenamefont
  {Langheld}, \citenamefont {Koziol}, \citenamefont {Adelhardt}, \citenamefont
  {Kapfer},\ and\ \citenamefont {Schmidt}}]{Langheld2022}%
  \BibitemOpen
  \bibfield  {author} {\bibinfo {author} {\bibfnamefont {A.}~\bibnamefont
  {Langheld}}, \bibinfo {author} {\bibfnamefont {J.~A.}\ \bibnamefont
  {Koziol}}, \bibinfo {author} {\bibfnamefont {P.}~\bibnamefont {Adelhardt}},
  \bibinfo {author} {\bibfnamefont {S.~C.}\ \bibnamefont {Kapfer}},\ and\
  \bibinfo {author} {\bibfnamefont {K.~P.}\ \bibnamefont {Schmidt}},\ }\href
  {https://doi.org/10.21468/SciPostPhys.13.4.088} {\bibfield  {journal}
  {\bibinfo  {journal} {SciPost Phys.}\ }\textbf {\bibinfo {volume} {13}},\
  \bibinfo {pages} {088} (\bibinfo {year} {2022})}\BibitemShut {NoStop}%
\bibitem [{\citenamefont {Adelhardt}\ and\ \citenamefont
  {Schmidt}(2023)}]{Adelhardt2023}%
  \BibitemOpen
  \bibfield  {author} {\bibinfo {author} {\bibfnamefont {P.}~\bibnamefont
  {Adelhardt}}\ and\ \bibinfo {author} {\bibfnamefont {K.~P.}\ \bibnamefont
  {Schmidt}},\ }\href {https://doi.org/10.21468/SciPostPhys.15.3.087}
  {\bibfield  {journal} {\bibinfo  {journal} {SciPost Phys.}\ }\textbf
  {\bibinfo {volume} {15}},\ \bibinfo {pages} {087} (\bibinfo {year}
  {2023})}\BibitemShut {NoStop}%
\bibitem [{\citenamefont {Song}\ \emph {et~al.}(2023)\citenamefont {Song},
  \citenamefont {Zhao}, \citenamefont {Qi}, \citenamefont {Rong},\ and\
  \citenamefont {Meng}}]{Song2023}%
  \BibitemOpen
  \bibfield  {author} {\bibinfo {author} {\bibfnamefont {M.}~\bibnamefont
  {Song}}, \bibinfo {author} {\bibfnamefont {J.}~\bibnamefont {Zhao}}, \bibinfo
  {author} {\bibfnamefont {Y.}~\bibnamefont {Qi}}, \bibinfo {author}
  {\bibfnamefont {J.}~\bibnamefont {Rong}},\ and\ \bibinfo {author}
  {\bibfnamefont {Z.~Y.}\ \bibnamefont {Meng}},\ }\href@noop {} {\bibinfo
  {title} {Quantum criticality and entanglement for 2d long-range heisenberg
  bilayer}} (\bibinfo {year} {2023}),\ \Eprint
  {https://arxiv.org/abs/2306.05465} {arXiv:2306.05465 [cond-mat.str-el]}
  \BibitemShut {NoStop}%
\bibitem [{\citenamefont {Adelhardt}\ \emph {et~al.}(2024)\citenamefont
  {Adelhardt}, \citenamefont {Koziol}, \citenamefont {Langheld},\ and\
  \citenamefont {Schmidt}}]{Adelhardt2024}%
  \BibitemOpen
  \bibfield  {author} {\bibinfo {author} {\bibfnamefont {P.}~\bibnamefont
  {Adelhardt}}, \bibinfo {author} {\bibfnamefont {J.~A.}\ \bibnamefont
  {Koziol}}, \bibinfo {author} {\bibfnamefont {A.}~\bibnamefont {Langheld}},\
  and\ \bibinfo {author} {\bibfnamefont {K.~P.}\ \bibnamefont {Schmidt}},\
  }\href {https://doi.org/10.3390/e26050401} {\bibfield  {journal} {\bibinfo
  {journal} {Entropy}\ }\textbf {\bibinfo {volume} {26}},\ \bibinfo {pages}
  {401} (\bibinfo {year} {2024})}\BibitemShut {NoStop}%
\bibitem [{\citenamefont {Koziol}\ \emph {et~al.}(2023)\citenamefont {Koziol},
  \citenamefont {Duft}, \citenamefont {Morigi},\ and\ \citenamefont
  {Schmidt}}]{Koziol2023}%
  \BibitemOpen
  \bibfield  {author} {\bibinfo {author} {\bibfnamefont {J.~A.}\ \bibnamefont
  {Koziol}}, \bibinfo {author} {\bibfnamefont {A.}~\bibnamefont {Duft}},
  \bibinfo {author} {\bibfnamefont {G.}~\bibnamefont {Morigi}},\ and\ \bibinfo
  {author} {\bibfnamefont {K.~P.}\ \bibnamefont {Schmidt}},\ }\href
  {https://doi.org/10.21468/SciPostPhys.14.5.136} {\bibfield  {journal}
  {\bibinfo  {journal} {SciPost Phys.}\ }\textbf {\bibinfo {volume} {14}},\
  \bibinfo {pages} {136} (\bibinfo {year} {2023})}\BibitemShut {NoStop}%
\bibitem [{\citenamefont {Smerald}\ \emph {et~al.}(2016)\citenamefont
  {Smerald}, \citenamefont {Korshunov},\ and\ \citenamefont
  {Mila}}]{Smerald2016}%
  \BibitemOpen
  \bibfield  {author} {\bibinfo {author} {\bibfnamefont {A.}~\bibnamefont
  {Smerald}}, \bibinfo {author} {\bibfnamefont {S.}~\bibnamefont {Korshunov}},\
  and\ \bibinfo {author} {\bibfnamefont {F.}~\bibnamefont {Mila}},\ }\href
  {https://doi.org/10.1103/PhysRevLett.116.197201} {\bibfield  {journal}
  {\bibinfo  {journal} {Phys. Rev. Lett.}\ }\textbf {\bibinfo {volume} {116}},\
  \bibinfo {pages} {197201} (\bibinfo {year} {2016})}\BibitemShut {NoStop}%
\bibitem [{\citenamefont {Smerald}\ and\ \citenamefont
  {Mila}(2018)}]{Smerald2018}%
  \BibitemOpen
  \bibfield  {author} {\bibinfo {author} {\bibfnamefont {A.}~\bibnamefont
  {Smerald}}\ and\ \bibinfo {author} {\bibfnamefont {F.}~\bibnamefont {Mila}},\
  }\href {https://doi.org/10.21468/SciPostPhys.5.3.030} {\bibfield  {journal}
  {\bibinfo  {journal} {SciPost Phys.}\ }\textbf {\bibinfo {volume} {5}},\
  \bibinfo {pages} {030} (\bibinfo {year} {2018})}\BibitemShut {NoStop}%
\bibitem [{\citenamefont {Saadatmand}\ \emph {et~al.}(2018)\citenamefont
  {Saadatmand}, \citenamefont {Bartlett},\ and\ \citenamefont
  {McCulloch}}]{Saadatmand2018}%
  \BibitemOpen
  \bibfield  {author} {\bibinfo {author} {\bibfnamefont {S.~N.}\ \bibnamefont
  {Saadatmand}}, \bibinfo {author} {\bibfnamefont {S.~D.}\ \bibnamefont
  {Bartlett}},\ and\ \bibinfo {author} {\bibfnamefont {I.~P.}\ \bibnamefont
  {McCulloch}},\ }\href {https://doi.org/10.1103/PhysRevB.97.155116} {\bibfield
   {journal} {\bibinfo  {journal} {Phys. Rev. B}\ }\textbf {\bibinfo {volume}
  {97}},\ \bibinfo {pages} {155116} (\bibinfo {year} {2018})}\BibitemShut
  {NoStop}%
\bibitem [{\citenamefont {Koziol}\ \emph {et~al.}(2019)\citenamefont {Koziol},
  \citenamefont {Fey}, \citenamefont {Kapfer},\ and\ \citenamefont
  {Schmidt}}]{Koziol2019}%
  \BibitemOpen
  \bibfield  {author} {\bibinfo {author} {\bibfnamefont {J.}~\bibnamefont
  {Koziol}}, \bibinfo {author} {\bibfnamefont {S.}~\bibnamefont {Fey}},
  \bibinfo {author} {\bibfnamefont {S.~C.}\ \bibnamefont {Kapfer}},\ and\
  \bibinfo {author} {\bibfnamefont {K.~P.}\ \bibnamefont {Schmidt}},\ }\href
  {https://doi.org/10.1103/PhysRevB.100.144411} {\bibfield  {journal} {\bibinfo
   {journal} {Phys. Rev. B}\ }\textbf {\bibinfo {volume} {100}},\ \bibinfo
  {pages} {144411} (\bibinfo {year} {2019})}\BibitemShut {NoStop}%
\bibitem [{\citenamefont {Rokhsar}\ and\ \citenamefont
  {Kivelson}(1988)}]{Rokhsar1988}%
  \BibitemOpen
  \bibfield  {author} {\bibinfo {author} {\bibfnamefont {D.~S.}\ \bibnamefont
  {Rokhsar}}\ and\ \bibinfo {author} {\bibfnamefont {S.~A.}\ \bibnamefont
  {Kivelson}},\ }\href
  {https://doi.org/https://doi.org/10.1103/PhysRevLett.61.2376} {\bibfield
  {journal} {\bibinfo  {journal} {Physical Review Letters}\ }\textbf {\bibinfo
  {volume} {61}},\ \bibinfo {pages} {2376} (\bibinfo {year}
  {1988})}\BibitemShut {NoStop}%
\bibitem [{\citenamefont {Read}\ and\ \citenamefont
  {Sachdev}(1990)}]{Read1990}%
  \BibitemOpen
  \bibfield  {author} {\bibinfo {author} {\bibfnamefont {N.}~\bibnamefont
  {Read}}\ and\ \bibinfo {author} {\bibfnamefont {S.}~\bibnamefont {Sachdev}},\
  }\href {https://doi.org/10.1103/PhysRevB.42.4568} {\bibfield  {journal}
  {\bibinfo  {journal} {Phys. Rev. B}\ }\textbf {\bibinfo {volume} {42}},\
  \bibinfo {pages} {4568} (\bibinfo {year} {1990})}\BibitemShut {NoStop}%
\bibitem [{\citenamefont {Moessner}\ \emph {et~al.}(2001)\citenamefont
  {Moessner}, \citenamefont {Sondhi},\ and\ \citenamefont
  {Chandra}}]{Moessner2001new}%
  \BibitemOpen
  \bibfield  {author} {\bibinfo {author} {\bibfnamefont {R.}~\bibnamefont
  {Moessner}}, \bibinfo {author} {\bibfnamefont {S.~L.}\ \bibnamefont
  {Sondhi}},\ and\ \bibinfo {author} {\bibfnamefont {P.}~\bibnamefont
  {Chandra}},\ }\href
  {https://doi.org/https://doi.org/10.1103/PhysRevB.64.144416} {\bibfield
  {journal} {\bibinfo  {journal} {Physical Review B}\ }\textbf {\bibinfo
  {volume} {64}},\ \bibinfo {pages} {144416} (\bibinfo {year}
  {2001})}\BibitemShut {NoStop}%
\bibitem [{\citenamefont {Fradkin}\ \emph {et~al.}(2004)\citenamefont
  {Fradkin}, \citenamefont {Huse}, \citenamefont {Moessner}, \citenamefont
  {Oganesyan},\ and\ \citenamefont {Sondhi}}]{Fradkin2004}%
  \BibitemOpen
  \bibfield  {author} {\bibinfo {author} {\bibfnamefont {E.}~\bibnamefont
  {Fradkin}}, \bibinfo {author} {\bibfnamefont {D.~A.}\ \bibnamefont {Huse}},
  \bibinfo {author} {\bibfnamefont {R.}~\bibnamefont {Moessner}}, \bibinfo
  {author} {\bibfnamefont {V.}~\bibnamefont {Oganesyan}},\ and\ \bibinfo
  {author} {\bibfnamefont {S.~L.}\ \bibnamefont {Sondhi}},\ }\href
  {https://doi.org/10.1103/PhysRevB.69.224415} {\bibfield  {journal} {\bibinfo
  {journal} {Phys. Rev. B}\ }\textbf {\bibinfo {volume} {69}},\ \bibinfo
  {pages} {224415} (\bibinfo {year} {2004})}\BibitemShut {NoStop}%
\bibitem [{\citenamefont {Knetter}\ and\ \citenamefont
  {Uhrig}(2000)}]{Knetter2000}%
  \BibitemOpen
  \bibfield  {author} {\bibinfo {author} {\bibfnamefont {C.}~\bibnamefont
  {Knetter}}\ and\ \bibinfo {author} {\bibfnamefont {G.~S.}\ \bibnamefont
  {Uhrig}},\ }\href {https://doi.org/10.1007/s100510050026} {\bibfield
  {journal} {\bibinfo  {journal} {The European Physical Journal B}\ }\textbf
  {\bibinfo {volume} {13}},\ \bibinfo {pages} {209} (\bibinfo {year}
  {2000})}\BibitemShut {NoStop}%
\bibitem [{\citenamefont {Knetter}\ \emph {et~al.}(2003)\citenamefont
  {Knetter}, \citenamefont {Schmidt},\ and\ \citenamefont
  {Uhrig}}]{Knetter2003}%
  \BibitemOpen
  \bibfield  {author} {\bibinfo {author} {\bibfnamefont {C.}~\bibnamefont
  {Knetter}}, \bibinfo {author} {\bibfnamefont {K.~P.}\ \bibnamefont
  {Schmidt}},\ and\ \bibinfo {author} {\bibfnamefont {G.~S.}\ \bibnamefont
  {Uhrig}},\ }\href {https://doi.org/10.1088/0305-4470/36/29/302} {\bibfield
  {journal} {\bibinfo  {journal} {Journal of Physics A: Mathematical and
  General}\ }\textbf {\bibinfo {volume} {36}},\ \bibinfo {pages} {7889}
  (\bibinfo {year} {2003})}\BibitemShut {NoStop}%
\bibitem [{\citenamefont {M\"uhlhauser}\ and\ \citenamefont
  {Schmidt}(2022)}]{Muehlhauser2022}%
  \BibitemOpen
  \bibfield  {author} {\bibinfo {author} {\bibfnamefont {M.}~\bibnamefont
  {M\"uhlhauser}}\ and\ \bibinfo {author} {\bibfnamefont {K.~P.}\ \bibnamefont
  {Schmidt}},\ }\href {https://doi.org/10.1103/PhysRevE.105.064110} {\bibfield
  {journal} {\bibinfo  {journal} {Phys. Rev. E}\ }\textbf {\bibinfo {volume}
  {105}},\ \bibinfo {pages} {064110} (\bibinfo {year} {2022})}\BibitemShut
  {NoStop}%
\bibitem [{\citenamefont {Koziol}\ \emph {et~al.}(2024)\citenamefont {Koziol},
  \citenamefont {Mühlhauser},\ and\ \citenamefont {Schmidt}}]{Koziol2024}%
  \BibitemOpen
  \bibfield  {author} {\bibinfo {author} {\bibfnamefont {J.~A.}\ \bibnamefont
  {Koziol}}, \bibinfo {author} {\bibfnamefont {M.}~\bibnamefont
  {Mühlhauser}},\ and\ \bibinfo {author} {\bibfnamefont {K.~P.}\ \bibnamefont
  {Schmidt}},\ }\href@noop {} {\bibinfo {title} {Order-by-disorder and
  long-range interactions in the antiferromagnetic transverse-field ising model
  on the triangular lattice -- a perturbative point of view}} (\bibinfo {year}
  {2024}),\ \Eprint {https://arxiv.org/abs/2402.10584} {arXiv:2402.10584
  [cond-mat.str-el]} \BibitemShut {NoStop}%
\bibitem [{\citenamefont {Takahashi}(1977)}]{Takahashi1977}%
  \BibitemOpen
  \bibfield  {author} {\bibinfo {author} {\bibfnamefont {M.}~\bibnamefont
  {Takahashi}},\ }\href {https://doi.org/10.1088/0022-3719/10/8/031} {\bibfield
   {journal} {\bibinfo  {journal} {Journal of Physics C: Solid State Physics}\
  }\textbf {\bibinfo {volume} {10}},\ \bibinfo {pages} {1289} (\bibinfo {year}
  {1977})}\BibitemShut {NoStop}%
\bibitem [{sup()}]{supplementary}%
  \BibitemOpen
  \href@noop {} {}\bibinfo {note} {See Supplemental Material at URL for a
  description of the perturbative approach, details on the analysis of the
  low-field limit, and comprehensive results obtained from the high-field
  series expansions.}\BibitemShut {Stop}%
\bibitem [{\citenamefont {C{\"o}ster}\ and\ \citenamefont
  {Schmidt}(2015)}]{Coester2015}%
  \BibitemOpen
  \bibfield  {author} {\bibinfo {author} {\bibfnamefont {K.}~\bibnamefont
  {C{\"o}ster}}\ and\ \bibinfo {author} {\bibfnamefont {K.~P.}\ \bibnamefont
  {Schmidt}},\ }\href
  {https://doi.org/https://doi.org/10.1103/PhysRevE.92.022118} {\bibfield
  {journal} {\bibinfo  {journal} {Physical Review E}\ }\textbf {\bibinfo
  {volume} {92}},\ \bibinfo {pages} {022118} (\bibinfo {year}
  {2015})}\BibitemShut {NoStop}%
\bibitem [{\citenamefont {Gelfand}\ and\ \citenamefont
  {Singh}(2000)}]{Gelfand2000}%
  \BibitemOpen
  \bibfield  {author} {\bibinfo {author} {\bibfnamefont {M.~P.}\ \bibnamefont
  {Gelfand}}\ and\ \bibinfo {author} {\bibfnamefont {R.~R.~P.}\ \bibnamefont
  {Singh}},\ }\href {https://doi.org/10.1080/000187300243390} {\bibfield
  {journal} {\bibinfo  {journal} {Advances in Physics}\ }\textbf {\bibinfo
  {volume} {49}},\ \bibinfo {pages} {93–140} (\bibinfo {year}
  {2000})}\BibitemShut {NoStop}%
\bibitem [{\citenamefont {Oitmaa}\ \emph {et~al.}(2006)\citenamefont {Oitmaa},
  \citenamefont {Hamer},\ and\ \citenamefont {Zheng}}]{Oitmaa2006}%
  \BibitemOpen
  \bibfield  {author} {\bibinfo {author} {\bibfnamefont {J.}~\bibnamefont
  {Oitmaa}}, \bibinfo {author} {\bibfnamefont {C.}~\bibnamefont {Hamer}},\ and\
  \bibinfo {author} {\bibfnamefont {W.}~\bibnamefont {Zheng}},\ }\href
  {https://doi.org/10.1017/CBO9780511584398} {\emph {\bibinfo {title} {Series
  Expansion Methods for Strongly Interacting Lattice Models}}}\ (\bibinfo
  {publisher} {Cambridge University Press},\ \bibinfo {year}
  {2006})\BibitemShut {NoStop}%
\bibitem [{\citenamefont {Baker}\ and\ \citenamefont
  {Graves-Morris}(1996)}]{Baker1996}%
  \BibitemOpen
  \bibfield  {author} {\bibinfo {author} {\bibfnamefont {G.~A.}\ \bibnamefont
  {Baker}}\ and\ \bibinfo {author} {\bibfnamefont {P.}~\bibnamefont
  {Graves-Morris}},\ }\href
  {https://doi.org/https://doi.org/10.1017/CBO9780511530074} {\emph {\bibinfo
  {title} {Pad{\'{e}} Approximants}}},\ \bibinfo {edition} {2nd}\ ed.,\
  Encyclopedia of Mathematics and its Applications\ (\bibinfo  {publisher}
  {Cambridge University Press},\ \bibinfo {year} {1996})\BibitemShut {NoStop}%
\bibitem [{\citenamefont {Guttmann}(1989)}]{Guttmann1989}%
  \BibitemOpen
  \bibfield  {author} {\bibinfo {author} {\bibfnamefont {A.~J.}\ \bibnamefont
  {Guttmann}},\ }in\ \href@noop {} {\emph {\bibinfo {booktitle} {Phase
  Transitions and Critical Phenomena}}},\ Vol.~\bibinfo {volume} {13},\
  \bibinfo {editor} {edited by\ \bibinfo {editor} {\bibfnamefont
  {C.}~\bibnamefont {Domb}}, \bibinfo {editor} {\bibfnamefont {M.~S.}\
  \bibnamefont {Green}},\ and\ \bibinfo {editor} {\bibfnamefont {J.~L.}\
  \bibnamefont {Lebowitz}}}\ (\bibinfo  {publisher} {Academic Press},\ \bibinfo
  {year} {1989})\BibitemShut {NoStop}%
\bibitem [{\citenamefont {Hasenbusch}(2019)}]{Hasenbusch2019}%
  \BibitemOpen
  \bibfield  {author} {\bibinfo {author} {\bibfnamefont {M.}~\bibnamefont
  {Hasenbusch}},\ }\href {https://doi.org/10.1103/PhysRevB.100.224517}
  {\bibfield  {journal} {\bibinfo  {journal} {Phys. Rev. B}\ }\textbf {\bibinfo
  {volume} {100}},\ \bibinfo {pages} {224517} (\bibinfo {year}
  {2019})}\BibitemShut {NoStop}%
\bibitem [{\citenamefont {Chester}\ \emph {et~al.}(2020)\citenamefont
  {Chester}, \citenamefont {Landry}, \citenamefont {Liu}, \citenamefont
  {Poland}, \citenamefont {Simmons-Duffin}, \citenamefont {Su},\ and\
  \citenamefont {Vichi}}]{Chester2020}%
  \BibitemOpen
  \bibfield  {author} {\bibinfo {author} {\bibfnamefont {S.~M.}\ \bibnamefont
  {Chester}}, \bibinfo {author} {\bibfnamefont {W.}~\bibnamefont {Landry}},
  \bibinfo {author} {\bibfnamefont {J.}~\bibnamefont {Liu}}, \bibinfo {author}
  {\bibfnamefont {D.}~\bibnamefont {Poland}}, \bibinfo {author} {\bibfnamefont
  {D.}~\bibnamefont {Simmons-Duffin}}, \bibinfo {author} {\bibfnamefont
  {N.}~\bibnamefont {Su}},\ and\ \bibinfo {author} {\bibfnamefont
  {A.}~\bibnamefont {Vichi}},\ }\href {https://doi.org/10.1007/JHEP06(2020)142}
  {\bibfield  {journal} {\bibinfo  {journal} {Journal of High Energy Physics}\
  }\textbf {\bibinfo {volume} {2020}},\ \bibinfo {pages} {142} (\bibinfo {year}
  {2020})}\BibitemShut {NoStop}%
\bibitem [{\citenamefont {Kos}\ \emph {et~al.}(2016)\citenamefont {Kos},
  \citenamefont {Poland}, \citenamefont {Simmons-Duffin},\ and\ \citenamefont
  {Vichi}}]{Kos2016}%
  \BibitemOpen
  \bibfield  {author} {\bibinfo {author} {\bibfnamefont {F.}~\bibnamefont
  {Kos}}, \bibinfo {author} {\bibfnamefont {D.}~\bibnamefont {Poland}},
  \bibinfo {author} {\bibfnamefont {D.}~\bibnamefont {Simmons-Duffin}},\ and\
  \bibinfo {author} {\bibfnamefont {A.}~\bibnamefont {Vichi}},\ }\href
  {https://doi.org/10.1007/JHEP08(2016)036} {\bibfield  {journal} {\bibinfo
  {journal} {Journal of High Energy Physics}\ }\textbf {\bibinfo {volume}
  {2016}},\ \bibinfo {pages} {36} (\bibinfo {year} {2016})}\BibitemShut
  {NoStop}%
\bibitem [{\citenamefont {Tarabunga}\ \emph {et~al.}(2022)\citenamefont
  {Tarabunga}, \citenamefont {Surace}, \citenamefont {Andreoni}, \citenamefont
  {Angelone},\ and\ \citenamefont {Dalmonte}}]{Tarabunga2022}%
  \BibitemOpen
  \bibfield  {author} {\bibinfo {author} {\bibfnamefont {P.~S.}\ \bibnamefont
  {Tarabunga}}, \bibinfo {author} {\bibfnamefont {F.~M.}\ \bibnamefont
  {Surace}}, \bibinfo {author} {\bibfnamefont {R.}~\bibnamefont {Andreoni}},
  \bibinfo {author} {\bibfnamefont {A.}~\bibnamefont {Angelone}},\ and\
  \bibinfo {author} {\bibfnamefont {M.}~\bibnamefont {Dalmonte}},\ }\href
  {https://doi.org/10.1103/PhysRevLett.129.195301} {\bibfield  {journal}
  {\bibinfo  {journal} {Phys. Rev. Lett.}\ }\textbf {\bibinfo {volume} {129}},\
  \bibinfo {pages} {195301} (\bibinfo {year} {2022})}\BibitemShut {NoStop}%
\bibitem [{\citenamefont {Patil}(2023)}]{Patil2023}%
  \BibitemOpen
  \bibfield  {author} {\bibinfo {author} {\bibfnamefont {P.}~\bibnamefont
  {Patil}},\ }\href@noop {} {\bibinfo {title} {Quantum monte carlo simulations
  in the restricted hilbert space of rydberg atom arrays}} (\bibinfo {year}
  {2023}),\ \Eprint {https://arxiv.org/abs/2309.00482} {arXiv:2309.00482
  [cond-mat.str-el]} \BibitemShut {NoStop}%
\bibitem [{\citenamefont {Schauß}\ \emph {et~al.}(2012)\citenamefont
  {Schauß}, \citenamefont {Cheneau}, \citenamefont {Endres}, \citenamefont
  {Fukuhara}, \citenamefont {Hild}, \citenamefont {Omran}, \citenamefont
  {Pohl}, \citenamefont {Gross}, \citenamefont {Kuhr},\ and\ \citenamefont
  {Bloch}}]{Schauss2012}%
  \BibitemOpen
  \bibfield  {author} {\bibinfo {author} {\bibfnamefont {P.}~\bibnamefont
  {Schauß}}, \bibinfo {author} {\bibfnamefont {M.}~\bibnamefont {Cheneau}},
  \bibinfo {author} {\bibfnamefont {M.}~\bibnamefont {Endres}}, \bibinfo
  {author} {\bibfnamefont {T.}~\bibnamefont {Fukuhara}}, \bibinfo {author}
  {\bibfnamefont {S.}~\bibnamefont {Hild}}, \bibinfo {author} {\bibfnamefont
  {A.}~\bibnamefont {Omran}}, \bibinfo {author} {\bibfnamefont
  {T.}~\bibnamefont {Pohl}}, \bibinfo {author} {\bibfnamefont {C.}~\bibnamefont
  {Gross}}, \bibinfo {author} {\bibfnamefont {S.}~\bibnamefont {Kuhr}},\ and\
  \bibinfo {author} {\bibfnamefont {I.}~\bibnamefont {Bloch}},\ }\href
  {https://doi.org/10.1038/nature11596} {\bibfield  {journal} {\bibinfo
  {journal} {Nature}\ }\textbf {\bibinfo {volume} {491}},\ \bibinfo {pages}
  {87–91} (\bibinfo {year} {2012})}\BibitemShut {NoStop}%
\bibitem [{\citenamefont {Zeiher}\ \emph {et~al.}(2015)\citenamefont {Zeiher},
  \citenamefont {Schau\ss{}}, \citenamefont {Hild}, \citenamefont {Macr\`{\i}},
  \citenamefont {Bloch},\ and\ \citenamefont {Gross}}]{Zeiher2015}%
  \BibitemOpen
  \bibfield  {author} {\bibinfo {author} {\bibfnamefont {J.}~\bibnamefont
  {Zeiher}}, \bibinfo {author} {\bibfnamefont {P.}~\bibnamefont {Schau\ss{}}},
  \bibinfo {author} {\bibfnamefont {S.}~\bibnamefont {Hild}}, \bibinfo {author}
  {\bibfnamefont {T.}~\bibnamefont {Macr\`{\i}}}, \bibinfo {author}
  {\bibfnamefont {I.}~\bibnamefont {Bloch}},\ and\ \bibinfo {author}
  {\bibfnamefont {C.}~\bibnamefont {Gross}},\ }\href
  {https://doi.org/10.1103/PhysRevX.5.031015} {\bibfield  {journal} {\bibinfo
  {journal} {Phys. Rev. X}\ }\textbf {\bibinfo {volume} {5}},\ \bibinfo {pages}
  {031015} (\bibinfo {year} {2015})}\BibitemShut {NoStop}%
\bibitem [{\citenamefont {Schauß}\ \emph {et~al.}(2015)\citenamefont
  {Schauß}, \citenamefont {Zeiher}, \citenamefont {Fukuhara}, \citenamefont
  {Hild}, \citenamefont {Cheneau}, \citenamefont {Macrì}, \citenamefont
  {Pohl}, \citenamefont {Bloch},\ and\ \citenamefont {Gross}}]{Schauss2015}%
  \BibitemOpen
  \bibfield  {author} {\bibinfo {author} {\bibfnamefont {P.}~\bibnamefont
  {Schauß}}, \bibinfo {author} {\bibfnamefont {J.}~\bibnamefont {Zeiher}},
  \bibinfo {author} {\bibfnamefont {T.}~\bibnamefont {Fukuhara}}, \bibinfo
  {author} {\bibfnamefont {S.}~\bibnamefont {Hild}}, \bibinfo {author}
  {\bibfnamefont {M.}~\bibnamefont {Cheneau}}, \bibinfo {author} {\bibfnamefont
  {T.}~\bibnamefont {Macrì}}, \bibinfo {author} {\bibfnamefont
  {T.}~\bibnamefont {Pohl}}, \bibinfo {author} {\bibfnamefont {I.}~\bibnamefont
  {Bloch}},\ and\ \bibinfo {author} {\bibfnamefont {C.}~\bibnamefont {Gross}},\
  }\href {https://doi.org/10.1126/science.1258351} {\bibfield  {journal}
  {\bibinfo  {journal} {Science}\ }\textbf {\bibinfo {volume} {347}},\ \bibinfo
  {pages} {1455} (\bibinfo {year} {2015})},\ \Eprint
  {https://arxiv.org/abs/https://www.science.org/doi/pdf/10.1126/science.1258351}
  {https://www.science.org/doi/pdf/10.1126/science.1258351} \BibitemShut
  {NoStop}%
\bibitem [{\citenamefont {Lienhard}\ \emph {et~al.}(2018)\citenamefont
  {Lienhard}, \citenamefont {de~L\'es\'eleuc}, \citenamefont {Barredo},
  \citenamefont {Lahaye}, \citenamefont {Browaeys}, \citenamefont {Schuler},
  \citenamefont {Henry},\ and\ \citenamefont {L\"auchli}}]{Lienhardt2018}%
  \BibitemOpen
  \bibfield  {author} {\bibinfo {author} {\bibfnamefont {V.}~\bibnamefont
  {Lienhard}}, \bibinfo {author} {\bibfnamefont {S.}~\bibnamefont
  {de~L\'es\'eleuc}}, \bibinfo {author} {\bibfnamefont {D.}~\bibnamefont
  {Barredo}}, \bibinfo {author} {\bibfnamefont {T.}~\bibnamefont {Lahaye}},
  \bibinfo {author} {\bibfnamefont {A.}~\bibnamefont {Browaeys}}, \bibinfo
  {author} {\bibfnamefont {M.}~\bibnamefont {Schuler}}, \bibinfo {author}
  {\bibfnamefont {L.-P.}\ \bibnamefont {Henry}},\ and\ \bibinfo {author}
  {\bibfnamefont {A.~M.}\ \bibnamefont {L\"auchli}},\ }\href
  {https://doi.org/10.1103/PhysRevX.8.021070} {\bibfield  {journal} {\bibinfo
  {journal} {Phys. Rev. X}\ }\textbf {\bibinfo {volume} {8}},\ \bibinfo {pages}
  {021070} (\bibinfo {year} {2018})}\BibitemShut {NoStop}%
\end{thebibliography}

\begin{thebibliography}{14}%
\makeatletter
\providecommand \@ifxundefined [1]{%
 \@ifx{#1\undefined}
}%
\providecommand \@ifnum [1]{%
 \ifnum #1\expandafter \@firstoftwo
 \else \expandafter \@secondoftwo
 \fi
}%
\providecommand \@ifx [1]{%
 \ifx #1\expandafter \@firstoftwo
 \else \expandafter \@secondoftwo
 \fi
}%
\providecommand \natexlab [1]{#1}%
\providecommand \enquote  [1]{``#1''}%
\providecommand \bibnamefont  [1]{#1}%
\providecommand \bibfnamefont [1]{#1}%
\providecommand \citenamefont [1]{#1}%
\providecommand \href@noop [0]{\@secondoftwo}%
\providecommand \href [0]{\begingroup \@sanitize@url \@href}%
\providecommand \@href[1]{\@@startlink{#1}\@@href}%
\providecommand \@@href[1]{\endgroup#1\@@endlink}%
\providecommand \@sanitize@url [0]{\catcode `\\12\catcode `\$12\catcode
  `\&12\catcode `\#12\catcode `\^12\catcode `\_12\catcode `\%12\relax}%
\providecommand \@@startlink[1]{}%
\providecommand \@@endlink[0]{}%
\providecommand \url  [0]{\begingroup\@sanitize@url \@url }%
\providecommand \@url [1]{\endgroup\@href {#1}{\urlprefix }}%
\providecommand \urlprefix  [0]{URL }%
\providecommand \Eprint [0]{\href }%
\providecommand \doibase [0]{https://doi.org/}%
\providecommand \selectlanguage [0]{\@gobble}%
\providecommand \bibinfo  [0]{\@secondoftwo}%
\providecommand \bibfield  [0]{\@secondoftwo}%
\providecommand \translation [1]{[#1]}%
\providecommand \BibitemOpen [0]{}%
\providecommand \bibitemStop [0]{}%
\providecommand \bibitemNoStop [0]{.\EOS\space}%
\providecommand \EOS [0]{\spacefactor3000\relax}%
\providecommand \BibitemShut  [1]{\csname bibitem#1\endcsname}%
\let\auto@bib@innerbib\@empty
%</preamble>
\bibitem [{\citenamefont {Takahashi}(1977)}]{Takahashi1977s}%
  \BibitemOpen
  \bibfield  {author} {\bibinfo {author} {\bibfnamefont {M.}~\bibnamefont
  {Takahashi}},\ }\href {https://doi.org/10.1088/0022-3719/10/8/031} {\bibfield
   {journal} {\bibinfo  {journal} {Journal of Physics C: Solid State Physics}\
  }\textbf {\bibinfo {volume} {10}},\ \bibinfo {pages} {1289} (\bibinfo {year}
  {1977})}\BibitemShut {NoStop}%
\bibitem [{\citenamefont {Knetter}\ and\ \citenamefont
  {Uhrig}(2000)}]{Knetter2000s}%
  \BibitemOpen
  \bibfield  {author} {\bibinfo {author} {\bibfnamefont {C.}~\bibnamefont
  {Knetter}}\ and\ \bibinfo {author} {\bibfnamefont {G.~S.}\ \bibnamefont
  {Uhrig}},\ }\href {https://doi.org/10.1007/s100510050026} {\bibfield
  {journal} {\bibinfo  {journal} {The European Physical Journal B}\ }\textbf
  {\bibinfo {volume} {13}},\ \bibinfo {pages} {209} (\bibinfo {year}
  {2000})}\BibitemShut {NoStop}%
\bibitem [{\citenamefont {Knetter}\ \emph {et~al.}(2003)\citenamefont
  {Knetter}, \citenamefont {Schmidt},\ and\ \citenamefont
  {Uhrig}}]{Knetter2003s}%
  \BibitemOpen
  \bibfield  {author} {\bibinfo {author} {\bibfnamefont {C.}~\bibnamefont
  {Knetter}}, \bibinfo {author} {\bibfnamefont {K.~P.}\ \bibnamefont
  {Schmidt}},\ and\ \bibinfo {author} {\bibfnamefont {G.~S.}\ \bibnamefont
  {Uhrig}},\ }\href {https://doi.org/10.1088/0305-4470/36/29/302} {\bibfield
  {journal} {\bibinfo  {journal} {Journal of Physics A: Mathematical and
  General}\ }\textbf {\bibinfo {volume} {36}},\ \bibinfo {pages} {7889}
  (\bibinfo {year} {2003})}\BibitemShut {NoStop}%
\bibitem [{\citenamefont {C{\"o}ster}\ and\ \citenamefont
  {Schmidt}(2015)}]{Coester2015s}%
  \BibitemOpen
  \bibfield  {author} {\bibinfo {author} {\bibfnamefont {K.}~\bibnamefont
  {C{\"o}ster}}\ and\ \bibinfo {author} {\bibfnamefont {K.~P.}\ \bibnamefont
  {Schmidt}},\ }\href
  {https://doi.org/https://doi.org/10.1103/PhysRevE.92.022118} {\bibfield
  {journal} {\bibinfo  {journal} {Physical Review E}\ }\textbf {\bibinfo
  {volume} {92}},\ \bibinfo {pages} {022118} (\bibinfo {year}
  {2015})}\BibitemShut {NoStop}%
\bibitem [{\citenamefont {M\"uhlhauser}\ and\ \citenamefont
  {Schmidt}(2022)}]{Muehlhauser2022s}%
  \BibitemOpen
  \bibfield  {author} {\bibinfo {author} {\bibfnamefont {M.}~\bibnamefont
  {M\"uhlhauser}}\ and\ \bibinfo {author} {\bibfnamefont {K.~P.}\ \bibnamefont
  {Schmidt}},\ }\href {https://doi.org/10.1103/PhysRevE.105.064110} {\bibfield
  {journal} {\bibinfo  {journal} {Phys. Rev. E}\ }\textbf {\bibinfo {volume}
  {105}},\ \bibinfo {pages} {064110} (\bibinfo {year} {2022})}\BibitemShut
  {NoStop}%
\bibitem [{\citenamefont {Gelfand}\ and\ \citenamefont
  {Singh}(2000)}]{Gelfand2000s}%
  \BibitemOpen
  \bibfield  {author} {\bibinfo {author} {\bibfnamefont {M.~P.}\ \bibnamefont
  {Gelfand}}\ and\ \bibinfo {author} {\bibfnamefont {R.~R.~P.}\ \bibnamefont
  {Singh}},\ }\href {https://doi.org/10.1080/000187300243390} {\bibfield
  {journal} {\bibinfo  {journal} {Advances in Physics}\ }\textbf {\bibinfo
  {volume} {49}},\ \bibinfo {pages} {93–140} (\bibinfo {year}
  {2000})}\BibitemShut {NoStop}%
\bibitem [{\citenamefont {Oitmaa}\ \emph {et~al.}(2006)\citenamefont {Oitmaa},
  \citenamefont {Hamer},\ and\ \citenamefont {Zheng}}]{Oitmaa2006s}%
  \BibitemOpen
  \bibfield  {author} {\bibinfo {author} {\bibfnamefont {J.}~\bibnamefont
  {Oitmaa}}, \bibinfo {author} {\bibfnamefont {C.}~\bibnamefont {Hamer}},\ and\
  \bibinfo {author} {\bibfnamefont {W.}~\bibnamefont {Zheng}},\ }\href
  {https://doi.org/10.1017/CBO9780511584398} {\emph {\bibinfo {title} {Series
  Expansion Methods for Strongly Interacting Lattice Models}}}\ (\bibinfo
  {publisher} {Cambridge University Press},\ \bibinfo {year}
  {2006})\BibitemShut {NoStop}%
\bibitem [{\citenamefont {Baker}\ and\ \citenamefont
  {Graves-Morris}(1996)}]{Baker1996s}%
  \BibitemOpen
  \bibfield  {author} {\bibinfo {author} {\bibfnamefont {G.~A.}\ \bibnamefont
  {Baker}}\ and\ \bibinfo {author} {\bibfnamefont {P.}~\bibnamefont
  {Graves-Morris}},\ }\href
  {https://doi.org/https://doi.org/10.1017/CBO9780511530074} {\emph {\bibinfo
  {title} {Pad{\'{e}} Approximants}}},\ \bibinfo {edition} {2nd}\ ed.,\
  Encyclopedia of Mathematics and its Applications\ (\bibinfo  {publisher}
  {Cambridge University Press},\ \bibinfo {year} {1996})\BibitemShut {NoStop}%
\bibitem [{\citenamefont {Guttmann}(1989)}]{Guttmann1989s}%
  \BibitemOpen
  \bibfield  {author} {\bibinfo {author} {\bibfnamefont {A.~J.}\ \bibnamefont
  {Guttmann}},\ }in\ \href@noop {} {\emph {\bibinfo {booktitle} {Phase
  Transitions and Critical Phenomena}}},\ Vol.~\bibinfo {volume} {13},\
  \bibinfo {editor} {edited by\ \bibinfo {editor} {\bibfnamefont
  {C.}~\bibnamefont {Domb}}, \bibinfo {editor} {\bibfnamefont {M.~S.}\
  \bibnamefont {Green}},\ and\ \bibinfo {editor} {\bibfnamefont {J.~L.}\
  \bibnamefont {Lebowitz}}}\ (\bibinfo  {publisher} {Academic Press},\ \bibinfo
  {year} {1989})\BibitemShut {NoStop}%
\bibitem [{\citenamefont {Rokhsar}\ and\ \citenamefont
  {Kivelson}(1988)}]{Rokhsar1988s}%
  \BibitemOpen
  \bibfield  {author} {\bibinfo {author} {\bibfnamefont {D.~S.}\ \bibnamefont
  {Rokhsar}}\ and\ \bibinfo {author} {\bibfnamefont {S.~A.}\ \bibnamefont
  {Kivelson}},\ }\href
  {https://doi.org/https://doi.org/10.1103/PhysRevLett.61.2376} {\bibfield
  {journal} {\bibinfo  {journal} {Physical Review Letters}\ }\textbf {\bibinfo
  {volume} {61}},\ \bibinfo {pages} {2376} (\bibinfo {year}
  {1988})}\BibitemShut {NoStop}%
\bibitem [{\citenamefont {Moessner}\ \emph {et~al.}(2001)\citenamefont
  {Moessner}, \citenamefont {Sondhi},\ and\ \citenamefont
  {Chandra}}]{Moessner2001news}%
  \BibitemOpen
  \bibfield  {author} {\bibinfo {author} {\bibfnamefont {R.}~\bibnamefont
  {Moessner}}, \bibinfo {author} {\bibfnamefont {S.~L.}\ \bibnamefont
  {Sondhi}},\ and\ \bibinfo {author} {\bibfnamefont {P.}~\bibnamefont
  {Chandra}},\ }\href
  {https://doi.org/https://doi.org/10.1103/PhysRevB.64.144416} {\bibfield
  {journal} {\bibinfo  {journal} {Physical Review B}\ }\textbf {\bibinfo
  {volume} {64}},\ \bibinfo {pages} {144416} (\bibinfo {year}
  {2001})}\BibitemShut {NoStop}%
\bibitem [{\citenamefont {Hasenbusch}(2019)}]{Hasenbusch2019s}%
  \BibitemOpen
  \bibfield  {author} {\bibinfo {author} {\bibfnamefont {M.}~\bibnamefont
  {Hasenbusch}},\ }\href {https://doi.org/10.1103/PhysRevB.100.224517}
  {\bibfield  {journal} {\bibinfo  {journal} {Phys. Rev. B}\ }\textbf {\bibinfo
  {volume} {100}},\ \bibinfo {pages} {224517} (\bibinfo {year}
  {2019})}\BibitemShut {NoStop}%
\bibitem [{\citenamefont {Chester}\ \emph {et~al.}(2020)\citenamefont
  {Chester}, \citenamefont {Landry}, \citenamefont {Liu}, \citenamefont
  {Poland}, \citenamefont {Simmons-Duffin}, \citenamefont {Su},\ and\
  \citenamefont {Vichi}}]{Chester2020s}%
  \BibitemOpen
  \bibfield  {author} {\bibinfo {author} {\bibfnamefont {S.~M.}\ \bibnamefont
  {Chester}}, \bibinfo {author} {\bibfnamefont {W.}~\bibnamefont {Landry}},
  \bibinfo {author} {\bibfnamefont {J.}~\bibnamefont {Liu}}, \bibinfo {author}
  {\bibfnamefont {D.}~\bibnamefont {Poland}}, \bibinfo {author} {\bibfnamefont
  {D.}~\bibnamefont {Simmons-Duffin}}, \bibinfo {author} {\bibfnamefont
  {N.}~\bibnamefont {Su}},\ and\ \bibinfo {author} {\bibfnamefont
  {A.}~\bibnamefont {Vichi}},\ }\href {https://doi.org/10.1007/JHEP06(2020)142}
  {\bibfield  {journal} {\bibinfo  {journal} {Journal of High Energy Physics}\
  }\textbf {\bibinfo {volume} {2020}},\ \bibinfo {pages} {142} (\bibinfo {year}
  {2020})}\BibitemShut {NoStop}%
\bibitem [{\citenamefont {Kos}\ \emph {et~al.}(2016)\citenamefont {Kos},
  \citenamefont {Poland}, \citenamefont {Simmons-Duffin},\ and\ \citenamefont
  {Vichi}}]{Kos2016s}%
  \BibitemOpen
  \bibfield  {author} {\bibinfo {author} {\bibfnamefont {F.}~\bibnamefont
  {Kos}}, \bibinfo {author} {\bibfnamefont {D.}~\bibnamefont {Poland}},
  \bibinfo {author} {\bibfnamefont {D.}~\bibnamefont {Simmons-Duffin}},\ and\
  \bibinfo {author} {\bibfnamefont {A.}~\bibnamefont {Vichi}},\ }\href
  {https://doi.org/10.1007/JHEP08(2016)036} {\bibfield  {journal} {\bibinfo
  {journal} {Journal of High Energy Physics}\ }\textbf {\bibinfo {volume}
  {2016}},\ \bibinfo {pages} {36} (\bibinfo {year} {2016})}\BibitemShut
  {NoStop}%
\end{thebibliography}

\newpage 
\clearpage

\end{document}